\documentclass[fleqn]{aa}  

\usepackage{supertabular,booktabs}  % For the big table
\usepackage{graphicx}
\usepackage{xcolor}
\usepackage{hyperref}
\usepackage{txfonts}
\usepackage{calligra}
\usepackage{calrsfs}
\usepackage[T1]{fontenc}
\usepackage[mathcal]{eucal}
\usepackage{longtable}
\usepackage{multicol}
\sfcode`A=1000 \sfcode`B=1000 \sfcode`C=1000 \sfcode`D=1000
\sfcode`E=1000 \sfcode`F=1000 \sfcode`G=1000 \sfcode`H=1000
\sfcode`I=1000 \sfcode`J=1000 \sfcode`K=1000 \sfcode`L=1000
\sfcode`M=1000 \sfcode`N=1000 \sfcode`O=1000 \sfcode`P=1000
\sfcode`Q=1000 \sfcode`R=1000 \sfcode`S=1000 \sfcode`T=1000
\sfcode`U=1000 \sfcode`V=1000 \sfcode`W=1000 \sfcode`X=1000
\sfcode`Y=1000 \sfcode`Z=1000

\newcommand{\vectX}{\bf {\it X}}
\newcommand{\vectTheta}{\bf {\it \Theta}}

\newcommand{\matrGamma}{\bf \Gamma}
\newcommand{\Lsun}{\hbox{L$_\sun$}}
\newcommand{\Msun}{\hbox{M$_\sun$}}

\newcommand{\gz}{geo-$z$}
\newcommand{\zs}{$z_{\rm spec}$}
\newcommand{\zg}{$z_{\rm geo}$}
\newcommand{\zp}{$z_{\rm phot}$}

\begin{document}

   \title{JWST's PEARLS: 119 multiply imaged galaxies behind MACS0416, lensing properties of caustic crossing galaxies, and the relation between halo mass and number of globular clusters at $z=0.4$}
   
   \titlerunning{A new JWST-based lens model for MACS0416}

   \author{Jose M. Diego \inst{1}\fnmsep\thanks{jdiego@ifca.unican.es}\fnmsep\thanks{Table B1 is only available in electronic form at the CDS via anonymous ftp to cdsarc.u-strasbg.fr (130.79.128.5) or via http://cdsweb.u-strasbg.fr/cgi-bin/qcat?J/A+A/}
       \and Nathan J. Adams \inst{2} 
       \and S. P. Willner \inst{3}   
       \and Tom Harvey \inst{2}       
       \and Tom Broadhurst \inst{4,5,6}      
\and Seth H. Cohen \inst{7} %%% seth.cohen@asu.edu
\and Rolf A. Jansen \inst{7} %%% rolfjansen.work@gmail.com
\and Jake Summers \inst{7} %%% jssumme1@asu.edu
\and Rogier A. Windhorst \inst{7} %%% Rogier.Windhorst@gmail.com
\and Jordan C. J. D'Silva \inst{8,9} %%% jordan.dsilva@research.uwa.edu.au
\and Anton M. Koekemoer \inst{10} %%% koekemoer@stsci.edu
\and Dan Coe \inst{10,11,12} %%% dcoe@stsci.edu
\and Christopher J. Conselice \inst{2} %%% conselice@gmail.com
\and Simon P. Driver \inst{8} %%% Simon.Driver@icrar.org
\and Brenda Frye \inst{13} %%% brendafrye@gmail.com
\and Norman A. Grogin \inst{10} %%% nagrogin@stsci.edu
\and Madeline A. Marshall \inst{14,15} %%% madeline_marshall@outlook.com
\and Mario Nonino \inst{16} %%% nnn.mario@gmail.com
\and Rafael Ortiz~III \inst{7} %%% rortizii@asu.edu
\and Nor Pirzkal \inst{10} %%% npirzkal@stsci.edu
\and Aaron Robotham \inst{8} %%% aaron.robotham@uwa.edu.au
\and Russell E. Ryan, Jr. \inst{10} %%% rryan.asu@stsci.edu
\and Christopher N. A. Willmer \inst{17} %%% cnawillmer@gmail.com
\and Haojing Yan \inst{18} %%% yanhaojing@gmail.com
\and Fengwu Sun \inst{17} %%% fengwusun@arizona.edu  (EAZY & MAGNIF)
\and Kevin Hainline  \inst{17} %%% (EAZY and MAGNIF)
\and Jessica Berkheimer \inst{7}
\and Maria del Carmen Polletta \inst{19}
\and Adi Zitrin \inst{20}
    }      
   \institute{Instituto de F\'isica de Cantabria (CSIC-UC). Avda. Los Castros s/n. 39005 Santander, Spain %1 Chema
%              \email{jdiego@ifca.unican.es2}
\and Jodrell Bank Centre for Astrophysics, Alan Turing Building, University of Manchester, Oxford Road, Manchester M13 9PL, UK  % 2) Nathan Adams
                                                                                                                                %    Tom Harvey
                                                                                                                                %    C. Conselice
        \and Center for Astrophysics. Harvard \& Smithsonian, 60 Garden St., Cambridge, MA 02138 USA   % 3)  Steve
\and Department of Physics, University of the Basque Country UPV/EHU, E-48080 Bilbao, Spain % 4 Tom-1
\and DIPC, Basque Country UPV/EHU, E-48080 San Sebastian, Spain                             % 5 Tom-2
\and Ikerbasque, Basque Foundation for Science, E-48011 Bilbao, Spain                       % 6 Tom-3  
\and School of Earth and Space Exploration, Arizona State University, Tempe, AZ 85287-1404, USA % 7)  Seth H. Cohen
                                                                                                %     Rolf A. Jansen
                                                                                                %     Jake Summers 
                                                                                                %     Rogier A. Windhorst
                                                                                                %     Rafael Ortiz~III 
\and International Centre for Radio Astronomy Research (ICRAR) and the International Space Centre (ISC), The University of Western Australia, M468, 35 Stirling Highway, Crawley, WA 6009, Australia                                                   % 8)  Jordan C. J. D'Silva -1
                                                                                                %     Simon P. Driver 
                                                                                                %     Aaron Robotham
\and ARC Centre of Excellence for All Sky Astrophysics in 3 Dimensions (ASTRO 3D), Australia    % 9)  Jordan C. J. D'Silva -2
\and Space Telescope Science Institute, 3700 San Martin Drive, Baltimore, MD 21218, USA         % 10)  Anton M. Koekemoer 
                                                                                                %     Dan Coe -1
                                                                                                %     Norman A. Grogin
                                                                                                %     Nor Pirzkal
                                                                                                %     Russell E. Ryan, Jr. 
\and Association of Universities for Research in Astronomy (AURA) for the European Space Agency (ESA), STScI, Baltimore, MD 21218, USA     % 11)  Dan Coe -2
\and Center for Astrophysical Sciences, Department of Physics and Astronomy, The Johns Hopkins University, 3400 N Charles St. Baltimore, MD 21218, USA % 12)  Dan Coe -3
\and Department of Astronomy/Steward Observatory, University of Arizona, 933 N Cherry Ave, Tucson, AZ, 85721-0009, USA % 13) Brenda Frye 
\and National Research Council of Canada, Herzberg Astronomy \& Astrophysics Research Centre, 5071 West Saanich Road, Victoria, BC V9E 2E7, Canada % 14) Madeline A. Marshall -1
\and ARC Centre of Excellence for All Sky Astrophysics in 3 Dimensions (ASTRO 3D), Australia                                                       % 15) Madeline A. Marshall -2
\and INAF-Osservatorio Astronomico di Trieste, Via Bazzoni 2, 34124 Trieste, Italy              % 16)   Mario Nonino 
\and Steward Observatory, University of Arizona, 933 N Cherry Ave, Tucson, AZ, 85721-0009, USA  % 17) Christopher N. A. Willmer
% Fengwu Sun
% Kevin Hainline
\and Department of Physics and Astronomy, University of Missouri, Columbia, MO 65211, USA       % 18)  Haojing Yan   
\and INAF - Institute of Space Astrophysics and Cosmic Physics (IASF Milano) Via Corti 12, I-20133 Milano, Italy % 19) Mari Polletta
\and Physics Department, Ben-Gurion University of the Negev, P.O. Box 653, Be’er-Sheva 84105, Israel % 20) Adi
          }
%   \date{Received September 15, 1996; accepted March 16, 1997}
 \abstract{
% context heading
% aims heading 
% methods heading (mandatory)
%  results heading (mandatory)
%  conclusions heading (optional), leave it empty if necessary 
 We present a new lens model for the $z=0.396$ galaxy cluster MACS J0416.1$-$2403 based on a previously known set of 77 spectroscopically confirmed, multiply imaged galaxies plus an additional set of 42 candidate multiply imaged galaxies from past HST and new JWST data.  The new galaxies lack spectroscopic redshifts but have geometric and/or photometric redshift estimates that are presented here. The new model predicts magnifications and time delays for all multiple images. The full set of constraints totals 343, constituting the largest sample of multiple images lensed by a single cluster to date.  Caustic-crossing galaxies lensed by this cluster are especially interesting. Some of these galaxies show transient events, most of which are interpreted as micro-lensing of stars at cosmological distances. These caustic-crossing arcs are expected to show similar events in future, deeper JWST observations. We provide time delay and magnification models for all these arcs. The time delays and the magnifications for different arcs are generally anti-correlated.
 In the major sub-halos of the cluster, the dark-matter mass from our lens model correlates well with the observed number of globular clusters, as expected from $N$-body simulations. This confirms earlier results, derived at lower redshifts, which suggest that globular clusters can be used as powerful mass proxies for the halo masses when lensing constraints are scarce or not available. 
   }

   \maketitle
%
%-------------------------------------------------------------------
\section{Introduction}
%%%%%%%%%%%%%%%%%%%%%%%%

%Intro JWST
The \textit{James Webb} Space Telescope (JWST) is providing unique insight into the distant Universe thanks to its superior capabilities (spatial resolution and sensitivity) in the infrared. The power of JWST increases even more when pointing it towards a massive galaxy cluster, acting as an additional lens in front of the telescope. Images from objects behind the lens can get amplified by factors O(10) if they are galaxy scale, to factors O(10,000) if they are the size of stars. Through these natural pinholes to the distant Universe we can see objects that are a few to $\approx 10$ magnitudes fainter than it would have been observed without the gravitational lens. That is, one can observe objects that (without the aid of lensing) would have had apparent magnitude AB $\approx 40$ in JWST's IR bands with just a few hours integration time. Put in perspective, for the most extreme magnification factors expected \citep[$\sim 10^4$, similar to the magnification estimated for Earendel, the farthest star detected through gravitational lensing and at $z\approx 6$, and other distant stars at $z>4$][]{Welch2022,Meena2023,Furtak2024} the combined effect of JWST plus the gravitational lens is equivalent to a JWST-like telescope but with a 600 meter diameter mirror. Precise modelling of these gravitational lenses is needed in order to identify and characterize these regions of maximum magnification, which in turn is needed in order to interpret the observations. \\

%Intro MACS0416
One of these lenses is the galaxy cluster MACS J0416.1$-$2403 \citep[MACS0416 hereafter;][]{Ebeling2001,Mann2012} at redshift $z=0.396$. 
This cluster is one of the six clusters studied in detail by the \textit{Hubble Space Telescope} (HST) as part of the Hubble Frontier Field (HFF) program \citep{Lotz2017}, and selected based on the presence of several strongly lensed galaxies \citep{Zitrin2013a}. As part of this program,  the central $\sim 10$ arcmin$^2$ region was observed in wavelengths ranging from 0.45 $\mu$m to 1.6 $\mu$m and to a depth of $\sim 28.5$ mag in the visible and infrared (IR) bands. The HFF data from MACS0416 have been extensively used by lens modellers \citep[][]{Jauzac2014,Jauzac2015,Johnson2014, Diego2015b, Kawamata2016, Hoag2016, Caminha2017, Richard2021,Bergamini2022,Diego2023b} in the search for high-redshift galaxies \citep{McLeod2015,Oesch2015,Kawamata2016,Ishigaki2018}, and for the discovery of some of the first lensed stars at $z>0.9$ \citep{Rodney2018,Chen2019,Kaurov2019}. 
The area observed by HST around MACS0416 was doubled thanks to the Beyond Ultra-deep Frontier Fields and Legacy Observations (BUFFALO) program \citep{Steinhardt2020}, although with shallower observations than in the HFF program.

%Intro about PEARLS and first results
More recently, JWST has observed MACS0416 as part of the Prime Extragalactic Areas for Reionization and Lensing Science \citep[PEARLS, ][]{Windhorst2023}, extending the wavelength coverage  to $\lambda \approx 5$\,$\mu$m, and with superior resolution and sensitivity in the IR range than previously observed by HST. The PEARLS observations of MACS0416 consisted of three epochs, which together with an additional fourth epoch from the CAnadian NIRISS Unbiased Cluster Survey \citep[CANUCS, ][]{Willot2017}, allowed to detect several transient events, out of which $\approx 10$ are suspected to be Extremely Magnified Objects (EMO), in particular EMO-stars, undergoing microlensing events near the regions of maximum magnification \citep{Yan2023}. These events are mostly due to the existence of two galaxies (dubbed Warhol and Spock) at  $z\approx 1$  that are crossing the cluster caustic. One of these events was studied in detail by \cite{Diego2023c} and corresponds to a double star being magnified by a millilens with mass smaller than $2.5\times10^6\, {\rm M}_{\odot}$, making it the smallest structure whose mass has been estimated through gravitational lensing above $z=0.3$, and with implications for dark matter (DM) models.
Earlier HST observations have revealed at least nine additional transient events in the Spock and Warhol galaxies \citep{Rodney2018,Chen2019,Kaurov2019,Kelly2023}, all of them consistent with being microlensing events, bringing the total number to a record 19 alleged microlensing events in a single cluster field.
Because of this wealth of flickering events, MACS0416 is referred to as the Christmas tree galaxy cluster and is so far the most remarkable cluster in terms of the number of transient events, allegedly due to microlensing.\\

%Intro about lensed stars
The low redshift of the two aforementioned caustic crossing galaxies, Warhol and Spock, correlates with the large number of microlensing transient events of distant stars detected in these galaxies. Without lensing, stars in these galaxies would have apparent magnitudes in the range 36--40. With magnification factors between $10^3$ and $10^4$ some of these stars could be detected by JWST \citep{MiraldaEscude1991}. At lower redshifts, lensing can more easily promote fainter but more abundant stars above the detection limit \citep[typically AB $\sim 29$, see also][]{Golubchik2023}. 
At $z\ga 2$, detecting such faint stars  would require more extreme and less likely magnification factors to compensate for the greater luminosity distance. 
Irrespective of the redshift, the stars that one expects to detect are always very luminous \citep[$L>10^4$\,\Lsun, see for instance][]{Kelly2018}. Among these, hot blue stars  can have luminosities exceeding $10^6$\,\Lsun. Red (cooler) stars are typically less luminous but are more often variable, facilitating their detection as transients. 

% Microlensing and ICL
These transient events are all found near the critical curve (CC) of the cluster and are the result of a temporary alignment between stars responsible for the intracluster light (ICL), and stars in the background galaxy, that increase their apparent brightness by a few magnitudes during the microlensing event. Other small structures in the cluster, such as globular clusters (GCs), can produce longer-lasting millilensing magnification when aligned with background stars. One such millilensed star, dubbed Mothra, was discovered in MACS0416 \citep{Diego2023c}. JWST's superior sensitivity and resolution allow  the best view of both the ICL and the GCs in MACS0416, and hence gain valuable insight into the observed micro- and millilensing events, as well as on the expected correlation between the spatial distribution of these events and the distribution of ICL and GCs. \\

% GCs
Earlier results based on observations and N-body simulations have shown a tight correlation between the number of GCs in a halo and its virial mass \citep{Blakeslee1997,Harris2017,Burkert2020,Valenzuela2021,Berkheimer2024}. This relation is satisfied also when relating the total virial mass of a halo and the total mass contained in the GCs \citep{Dornan2023}. This relation has been verified multiple times at low redshifts ($z\la 0.2$) with observations, although in most cases, the masses are derived from kinematics, X-ray or weak lensing. 
On galaxy cluster scales and at higher redshifts, the first results from JWST on SMACS0723 ($z=0.39$) have shown the strong correlation between the distributions of mass, GCs, and ICL \citep{Lee2022,Faisst2022,Montes2022,Diego2023d}.  This is observed also in highly irregular galaxy clusters such as A2744 at $z=0.308$ \cite{Harris2023}. MACS0416, with its wealth of strong lensing constraints and relatively deep observations, allows us to explore the correlation between the distribution of mass and the spatial distribution of GCs in greater detail. Properly characterizing the abundance and distribution of GCs in gravitational lenses is also important in order to correctly interpret past and future transient events near the region of high magnification, since GCs can introduce significant distortions into the magnification pattern.\\

%Preamble to rationale
% The study of the transients can be used to constrain the abundance of these very luminous stars at high redshift \cite{Diego2023b}, and also to study models of DM on the smallest scales. 
The study of the topics described above (high-z stars, magnification and time delay near CCs, or the relation between total mass and distribution of GCs) is only possible if accurate lens models are available. Lens models are also needed to correctly interpret these events, for instance by setting a constraint on their physical size that can later be used to identify some objects near CCs as stars. 

 %Rationale of the paper
This paper presents a new lens model derived using new lensing constraints from the latest observations of MACS0416 with JWST. The paper focuses on the lensing properties (magnification and time delay) of the region of maximum magnification around galaxies that are crossing caustics in order to provide the proper context for the observed transients and future ones that will continue to be discovered in this cluster. We also pay special attention to the population of GCs found in this cluster. Their distribution can be used as a mass proxy in regions of the lens plane where lensing constraints are rare or absent. Also, GCs can act as millilenses, so it is important to understand their distribution and be able to estimate the probability of the alignment between GCs and the background luminous stars that can be magnified by these GCs. 
The paper is organised as follows.
Section~\ref{s:PEARLS} describes the observations carried out during the PEARLS program. 
The lensing constraints (including the new ones discovered by JWST) used to derive the lens model are discussed in Section~\ref{Sect_Constraints}.
Section~\ref{s:WSLAPplus} presents the lens model, and Section~\ref{s:EMO} focuses on the lensing implications for galaxies crossing caustics and the possibility of detecting stars in lensed arcs.
Section~\ref{s:GC} discuses GCs as mass tracers and lensing perturbers.
Section~\ref{sect_discussion} considers in more detail the complexity of the lens model around the Spock arc, one of the transients in \cite{Yan2023}, and the lensing distortions introduced by GCs.
Finally, Section~\ref{s:concl} summarizes our conclusions.

We adopt a standard flat cosmological model with $\Omega_M=0.3$ and $h=0.7$. At the redshift of the lens ($z=0.396$), and for this cosmological model, one arcsecond corresponds to 5.39~kpc.
%while at $z_s=1.0054$, one arcsecond corresponds to 8.13 kpc. 
Unless otherwise noted, magnitudes are given in the AB system \citep{Oke1983}. Common definitions used throughout the paper are the following. ``Macromodel''  refers to the global lens model derived in Section~\ref{s:WSLAPplus}. The main caustic and main CCs are from this model. ``Micromodel'' means perturbations introduced by microlenses in the lens plane (stars and remnants), and  ``microcaustics'' means regions in the source plane of divergent magnification associated with  microlenses. The term millilens is used for objects in the mass range $10^4$--10$^6\, \Msun$ (typically GCs and dwarf galaxies in the galaxy cluster).
%Finally, when referring to the effective temperature of a star, we simply denote this temperature by $T$. 

\section{New JWST observations of MACS0416}
\label{s:PEARLS}
%%%%%%%%%%%%%%%%%%%%%%%%%%%%%%%%%%%%%%%%%%%%%
MACS0416 was observed in three different epochs by JWST during October 2022 and February 2023 as part of the PEARLS program. Observations were done in 8 filters, F090W, F115W, F150W, F200W, F277W, F356W, F410M, and F444W. Integration times in each filter were between 2920 seconds and 3779 seconds reaching $5\sigma$ limiting magnitudes between 28.45 and 29.9 per epoch and for point sources. Details of the observations for this cluster can be found in \cite{Windhorst2023}. The data processing for this cluster is described in detail by \cite{Yan2023}. The only special steps involved  removing artifacts \citep[$1/f$ noise, pedestal, and wisps, that were removed with {\small ProFound}][]{Robotham2017,Robotham2018} and aligning to {\it Gaia} DR3. As part of the process, the available data from the HFF and BUFFALO programs with HST (including data taken with the relevant bluer filters, F435W, F606W, and F814W) were realigned to the same {\it Gaia} data. In the central region of the cluster, HST data have similar depth to JWST thanks to the considerably longer exposures, but it proves very valuable because it extends the wavelength coverage down to 0.4 $\mu$m, providing useful and complementary photometric measurements at these wavelengths. \\

 The three epochs from the PEARLS program are taken at different position angles and cover an area of $\approx 2.8\times2.8$ arcminute$^2$ in the cluster core field, with additional area being covered further away by the second module. This work focuses only in the central cluster core field which is the only one containing lensing constraints. A portion of the area observed by JWST in this cluster core region is shown in Fig.~\ref{Fig_Arcs}. The area shown in this figure contains all multiple images identified in the data and marked with circles. This image (and others in this paper) is a color composition made after combining previous HST data from the HFF program with the new JWST data. The blue channel is a combination of HST's F435W, F606W, F814W, and JWST's F090W. The green channel combines HST's F105W, and F160W with JWST's F115W, F150W, and F200W. The red channel combines JWST's F277W, F356W, F410M, and F444W.

%%%%%%%%%%%%%%%%%%%%%%%%%%%%%%%%%%%%%%%%%%%%%%%%%%%%%%%%%%%%%
\section{Lensing constraints}\label{Sect_Constraints}
%%%%%%%%%%%%%%%%%%%%%%%%%%%%%%%%%%%%%%%%%%%%%%%%%%%%%%%%%%%%%

Existing strong-lensing constraints comprise 77 multiply lensed galaxies that produce 223 individual images. \cite{Diego2023b} (their Table~A.1) gave a complete list compiled from the literature \citep[][]{Jauzac2014,Johnson2014, Diego2015b, Kawamata2016, Caminha2017, Richard2021,Bergamini2022}.  All galaxies in that list have spectroscopic redshifts.
Additional lensed system candidates that lack spectroscopic redshift were identified in HST images in the references above. This set is enlarged with more system candidates identified with the new JWST data. We refer to this second set of constraints as the non-spectroscopic sample.  All available lensing constraints are listed in Table~\ref{tab_arcs}.  These include some of the missing counterimages in the spectroscopic sample, identified in the new JWST data, and marked with \dag\, in the table.\\

Identification of new lensed candidates in JWST images is done in the usual manner. Families of multiple images with compatible photo-z (when available), color, morphology, parity, and consistency with the previous lens model by \cite{Diego2023b} are identified in the new JWST images. Based on the large mismatch between the geometric and photometric redshifts in some systems, we expect a few of these new candidates ($\approx 4$) to be false positives (see discussion at the end of Section~\ref{Sect_GeoZ}). These low confidence systems are not used as lensing constraints. 
In Table~\ref{tab_arcs} the confidence of all systems is ranked with A for the spectroscopically confirmed systems, B for systems without spectroscopy but reliable according to the criteria listed above, C for systems that are less reliable (for instance, lacking sufficient morphological information), and D for the least reliable systems (for instance, the lens model is not accurate enough around these systems). Images with score C and D are not used as lensing constraints but are added here as potentially interesting targets for spectroscopic follow-up. 

\begin{figure*} 
  \includegraphics[width=\linewidth]{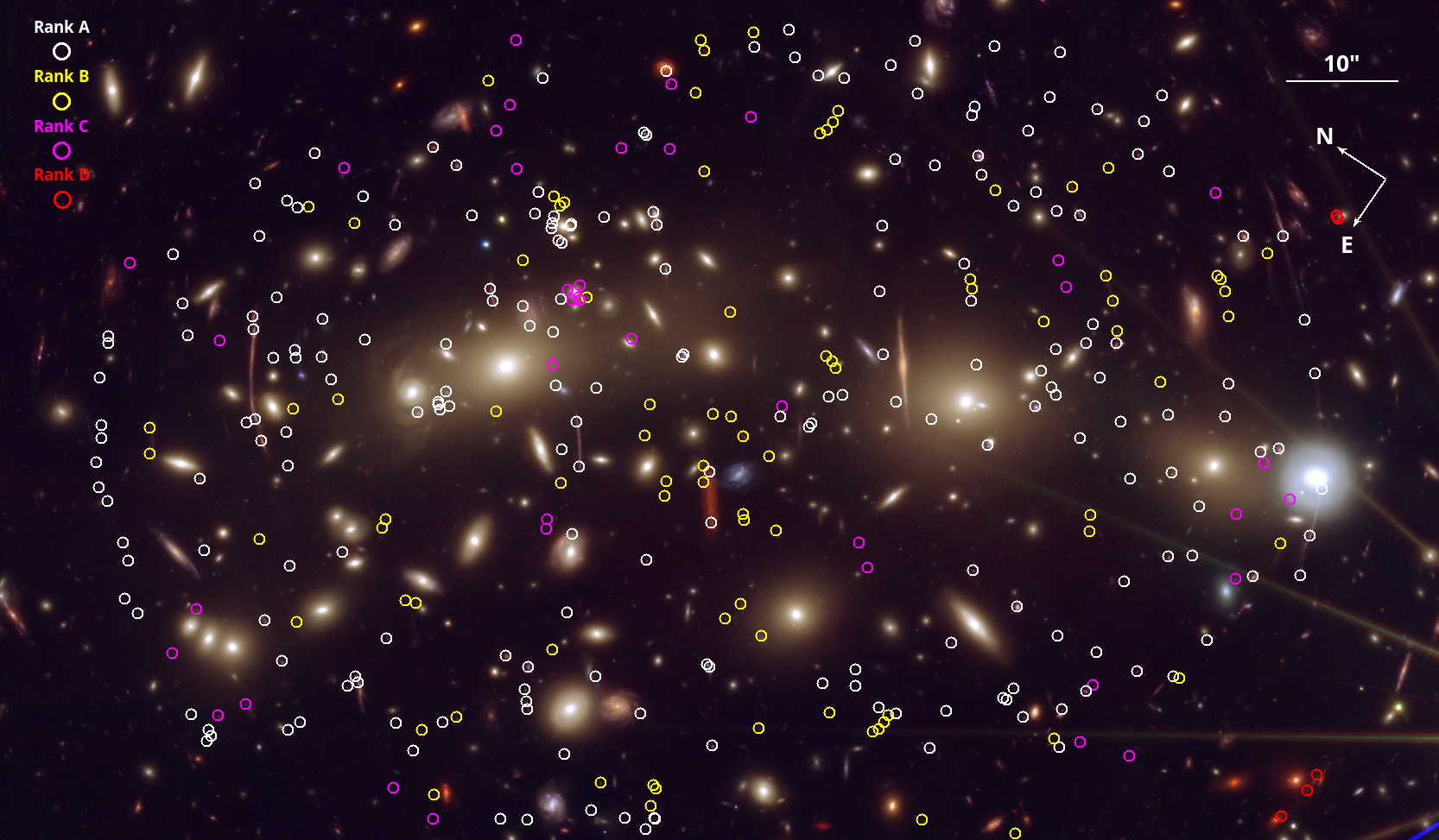}  
      \caption{Color image of MACS0416.  Scale and orientation are indicated. 
       White circles mark previously known counterimages with spectroscopic redshifts from \cite{Richard2021,Bergamini2022}. Yellow circles mark reliable (rank B) HST- and JWST-identified counterimage candidates that lack spectroscopic confirmation. Magenta and red circles mark the least reliable counterimage candidates with rank C and D respectively. 
         }
         \label{Fig_Arcs}
\end{figure*}

All arcs listed in Table~\ref{tab_arcs} are shown in Fig.~\ref{Fig_Arcs}. The systems with spectroscopic confirmation are marked with white circles, while system candidates without spectroscopic confirmation are marked with yellow circles.  

%Figure made by "screenshot". DS9 files in Dell /JWST/MACS0416
% Data/CropImages/PowerR6.fits PowerG6.fir & PowerB6.fits (ASINHH)
% FIG1_SPECT_circles_Rectangle_NoLabels.reg
% BUFFALO22_Bergamini22_JWST_Sept2023_MASTER_deg_V10_OK_Missing3rd_NoLabels.reg

%%%%%%%%%%%%%%%%%%%%%%%%%%%%%%%%%%%%%%%%%%%%%%%%%%%%%%%%%%%%%
\section{Lens model}\label{s:WSLAPplus}
%%%%%%%%%%%%%%%%%%%%%%%%%%%%%%%%%%%%%%%%%%%%%%%%%%%%%%%%%%%%%
\subsection{Lens model method}

The lens model is based on the code WSLAP+ \citep{Diego2005,Diego2007,Sendra2014,Diego2016}. This is a hybrid type of model that combines a free-form decomposition of the smooth, large-scale cluster component with a small-scale contribution from cluster galaxies. The code combines weak (when available) and strong lensing in a natural way with changes in the large-scale and small-scale components equally affecting the strong- and weak-lensing observables. For this work, only strong-lensing constraints are used.

As a brief description, WSLAP+ solves a system of linear equations that can be represented in compact form as
\begin{equation}
\vectTheta = \matrGamma \vectX\quad, 
\label{eq_lens_system} 
\end{equation} 
where $\vectTheta$ is a vector containing the observed sky-plane positions of the multiple images, $\vectX$ is a vector containing the unknown source-plane positions of the background sources and the model mass distribution in the lens-plane, and $\matrGamma$ is a matrix representing the ray-trace calculations that predict observed positions, $\vectTheta'$,  given $\vectX$ for a fiducial mass. The solution $\vectX$ is found by iteratively minimising a function of $\vectTheta'-\vectTheta$ with the constraint $\vectX>0$ (that is, masses and positions are positive, where the origin for the positions is the bottom-left corner of the area being considered, hence by definition all coordinates within this area are positive; \citealt{Diego2005}). In practice, the lens-plane mass is divided into $N_c$ grid points or cells.  These cells are associated with a smooth matter distribution representing the sum of DM, intra-cluster medium (gas and stars), and any other smooth component. These cells have pre-selected positions and sizes chosen by the modeller.  Each of the $N_s$ source galaxies is represented by two variables in $\vectX$, and WSLAP+ allows ``layers'' representing independent mass normalizations for different groups of member galaxies. The number of layers, $N_{\rm l}$, can vary between 1 when all member galaxies are included in one single layer, to the total number of member galaxies. In practise, the number of layers is usually between 1 and 5.  With these components, vector $\vectX$ has dimension
\begin{equation}
N_{\rm X}=N_{\rm c} + 2N_{\rm s} + N_{\rm l}\quad. 
\label{eq_Nx}
\end{equation}
If there are $N_{\rm sl}$  strong-lensing observables, the vector $\vectTheta$ has dimension $N_{\Theta }=2N_{\rm sl}$ because each multiple image contributes two position constraints, $\theta_x$ and $\theta_y$, and distances (or redshifts) are considered known. Hence, the matrix $\matrGamma$ has dimension $N_{X}\times 2N_{\rm sl}$ and is also known. 

Our MACS0416 models used different sets of constraints and different numbers and distributions of cells as described below.  All models used three layers in addition to the dark-matter layer. Layer 1 contained the two brightest cluster galaxies (BCGs); Layer 2 contained the remaining most-prominent cluster galaxies ($444$ galaxies brighter than magnitude $23.5$) with the bulk of them identified as members of the red sequence (414 galaxies, out of which 298 are also spectroscopically confirmed by MUSE) and an additional 30 galaxies confirmed spectroscopically by MUSE \citep{Caminha2017} but not included in our red sequence sample. For the spectroscopic galaxies we considered the range $0.39<z<0.41$ in order to account for the large velocity dispersion (and likely elongation along the line-of-sight) of this cluster; and Layer 3 contained a few $z=0.12$ foreground galaxies. This third layer contains only 3 galaxies but is included here because these foreground galaxies are near elongated arcs and play a role in their distortion. We expect a negligible contribution from the smooth component associated with these three galaxies so the smooth component at $z=0.12$ is not included in our model. We also expect a negligible contribution from small member galaxies not included in Layer 2. 
%
%Only galaxies spectroscopically confirmed by MUSE, either as cluster members or in the  $z=0.12$ group, were included. 
We assigned each galaxy's mass distribution to be proportional to the galaxy's surface brightness in JWST's F356W filter. This filter has the best sensitivity in the long wavelength channel of JWST.  The proportionality constant could differ for the three layers and  was readjusted as part of the optimisation process. The large scale mass around the member galaxies is modelled by the Gaussians distributed in the grid.
\citet{Diego2005} gave a complete description of the iterative solution process, and \citet{Sendra2014} described the model's convergence and performance based on simulated data. For a direct comparison with other lens modelling techniques see \cite{Meneghetti2017}. 

\subsection{Lens model with spectroscopic systems only}
%---------------------------------------------------------
First, we derive a solution using only image systems with spectroscopic redshifts as constraints. This initial trial used a regular $20\times20$ grid for the smooth mass component. At this resolution, a cell has a dimension of 45.3 kpc on a side. Adopting a regular grid is equivalent to assuming no prior information for the mass distribution because all grid points contribute similarly to the system of equations in Eq.~\ref{eq_lens_system}.  The next solution used an adaptive grid with more grid points around the densest regions of the first solution, to increase the resolution in the central region. The cell size varies ranging from $\approx 15$ kpc near the two BCGs to $\approx 100$ kpc in the outer regions of the field. This applies a prior to the mass distribution in which more mass is expected in regions where the distribution of grid points is denser. The prior helps to stabilize the solution and reduces the scatter from the choice of the initial grid.  Because the largest source of variability in the derived models is the choice of grid, we obtained multiple solutions with different grid configurations. 
Solutions obtained with an adaptive grid having a number of grid points smaller than the regular grid number of grid points ($N<400$) do  not reproduce some arcs as well as solutions obtained with an adaptive grid having a number of grid points $N>600$ result in artifacts (spurious large scale mass structures) in the regions of the cluster  with no lensing constraints. 
We find that solutions obtained with a multiresolution grid of 495 grid points offers robust and more stable solutions, that do no introduce artifacts in the unconstrained boundaries of the field of view.  Small variations ($\delta N \approx 20$) in the number of grid points around $N\approx 500$ result in almost indistinguishable solutions.
%The CCs for this model are shown in Fig.~\ref{Fig_CritCurves}
%With the new multiresolution grid, and using only the lensed systems with spectroscopic redshift, we derive a new solution. 
%We also vary the initial condition, $X_o$ and redshifts of the systems with photometric or geometric redshifts. 

\subsection{Geometric redshifts}\label{Sect_GeoZ}
%-------------------------------

%%Figure made y Dell /JWST/MACS0416/PhotoZ/PlotZ.pro
\begin{figure} 
   \includegraphics[width=9cm]{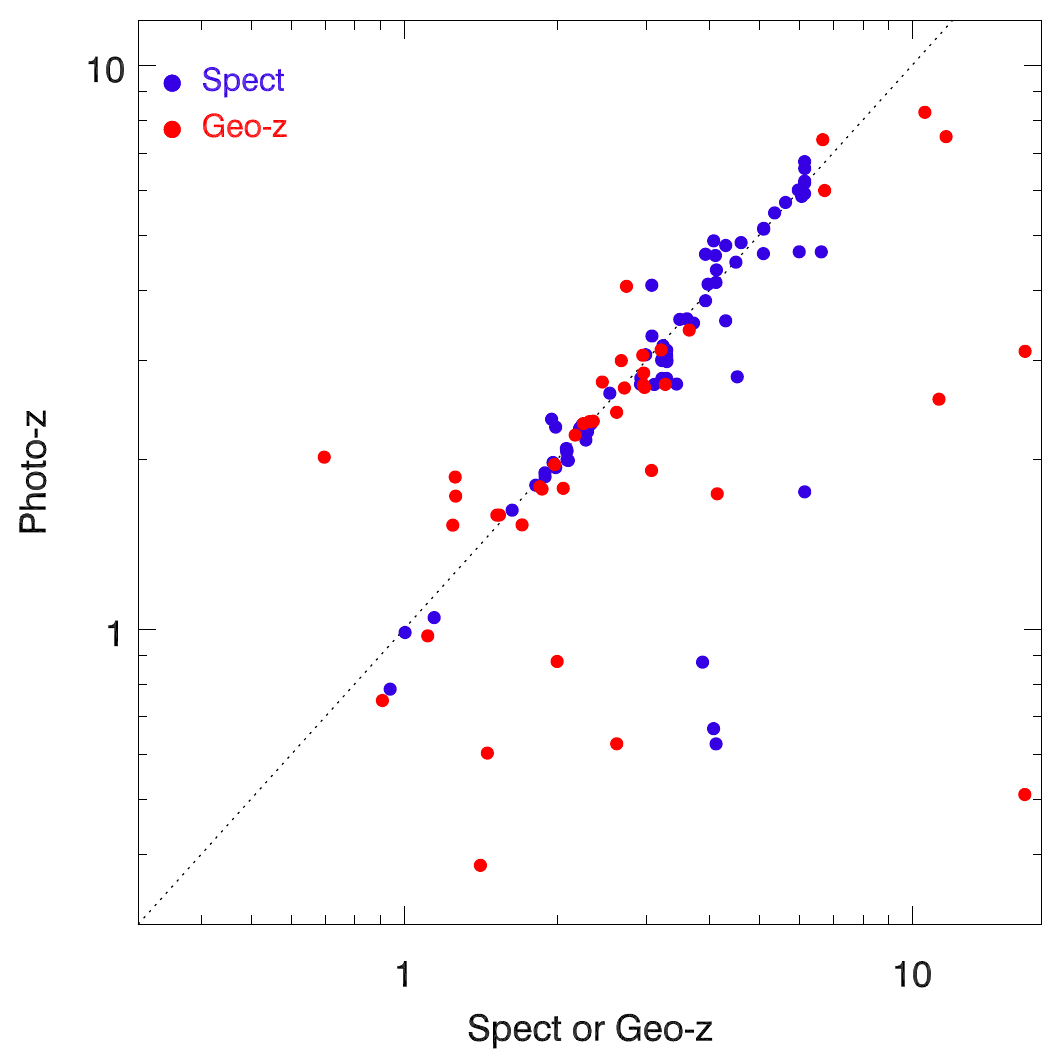}   
      \caption{Comparison of photometric redshifts with spectroscopic (blue symbols) or geometric (red symbols) redshifts for the lensed systems in Table~\ref{tab_arcs}. For the photometric redshifts, we chose the value (from the multiple counterimages of the same system)  closest to the spect-$z$ or geo-$z$.  The dotted line shows equality. The comparison between spect-$z$ and geo-$z$ is not shown but it follows almost perfectly the dotted line (see Table~\ref{tab_arcs}).
%    Filled symbols represent sources having several filters with sufficient signal for SED fitting. Open symbols represent faint sources where forced photometry was applied. 
         }
         \label{Fig_GeoZ}
\end{figure}

Once a lens model is created from the set of spectroscopic systems, it predicts ``geometric redshifts'' (geo-z or \zg) for lensed systems with and without spectroscopic redshifts. We used the solution obtained with the multiresolution grid described above to derive \gz\ values for all arcs following the same procedure as \cite{Diego2023d,Diego2023a}. For systems with spectroscopic redshifts, the lens model predicts \zg\ close to \zs\ as shown in Table~\ref{tab_arcs}. Therefore \zg\ is likely to be accurate for candidate image systems that lack spectroscopic confirmation and that appear near regions where a spectroscopic system is acting as a constraint.

As a consistency test, we compare geo-z with photometric redshifts for systems that lack \zs, and results are shown in Fig.~\ref{Fig_GeoZ}.
% Text from Nathan ->
The photometric redshifts were obtained from or following the methods of \citet{Adams2023}. In brief, the photo-$z$'s were calculated from JWST photometry, extracted in 0\farcs16-radius apertures, and aperture corrected based on PSF models. The photometry was then run through the EAZY phot-$z$ code \citep{Brammer2008} using the template combination  from \citet{Larson2022}. In MACS0416, these photo-$z$'s recovered 235/261 (90\%) of known MUSE spectroscopic redshifts  \citep{Caminha2017}. Approximately half of our lensed sources are in the \citet{Adams2023} catalogs, while the rest are blended or contaminated with foreground cluster galaxies or the intra-cluster light. These required manual aperture placement to obtain \zp\ estimates.
% <- text from Nathan

To first order, there is a good correspondence between all three redshifts, but there are several examples of significant mismatches. (All results for all systems are given in Table~\ref{tab_arcs}.) Many of the outliers in the photometric redshift concentrate below $z_{phot}\approx 1$. This may be simply because these sources are fainter or contaminated by intracluster light as suggested by the relatively large concentration of outliers near $z_{phot}\approx 0.4$ (the redshift of the cluster). The number and distribution of mismatches between geo-z and photo-z are comparable to the mismatch between photo-z and the spectroscopic redshift. Focusing on the biggest outliers, there are $\approx 4$ red points (geo-z) in Fig.~\ref{Fig_GeoZ} that are more offset (when compared with the $z_{phot}$--spect-z outliers or blue points outside the perfect correlation). These could be false positives in our sample (that is, incorrectly identified lensed systems). Two of these systems are already ranked D in Table~\ref{tab_arcs} as unreliable. These are systems 113 ($z_{geo}=0.69$, $z_{phot}=2.02$) and 114 ($z_{geo}=16.67$, $z_{phot}=0.51$) in Table~\ref{tab_arcs}. A third suspicious system is 116 ($z_{geo}=16.69$, $z_{phot}=3.11$) which is ranked C (and hence not used as a constraint) and with very discrepant geo-z and photo-z. System 81 ($z_{geo}=11.3$, $z_{phot}=2.56$), in this case two counterimages form a pair of symmetric features near the expected position of the CC and are ranked B and used as constraints in the full lens model. The third counterimage candidate is very faint, ranked C and hence not used as a constraint. Moreover, when using only the two images ranked B, we derive $z_{geo}= 2.8_{-1.1}^{+4.4}$, in better agreement with $z_{phot}$, thus confirming our low confidence on the third image of system 81.
%The discrepant redshift for this system can easily be explained if this third counterimage is not the true one, since in this particular case most of the constraining power on the geo-z comes from the third counterimage. 
Even if the adopted redshift for System 81 is wrong, the fact that we only use the relatively close pair 81.a and 81.b as constraints results in very little weight in the final model, since this pair of images is already very close in the source plane for a wide range of redshifts, with a separation of just $\approx 1.5$ arcsec in the image plane. Despite the suspicion that the high geo-z of System 81 may be biased (when considering the third counterimage candidate), we opt for including this system with the high-z estimate of the redshift, in order to maintain consistency with the other rank B images.
We conclude from this comparison that for systems ranked B, $z_{geo}$ is a reliable estimate of the redshift of the system. Errors for the geo-z estimations are listed in Table~\ref{tab_arcs}. A more detailed study of the errors in geo-z from estimates obtained with WSLAP+ can be found in \cite{Diego2023a} where it is demonstrated, with a  bootstrap analysis, how the errors are representative of the true errors (see their figure 8), except for systems in the outskirts of the lens where the number density of lensing constraints is low. 

Since $z_{geo}$ are by construction correlated with a previously derived lens model (and may be affected by biases in that model), using high-fidelity $z_{phot}$ for the lensed galaxies is in general a preferable option than using $z_{geo}$ for the same galaxies when deriving lens models. However, for our particular case, we are interested in having the most reliable redshift that allow us to predict magnification and time delays for counterimages of sources at those redshifts. Given the large number of spectroscopic systems available in MACS0416, the uniform and dense spatial distribution of the lensed images, and the relatively poor photometry (most counterimage candidates are faint) affecting the $z_{phot}$ estimates, $z_{geo}$ estimates are more reliable for this particular case and are used in the next sections to derive the final lens model and the derived lensing properties of all arcs. This option is also the safest, since an inaccurate $z_{phot}$ can introduce a large bias in the derived model, while $z_{geo}$ estimates are expected to introduce smaller biases since they produce lens models which are consistent with the model derived using spectroscopic systems only. \\

As a final note, during the last phases of the refereeing process, \cite{Grego2024} publish new spectroscopic redshifts for 13 of the systems candidates in Table~\ref{tab_arcs}. For these 13 systems, we find that our previously derived geometric redshifts are accurate estimates of the true redshift (the comparison is shown in Figure~\ref{Fig_Zgeo_vs_MUSE}) with a dispersion in the error of 6.8\%, but with a median in the distribution of errors of less than 1\% error.

\subsection{Lens model with all systems}
%---------------------------------------------------------
A final lens model comes from adopting \zg\ for candidate image systems that lack \zs and with rank A or B. We refer to this sample of lensing constraints as the full sample.  The model, based on the 495 multiresolution grid and the full sample, does not differ much from the previous one because the model was already well constrained with the spectroscopic sample and the new systems have, by construction,  redshifts consistent with the spectroscopic model. However, the new constraints help to reduce uncertainties, particularly in regions where the density of spectroscopic systems is low.
Regarding the precision in the predicted positions of the counterimages, this model has a root mean square (RMS) between the observed and predicted positions in the image plane of 1.07". When considering only the images with spectroscopic redshift, the RMS is 0.91", while images with geo-z have an RMS of 1.36". These values are typical of lens models derived with WSLAP+.

\begin{figure} 
  \includegraphics[width=9cm]{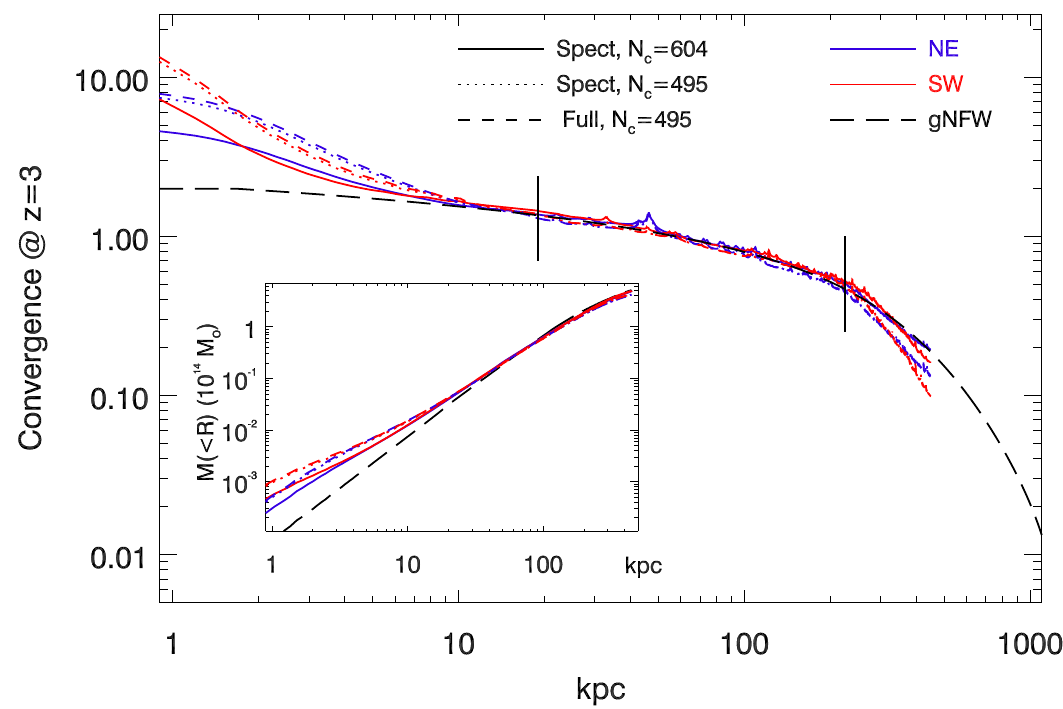}   
      \caption{Mass profile of the lens models.
       Blue curves correspond to  profiles centered on the NE BCG while red curves are for  profiles centered at the SW BCG. 
Solid lines represent the models derived using only spectroscopically confirmed (rank A in Table~\ref{tab_arcs}) galaxies as constraints and an adaptive grid with 604 grid points for the smooth DM. Dotted lines show the corresponding solution for an alternative model with the same constraints (rank A) but with only 495 grid points. Dashed lines represent the same 495-point grid but with the full sample of lensed systems ranked A and B (geometric redshifts) as constraints. The mass profiles are given in dimensionless $\kappa$ units assuming a source at $z=3$. The long-dashed line shows a gNFW profile with parameters (Eq.~\ref{eq:NFW}) $\gamma=0.9$, $\alpha=2$, $\beta=3$, and scale radius $r_s= 640$\,kpc. The two vertical black segments indicate the distance interval where the profiles are constrained by lensing. 
      The inset shows the integrated mass as a function of radius in units of $10^{14}\, \Msun$  for the same three profiles shown in the main panel. The black long-dashed line is the integrated profile when the center is chosen as the midpoint between the two BCGs (RA = $4^{\rm hr} 16^{\rm m} 08\fs428$, Decl. = $-24^\circ 04' 21\farcs0$). 
      % For comparison, we show as light blue and green the profiles from \cite{Bergamini2022}. 
         }
         \label{Fig_Profile}
\end{figure}

Fig.~\ref{Fig_Profile} shows the resulting mass profiles centered on each of the two groups (BCGs). The biggest differences between models come from the choice of grid configuration rather than from the set of constraints. Nevertheless, differences are minuscule in the region constrained by lensing.  

%%Figure made in Toshiba, "scripts/Profile_MACS0416_JWST.pro" 

All models show a striking similarity between the NE and SW halos, consistent with other recent models \cite{Bergamini2022,Diego2023b}. The profiles can be well reproduced by the same generic Navarro--Frenk--White profile \citep{NFW,Hernquist1990,Zhao1996,Wyithe2001,Nagai2007}:
\begin{equation}
\rho(r)=\frac{\rho_o}{(r/r_s)^\gamma[(1+(r/r_s)^\alpha]^{(\beta-\gamma)/\alpha}}
\label{eq:NFW}
\end{equation}
where $\gamma$, $\alpha$, and $\beta$ are the inner, intermediate, and outer slopes, and $r_s$ is the scale radius. 
In all models, the inner 10\,kpc is dominated by the baryonic component of the BCGs surrounded by a shallow  convergence profile out to 200\,kpc. This type of shallow profile extending beyond 100\,kpc has been observed in other merging systems similar to MACS0416, for example A370 \citep{Lagattuta2017,Diego2018b}. \\

In our case, the maximum radius constrained by lensing is 200\,kpc, and the mass interior to that $M({<}200\rm\, kpc) =1.72 \times 10^{14}$\,\Msun\ for the NE halo and $M({<}200\rm\, kpc) =1.77\times 10^{14}$\,\Msun\ for the SW halo. 
In a larger volume, \cite{Bergamini2022} found masses (from their best-fitting model) $M({<}400\rm\, kpc) {\approx} 3.4\times 10^{14}$\,\Msun\, for both halos. At this distance, we find $M(<400\, {\rm kpc}) =3.81 \times 10^{14} \Msun$ and $M(<400\, {\rm kpc}) =3.93 \times 10^{14} \Msun$ for the NE and SW halos, respectively.  Earlier results derived using the same WSLAP+ code, BUFFALO data from HST, the spectroscopic systems only, and including weak-lensing measurements, obtained a mass $M({<}400~\rm kpc) = 3.6\times10^{14}$~\Msun\ \citep{Diego2023b} for their best model, also in reasonable agreement with our results. \\
For the compact components we find a mass of $2.28\times10^{12}$ \Msun\, for Layer 1 (BCG1+BCG2), $1.03\times10^{13}$ \Msun\, for Layer 2 (red-sequence and spectroscopic member galaxies) and  $3.31\times10^{9}$ \Msun\, for Layer 3 (foreground galaxies).

%Figure made by "screenshot". DS9 files in Dell /JWST/MACS0416 and JWST/MACS0416/CritCurves_JWST_V2.odp
%RGB /home/jdiego/JWST/MACS0416/Data/CropImages/LR6.fits LG6.fits & LB6.fits
%/home/jdiego/JWST/MACS0416/LensModels_JWST/CritCurve_Case3_z0p9397.ctr
%/home/jdiego/JWST/MACS0416/LensModels_JWST/CritCurve_Case3_z2p091.ctr
%/home/jdiego/JWST/MACS0416/LensModels_JWST/CritCurve_Case3_z6p145.ctr
%%%/home/jdiego/JWST/MACS0416/Fig_CritCurves.reg
%/home/jdiego/JWST/MACS0416/Tables/Regions_CaustCrossGals_Boxes.reg
%%%/home/jdiego/JWST/MACS0416/Tables/Regions_CaustCrossGals.reg
\begin{figure*} 
    \includegraphics[width=18.0cm]{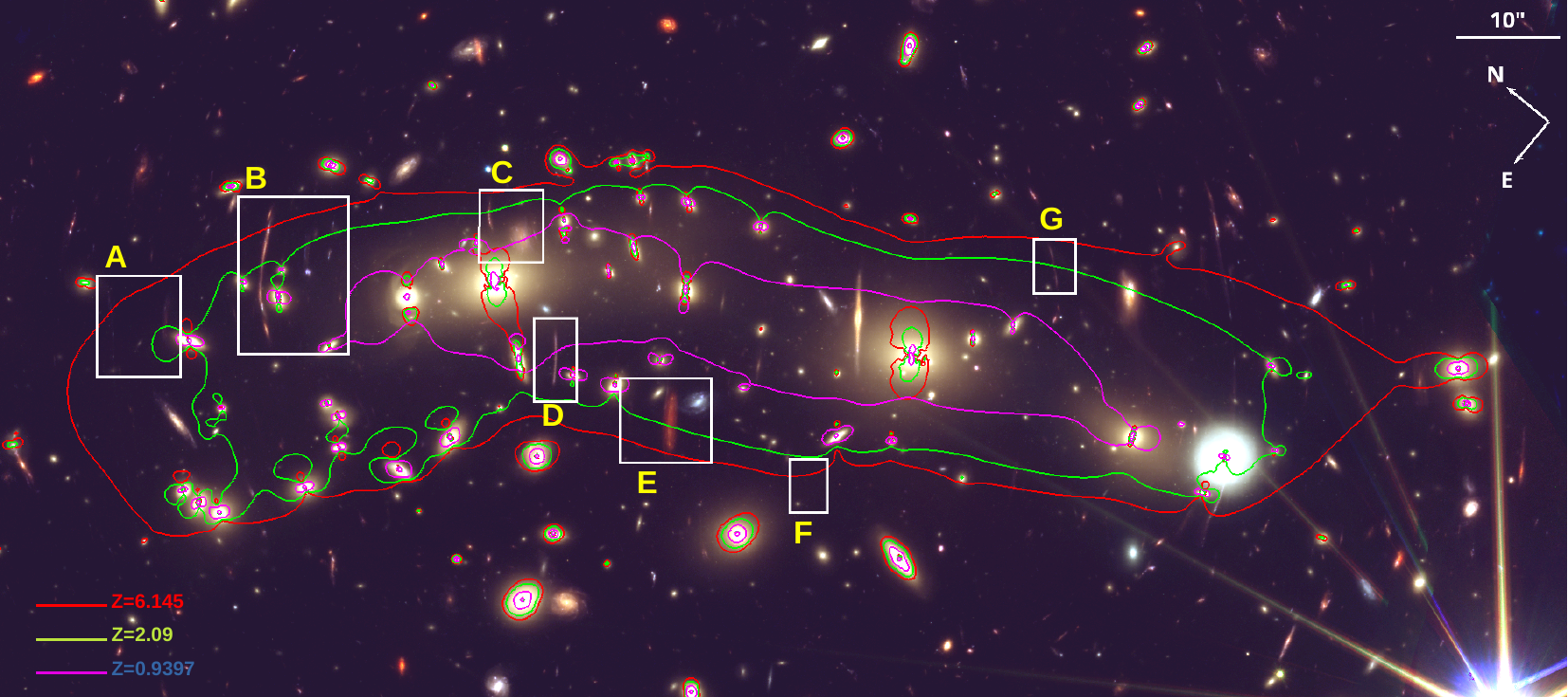}
        \includegraphics[width=18.0cm]{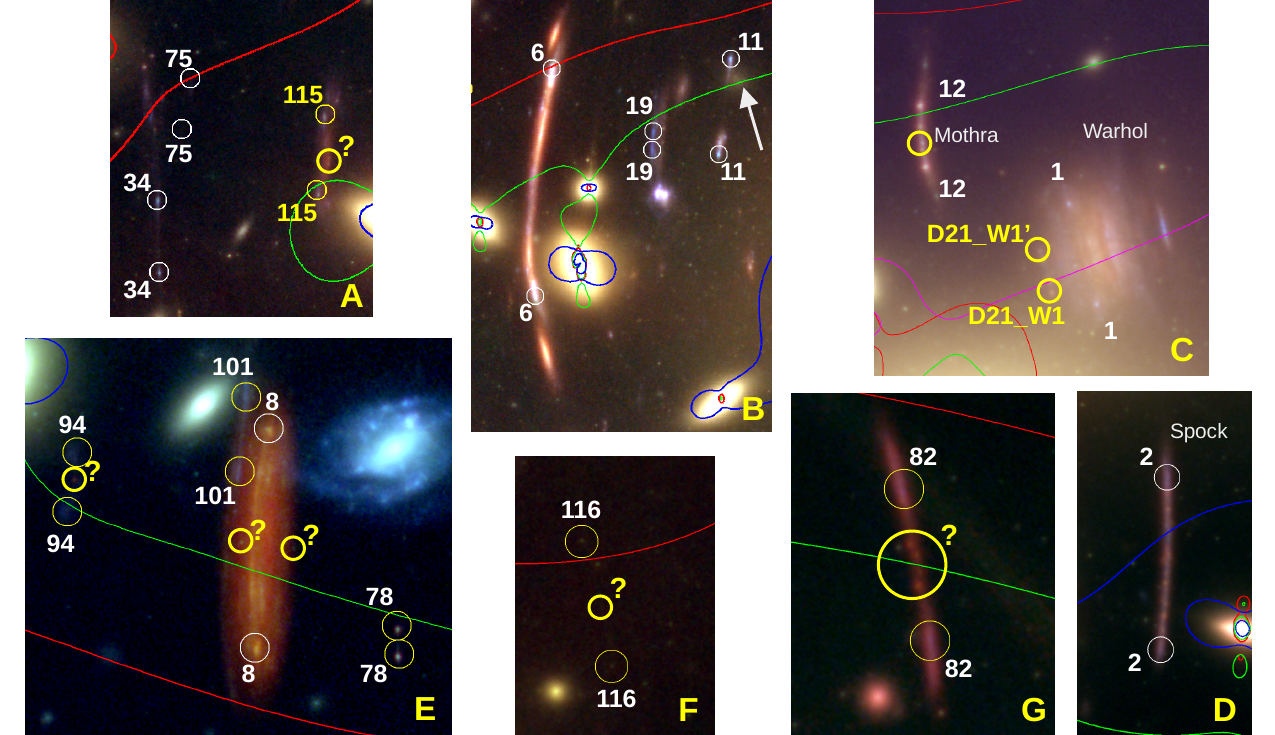}
      \caption{Critical curves from the lens model superposed on the MACS0416 color image.  Image scale and orientation are indicated. The red curve is for sources at $z=6.145$ (systems 73 and 74), the green curve is for  $z=2.091$ (System 12), and the magenta curve is for  $z=0.9397$ (System 1). The image colors have been stretched to better show the distribution of ICL. Panels A through G show enlarged images of areas indicated by white rectangles in the main image. These areas contain some of the most prominent arcs crossing a caustic, where candidate EMOs (marked by thick yellow circles) can be found in JWST images. The CCs in panels A--G are the same as in the larger plot. The numbers in each panel indicate the system ID, and white (rank A) or yellow  (rank B) thin circles indicate system counterimages. The lensing properties of these systems are described in more detail in Section~\ref{s:EMO}. Galaxies Warhol (System 1), Spock (System 2), and Mothra (System 12) are shown in panels C and D.  
         }
         \label{Fig_CritCurves}
\end{figure*}
We also find very similar magnification patterns to earlier models. The CCs for three different source redshifts are shown in Fig.~\ref{Fig_CritCurves}. In the same image, some of the arcs that are crossing caustics at their respective redshifts are shown in the zoomed rectangular regions. The white (rank A) and thin yellow (rank B) circles mark the position of the lensing constraints from these arcs used to derive the lens model. The thick yellow circles mark the positions of unresolved sources close to the CC. The source labeled D21\_W1 is a peculiar transient presented in \cite{Yan2023} and discussed in more detail later. Since the magnification predicted by the lens model at these positions is very large (typically hundreds), the unresolved nature of these sources at these magnification factors imply that these objects are good candidates to be EMO-stars. Some of the objects marked with a thick yellow circle have indeed shown variability, which is a fundamental prediction for EMO-stars moving across a web of microcaustics. JWST is expected to detect many more microlensing events in this cluster, in particular from the low redshift galaxies Warhol and Spock \cite{Diego2023b,Yan2023}.  Section~\ref{s:EMO}  discusses in more detail the lensing properties of the highly magnified arcs shown in panels A--G of Fig.~\ref{Fig_CritCurves}.

We conducted a search for dropout galaxies intersecting the CC at high redshift but found no convincing candidate although a few very faint and stretched features are found, which are high-redshift galaxy candidates intersecting the cluster caustic. One such candidate is shown in panel F of Fig.~\ref{Fig_CritCurves} (system candidate 116 and barely visible in JWST images). The CC in that panel is at $z=6.15$, suggesting that this arc must be at $z>6$

The solution presented above relies on geo-z estimates. Earlier geo-z predictions based on WSLAP+  models have been confirmed with subsequent spectroscopic follow up. For instance System 25 in \cite{Diego2015b} predicted at $z=1.05$ and later measured at $z=0.94$ \citep[System 12 in][]{Caminha2017}, System 10 in \cite{Diego2016b} at a predicted redshift of $z=0.78$ and later measured at $z=0.73$ \citep[B4 in Table 3 of][]{Caminha2016}, systems 7 and 19, proposed in \cite{Diego2018b} as being part of the same system with a predicted redshift of $z=2.8$ and later confirmed by \cite{Lagattuta2017} as being indeed the same system and at $z=2.7512$, or System 11 in \cite{Diego2020} predicted at $2.2\lesssim z\lesssim3.1$ (see their Figure 5) and measured at $z=2.1887$ in \cite{Caminha2023} (their System 23). 
When reliable photo-z are not available because of insufficient photometric information or the sources are too faint, geo-z are in general a safer option since they are derived from a model that relies on systems having spectroscopic redshifts. Because of this, solutions obtained with spectroscopic plus geo-z systems are often very similar to solutions obtained with spectroscopic systems only. The similarity between the profiles discussed earlier demonstrates this (Figure~\ref{Fig_Profile}). Earlier results based on WSLAP+ and JWST data from PEARLS also show a high degree of similarity between both type of solutions \citep{Diego2023a}. In particular, in \citep{Diego2023a} it is shown with a bootstrap method how the errors in the geo-z obtained by WSLAP+ are representative of the true errors, except in regions where the spectroscopic constraints are the most scarce, such as in the outskirts of the region constrained by strong lensing. Given the very high number density of spectroscopic systems surrounding the non-spectroscopic systems (see Figure~\ref{Fig_Arcs}) we expect the geo-z predictions for MACS0416 to be better than in earlier work, and hence very reliable. 

\section{Magnification and time delay for known caustic-crossing galaxies}\label{s:EMO}
%%%%%%%%%%%%%%%%%%%%%%%%%%%%%%%%%%%%%%%%%%%%%%%%%%%%%%%%%%%%%%%%%%

\cite{Yan2023} found eleven transients in three highly magnified galaxies (Warhol, Spock, and Mothra) using the new JWST PEARLS+CANUCS data. Most of these transients are consistent with being microlensed stars.  Previously, nine transient events were identified in Warhol and Spock in HST observations \cite{Rodney2018,Chen2019,Kaurov2019,Kelly2023}. The lensed star in the Mothra arc was also seen in previous HST data, but no flux variations were noticed in the HST data so prior to JWST observations it was not recognized as a lensed star.
All these transients are expected to be very luminous stars ($L>10^4$\,\Lsun) that temporarily increase their observed flux as a microlens from the galaxy cluster intersects the line of sight. Interpreting these events requires knowing their magnifications in the absence of microlensing, that is, the macromodel magnifications at the stars' positions. The lens model predicts the magnifications, but the accuracy depends on whether sufficient local constraints on the model are available. A feature of highly magnified stars is that they usually produce two counterimages, one on each side of the CC. Given the difference in paths and lensing potential, the two counterimages suffer different time delays, which are predicted by the lens model. The delays are important when a star exhibits intrinsic variability (a SN being the most extreme example) to predict the time when the trailing counterimage will appear. \\

Rather than presenting lens-model-predicted magnifications and time delays for specific events, it is more useful to present predictions for a range of distances from the CC because new transients can appear anywhere along these arcs. This needs to be done for each lensed galaxy individually because differences in redshift and lensing potential change the model predictions. 
Also, this approach reduces the uncertainties due to imperfect lens modelling that often places the CC at positions a fraction of an arcsecond away from the true (unknown) position. Even though these offsets are small, they are sufficiently large for accurately predicting the magnification of unresolved sources very close to the CC. In these situations, a fit to the expected scaling laws combined with a model independent estimation of the distance to the CC (for instance from symmetry arguments between a pair of counterimages) results in more accurate predictions for the magnification and time delays.
In this section, we present models for the magnification and time delay along 12 prominent arcs that are crossing the cluster caustic at their corresponding redshifts.

\begin{table}[h!]
\caption{Lensing properties of caustic-crossing galaxies. $T_0$ is expressed in years/arcsec, while $\mu_0^{+}$ and $\mu_0^{-}$ are given in arcseconds.  The caustic-crossing arcs are identified in  Fig.~\ref{Fig_CritCurves}.}\label{Tab_1}
\begin{center}
\small
\begin{tabular}{ |c|ccccccc| } 

\hline

{\small ID} & $z$ & $T_0$ & $\mu_0^{-}$ & $\mu_0^{+}$ & $\sigma({T_0})$ & $\sigma({\mu_0^{-}})$ &  $\sigma({\mu_0^{+}})$ \\
\hline
 1  & 0.939  & 3.14 &  12.16 &  19.94 &  0.23 &   0.39 &   1.01\\
 2  & 1.005  & 0.38 &  49.40 &  16.04 &  2.26 &   2.78 &   0.52\\
 6  & 1.895  & 0.47 &  16.18 &  27.15 &  1.22 &   1.62 &   0.47\\
 8  & 1.953  & 0.22 &  26.33 &  30.82 &  0.38 &   0.10 &   0.20\\
12  & 2.091  & 0.53 &  68.02 &  50.10 &  1.23 &   0.12 &   0.50\\
19  & 2.243  & 0.13 &  33.28 &  31.24 &  6.95 &   0.50 &   0.36\\
34  & 3.235  & 0.03 &  188.7 &  118.6 &  4.61 &   1.46 &   0.67\\
75  & 6.629  & 0.12 &  84.37 &  63.02 &  3.68 &   0.16 &   0.43\\
82  & 1.98   & 0.35 &  26.85 &  29.73 &  1.73 &   0.16 &   0.12\\
94  & 1.26   & 0.75 &  12.21 &  13.67 &  0.75 &   3.17 &   1.71\\
115 & 2.62   & 0.14 &  20.87 &  41.98 &  3.60 &   0.56 &   1.28\\
116 & 8--16  & 0.44 &  26.72 &  33.81 &  1.20 &   0.26 &   0.29\\
\hline
\end{tabular}
\end{center}
% \tablefoot{Numbers represent the total of the temperature--magnification diagrams in Figures~\ref{Fig_Nstar_T_dist_F814W}, \ref{Fig_Nstar_T_dist_F200LP}, and \ref{Fig_Nstar_T_dist_F200W}. The slope $\delta$ in Eq.~\ref{eq_dNdT} controls the relative abundance of red SGs vs. blue SGs. F814W and F200LP are HST filters, while F200W is a JWST filter.}
\end{table}

The magnification along an arc and as a function of distance to the CC can be fitted with the canonical law for CCs $\mu_{\rm th}=\mu_0/d_{\rm cc}$. Here $\mu_0$ is a constant (expressed in arcseconds) that depends on the redshift of the source and position along the CC, and $d_{\rm cc}$ is the distance to the CC expressed in arcseconds. The time delay has a similar scaling with distance,  $\Delta T_{\rm th}=T_0*d_{\rm cc}$, where $T_0$ is another constant expressed in years per arcsecond. For a given source redshift, the constant $T_0$ varies also from position to position along the CC at that redshift. The time delay $\Delta T$ is measured between two counterimages of the same source position on each side of the CC and separated by a distance $d = 2d_{\rm cc}$. 
These scalings in $\mu$ and $\Delta T$ are  expected for a lens with an isothermal-sphere density distribution, and they normally reproduce well the observed magnifications and time delays in real lenses. 

Departures from the linear scalings with $d_{\rm cc}$ are seen when substructures  perturb the deflection field near the CCs. To account for these possible perturbations, we computed the deviation of the observed magnifications from these simple scalings. We did so at three fixed distances from the CC, 0\farcs6, 1\arcsec, and 2\arcsec\ and on either side of the CC, i.e., for images with positive and negative parity. The deviation from the scaling is computed as:
\begin{equation}
\sigma_{\mu} = \sum_i \left ( \frac{|\mu_{\rm pred}-\mu_{\rm th}|}{\mu_{\rm th}} \right )_i 
\end{equation}
for the magnification, and similarly for the time delay
\begin{equation}
\sigma_{\Delta T} = \sum_i \left( \frac{|\Delta T_{\rm pred}-\Delta T_{\rm th}|}{\Delta T_{\rm th}} \right)_i \quad.
\label{Eq_DeltaT}
\end{equation}
The index $i$ runs over three separations ($i=1$, 2, 3 for $d_{\rm cc}=0\farcs6$, 1\arcsec, and 2\arcsec, respectively). In the equations above, the subscript "pred" refers to the predicted magnification while the subscript "th" refers to the expected scaling laws. For convenience, we took $\mu_0$ as the magnification measured at 1\arcsec, and similarly we took $T_0$ as the time delay measured when $d_{\rm cc}$ is 1\arcsec. With this definition, the terms where $i=2$ (that is, where $d_{\rm cc}=1\arcsec$) in the equations above are identically zero. 
Because the magnification can scale differently depending on the parity of the image, for each arc we derived two normalization factors, $\mu_0^{+}$ and  $\mu_0^{-}$ for the sides with positive and negative parity,  respectively (both following laws $\mu=\mu_0^{+}/d_{\rm cc}$ and $\mu=\mu_0^{-}/d_{\rm cc}$). The time delay is simpler because the image with positive parity always arrives first, so the relative time delay is given by the delay of the image with negative parity with respect to the image with positive parity. Hence only one normalization factor, $T_0$, is needed for each arc. 

Using the three fixed relative separations to the CC at the position of each of the 12 arcs, we computed  $\mu_0^{+}$,  $\mu_0^{-}$, $T_0$, 
$\sigma_{\mu^{+}}$,$\sigma_{\mu^{-}}$, and $\sigma_{\Delta T}$ for the 12 caustic-crossing arcs. 
The results are summarized in Table~\ref{Tab_1}, and the 12 arcs are shown in panels A--G of Fig.~\ref{Fig_CritCurves}. 
%For convenience, the ID shown in the first column of Table~\ref{Tab_1} is also indicated in Fig.~\ref{Fig_CritCurves}.
The values listed in this table are derived from measured distances to the lens model CC and not from observed sources along these arcs. Hence, they are not affected by the small offsets (typically a fraction of an arcsecond) between the predicted positions of the CC or lensed arcs and the true (unknown in the case of the CC) positions. Very small and local modifications in the lens model (to small to be captured by the grid of Gaussians in our model) can place the predicted CC or arcs at the true position while leaving the values listed in  Table~\ref{Tab_1} virtually unchanged. When computing distances, these are measured in a direction perpendicular to the CC. The arcs studied in this section are all at directions which are almost orthogonal to the CC, so distances computed this way should be accurate enough. 
A simple estimation of the uncertainty in the magnification and time delay can be obtained from the $\sigma$ values listed in  Table~\ref{Tab_1}. Relative uncertainties for the magnification are between a few percent to $\approx 20\%$. For time delays, the uncertainty in $\sigma$ is larger, specially for arcs with small time delay amplitudes, $T_o$, for which the denominator in Eq.~\ref{Eq_DeltaT}, can be very small.  These uncertainties are for one lens model only and does not account for uncertainty in the lens model itself which typically is larger and can exceed 50\% very close to the CC \cite[see for instance][]{Meneghetti2017}. Since the new candidate systems lack spectroscopic confirmation we avoid the lengthy exploration of uncertainties in the lens model, which has been done in the past for the spectroscopic sample. 

Below we briefly discuss some of the relevant arcs in Table~\ref{Tab_1} and shown in Fig.~\ref{Fig_CritCurves}: \\

% Figure made by Toshiba WSLAPplus/scripts/Plot_TimeDelay_MagnificationRelation.pro
\begin{figure} 
   \includegraphics[width=9cm]{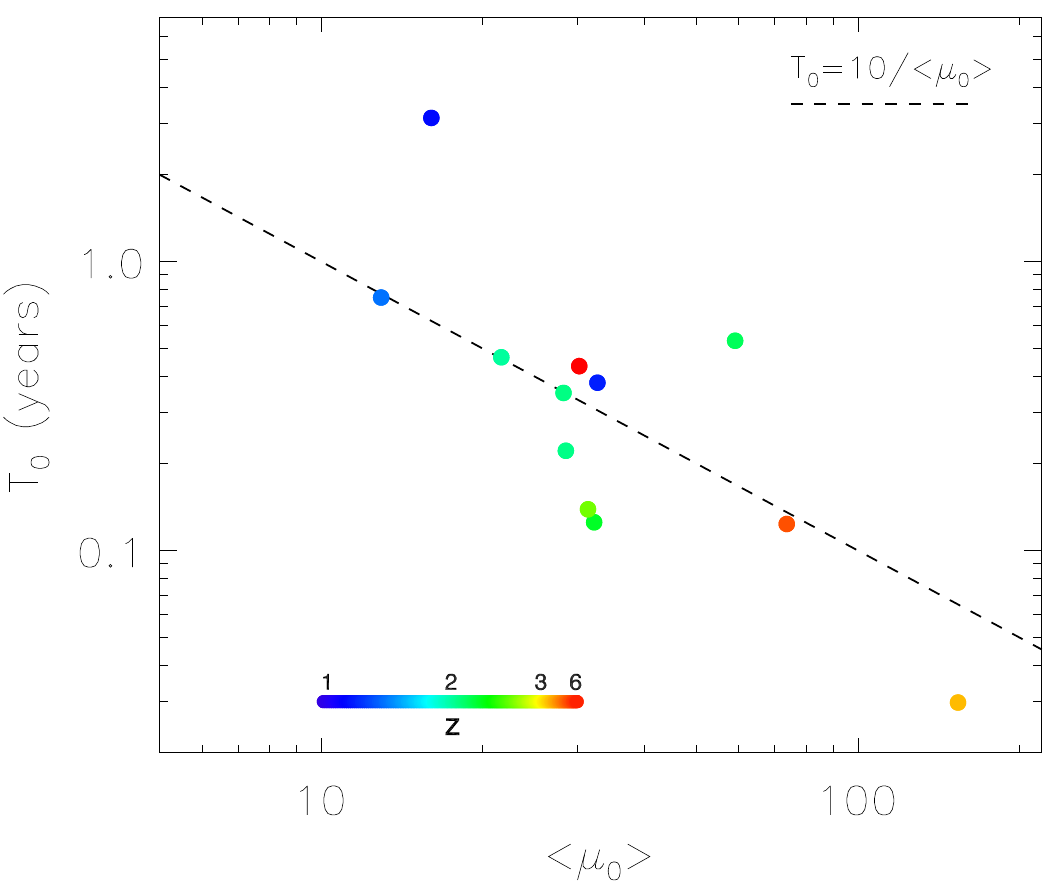}   
      \caption{Magnification versus time delay normalization factors for the 12 caustic-crossing galaxies in MACS0416. 
      Each point represents one of the 12 galaxies color-coded by their redshift. Low-redshift arcs normally appear near the center of the cluster, which gives larger Shapiro delays and steeper potentials. That largely explains the larger delays for smaller magnification factors.  
      The x-axis shows the average of the two normalization factors $\langle\mu_0\rangle=(\mu_0^{+}+\mu_0^{-})/2$ for each arc. The dashed line shows the expected scaling for an isothermal model.
         }
         \label{Fig_Mu_vs_dT}
\end{figure}
$\bullet$ {\it System 1} corresponds to the Warhol galaxy (panel C in Fig.~\ref{Fig_CritCurves}). This galaxy is extreme on several levels. It has the lowest redshift among all the lensed galaxies and is the closest to the BCG. The BCG's proximity gives System~1 the most  (alleged) microlensing events: five seen by HST plus seven seen by JWST.  Microlensing events are relatively short-lived, with durations between a few days to a few weeks, depending on several factors such as relative velocity, radius of the background star, mass of the microlens, or redshift of the background star \citep{Kelly2018,Diego2018,Diego2024b}.
It also has the largest time-delay factor with $T_0=3.14$ years (Table~\ref{Tab_1}, again mostly due to proximity to the BCG, which results in a large gradient in the lensing potential and hence a large contribution from the Shapiro term of the time delay, see Eq.~\ref{Eq_dT} below). This system contained the brightest event found by \cite{Yan2023}, denoted D21\_W1, which reached AB mag 27.1 in F277W during PEARLS's epoch~2. This event was located at $d_{\rm cc} = 0\farcs43$ in the negative-parity region, where the lens model predicts macromodel magnification $\mu=28$. The unusual brightness of the event at such a small macromodel magnification suggests the microlensing event may not have been a typical one.  Section~\ref{sect_discussion} discusses some possibilities. \\
%Additional observations will be needed in order to confirm the nature of  D21\_W1. \\
%The large time delay for  D21\_W1 does not rule out the possibility that such an event was a genuine microlensing event but offers an alternative explanation.  \\

$\bullet$ {\it System 2} corresponds to the Spock galaxy (panel D in Fig.~\ref{Fig_CritCurves}). \cite{Rodney2018,Diego2023b} described the lensing properties of this arc and discussed the difficulty of modelling the CC around the arc's position. Near this arc, there are several cluster galaxies with slightly different redshifts (a bimodal distribution), there is a $z=0.12$ foreground galaxy to the NW, and there are no other constraints for $z=1$. These properties make  modelling this portion of the lens plane challenging and uncertain. \cite{Rodney2018}  described how different models predict CCs that intersect the elongated Spock arc in different positions. There may be more than one caustic that crosses this arc, and if so, the distribution of magnifications and time delays would depart from the assumed scaling laws. (Section~\ref{sect_discussion} has further discussion.) This departure would be more evident on the side with negative parity, where member galaxies contribute more to the deflection, as shown by the large value of $\sigma_{\mu_0}^{-}$ in Table~\ref{Tab_1}. The time delay would also be affected, resulting in a poor fit to the $\Delta T=T_0*d_{\rm cc}$ scaling law. \\

$\bullet$ {\it System 6} is the most prominent arc in MACS0416 and one of the longest (panel B in Fig.~\ref{Fig_CritCurves}). No evidence of individual lensed stars has been found so far, but this arc is a prime target for future observations. Its higher redshift ($z=1.895$) makes it more challenging to identify lensed stars than in the Spock or Warhol galaxies because larger magnification is required to make any given star detectable. The macromodel magnification is also relatively modest, $\mu_0^{-}=16$ and $\mu_0^{+}=27$, partially explaining the lack of success at finding EMO-stars in this arc. 
Southwest of System~6 is System~11 (panel B in Fig.~\ref{Fig_CritCurves}), which is close to the caustic and shows a clear asymmetry between the two counterimages, with the west one being partially demagnified. This is possible if a small perturber is near that counterimage.  A faint galaxy is barely visible at that location (and is marked with a white arrow in Fig.~\ref{Fig_CritCurves}). This galaxy is likely the one responsible for the distortion in the magnification.   
\\

$\bullet$ {\it System 8} is a
%($z=1.953$). This 
colorful and very red arc containing two candidate EMO-stars (panel~E in Fig.~\ref{Fig_CritCurves}).  The two candidates are very close to the expected position of the CC. An alternative is that these two objects could be GCs in the cluster. Further evidence, for instance flux variability, is needed to confirm these candidates: GCs are not expected to vary, but EMO-stars do so because of both the ubiquitous microlenses and intrinsic variability. System~8 has a redshift similar to that of System~6, but it is crossing a more powerful caustic. This makes it a better candidate to search for EMO-stars. Next to System 8 is another caustic-crossing galaxy, System~101, at lower redshift but without any obvious EMO-stars.\\

$\bullet$ {\it System 12} is 
%($z=2.09$). This corresponds to 
the Mothra galaxy. This galaxy (panel C of Fig.~\ref{Fig_CritCurves}) is crossing one of the most powerful portions of the caustic, making it a great candidate for detecting EMO-stars. So far, though, only one EMO-star (Mothra) has been detected. Mothra's counterimage remains elusive, although it may be marginally detected in JWST data.
In their detailed study, \cite{Diego2023c} found that Mothra is most likely to be a double star milli-lensed by an undetected halo of mass ${\sim} 10^4$--10$^6$\,\Msun.
\citeauthor{Diego2023c} estimated Mothra to be only $d_{\rm cc}=0\farcs07$ from the CC  (this conclusion is reached in a model independent way as detailed in that work). At that distance, our lens model gives macro-model $\mu= 972$ and 716 for the counterimages with negative (the observed image) and positive parity,  respectively, before accounting for milli- or micro-lensing. Both \cite{Diego2023c} and \cite{Yan2023} found that the red component of the Mothra binary is intrinsically variable, a common feature of red supergiant stars. Deeper observations should reveal the positive counterimage, which should exhibit matching time variations in the red component with a  delay of just 13.5 days. (Photons from the observed counterimage arrive first.) \\

$\bullet$ {\it System 19}
%($z=2.243$). This galaxy 
is crossing a moderately powerful caustic. It shows two resolved, blue counterimages meeting at the CC (panel B of Fig.~\ref{Fig_CritCurves}). The resolved nature of these  counterimages and their length, $L\approx 0\farcs2$, indicate that the source must have a size $R\approx 8$\,pc.  It is most likely a star-forming region or a young GC. This source probably harbors a multitude of luminous stars, making this arc in principle a good candidate to detect individual micro-lensing events. However, as noted by \cite{Dai2021}, microlensing events in regions of high stellar density are harder to identify because they have smaller flux changes than the same micro-lensing of an isolated star. \\

$\bullet$ {\it System 34} (panel A of Fig.~\ref{Fig_CritCurves})
%At $z=3.235$ this galaxy 
is in principle a challenging target for EMO-stars because of its high redshift (Table~\ref{Tab_1}). However, this system is crossing the most powerful portion of the caustic among all galaxies listed in Table~\ref{Tab_1}, making it an interesting target. On the negative side, only the outer portion of the galaxy is touching the caustic, thus reducing the probability of seeing an EMO-star in this arc. The time delay for this arc is  small (although with large uncertainty): an intrinsically variable source located in the middle of one of the two blue images near the CC would have its counterimage arrive just 11 days later
%, so both sources would change in flux almost simultaneously 
(ignoring  micro-lensing, which could be different in each counterimage).\\

$\bullet$ {\it System 75}
%($z=6.629$, 
(panel A of Fig.~\ref{Fig_CritCurves}) was barely detected in the JWST images. It was originally identified with MUSE \citep{Vanzella2021,Bergamini2022}, thanks to emission lines. An EMO-star in this arc would be the most distant one ever observed, surpassing the current record holder, Earendel at $z\approx 6$ \citep{Welch2022}. No EMO-candidate was found near the CC position, but the relatively large values of $\mu_0^{-}$ and $\mu_0^{+}$ offer hope for a future detection of an EMO-star in this arc during a microlensing boost in its magnitude. \\

$\bullet$ {\it System 82} (panel G of Fig.~\ref{Fig_CritCurves})
%. At an estimated redshift of $z\approx2$ this arc 
has a good chance of revealing some of its brightest stars but has not yet done so. Two pairs of symmetric, unresolved counterimages are easily identified near the CC. (One close pair is inside the yellow circle in panel~G.) The outer pair is at  $d_{\rm cc}\approx 0\farcs3$, resulting in predicted macromodel magnifications of 89 and 99 for the counterimages with negative and positive parity, respectively. The time delay between these counterimages is 38 days. No  variability was seen in the PEARLS+CANUCS epochs, so these images are probably not microlensed and therefore are magnified by factors close to the macromodel values.  These magnifications are insufficient to make a red supergiant star detectable at $z=2$, even if caught during an outburst. The source in this case is probably a compact star-forming region with $R<250$\,pc. The inner pair is at $d_{\rm cc}\approx 0\farcs1$, resulting in macromodel magnifications three times larger than for the outer pair (270 and 300) and time delays three times shorter ($\approx$13~days). The constraint on the source size is likewise three times stronger, $R<83$\,pc. \\

%%Figure made by "screenshot". DS9 files in Dell /JWST/MACS0416/Data/CropImages/ 
%% Use Combo_*_V2.fits files for the RGB image.
\begin{figure*} 
   \includegraphics[width=18cm]{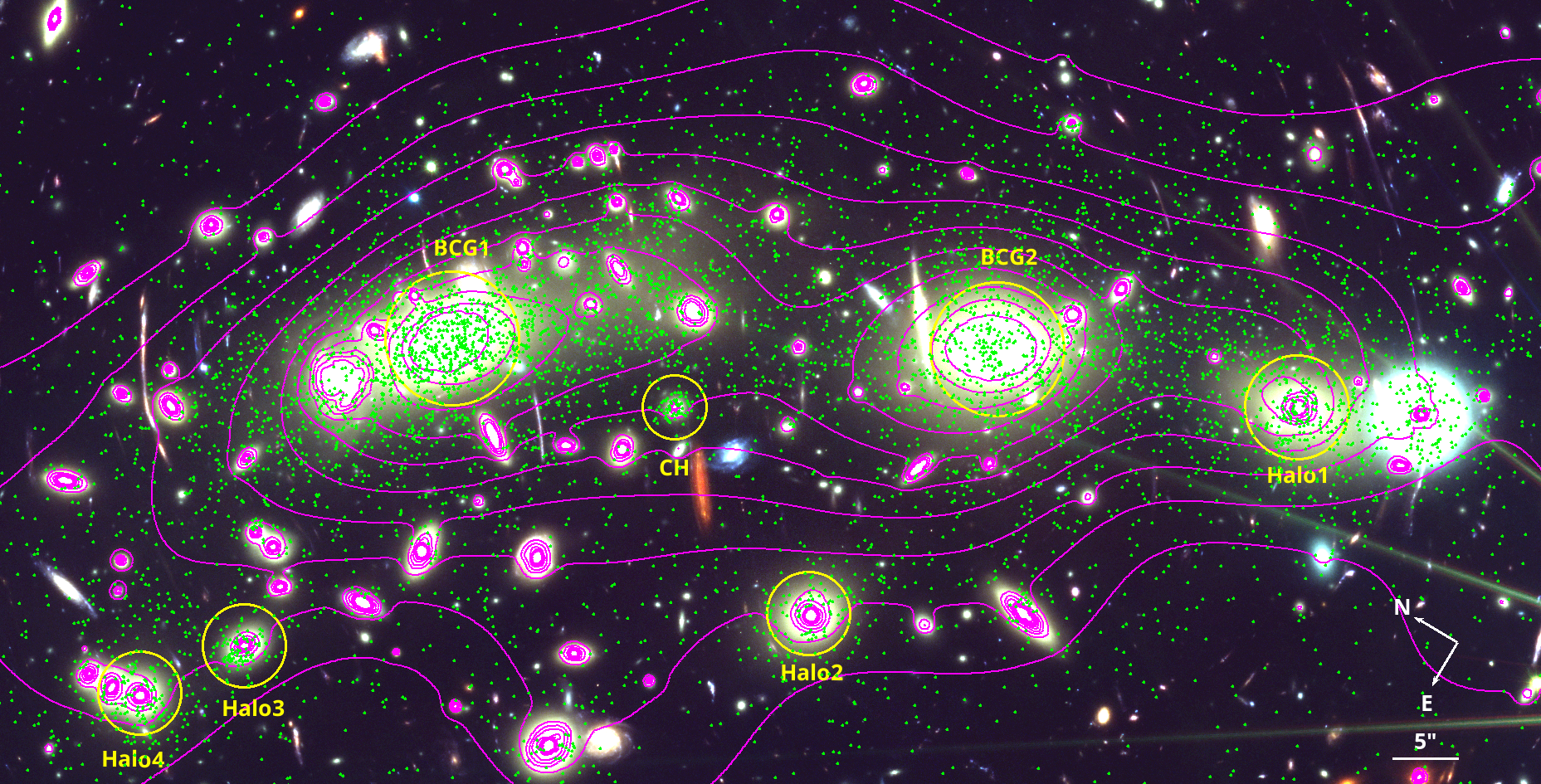}
      \caption{Globular clusters in MACS0416 (green dots) superposed on the JWST image. The seven yellow circles mark the positions of massive halos for which the lens model places good constraints on the halo masses. The size of each circle indicates the radius within which the mass was computed. The magenta contours are the convergence map, $\kappa$, from the lens model computed at $z=3$ and in intervals of 0.1, starting at $\kappa=0.5$. 
      %Note the good correspondence between the GC distribution, the ICL and the lens model distribution.
         }
         \label{Fig_GC_All}
\end{figure*}

$\bullet$ {\it System 94} (panel E of Fig.~\ref{Fig_CritCurves})
%. At relatively low redshift ($\gz \approx 1.3$), this galaxy 
crosses the caustic at one of the weakest points. That means EMO-stars can be found only very close to the CC even though the system has a low redshift. There is a red, unresolved source  halfway between the arc's counterimages, making the source a good candidate to be a pair of unresolved counterimages from a star located  $<$1\,pc from the cluster caustic. In this case, the net magnification (from both counterimages) must be $\mu>1500$ and the time delay between the two counterimages (assuming intrinsic variability) would be very short, $\Delta T < 4$ days, but these fluctuations would be observed at the same red dot but separated by regular intervals of $\Delta T$ days.\\

$\bullet$ {\it System 115} (panel A of Fig.~\ref{Fig_CritCurves})
%. As in system 82, this arc 
shows two pairs of resolved feature on either side of the CC. (As in System~82, one pair is inside the yellow circle in panel~A.) The situation is very similar to System 82 discussed above, but the System~115 caustic is significantly less powerful. Therefore the upper limits on the sizes are larger than the ones in System~82.\\

$\bullet$ {\it System 116} (panel F of Fig.~\ref{Fig_CritCurves})
was barely detected in JWST images. A long, elongated arc runs almost perpendicular to the CC, as expected for a strongly lensed galaxy in this position of the image plane. The geometric redshift is poorly constrained, but the photometric redshift is consistent with  $z\approx 8$. Like System 75, this arc is an excellent candidate to search for very high redshift EMO-stars, although it is even more challenging than System 75 given the higher redshift and smaller $\mu_0^{+}$ and $\mu_0^{-}$. Detecting an EMO-star in this arc will require a considerable boost from micro-lensing. No candidates were found in the current observations. \\
 
The model parameters discussed above imply a relation between the normalization constants $T_0$ and $\mu_0$. In general, the time delay
\begin{equation}
\label{Eq_dT}
\Delta T = \frac{D_sD_d}{D_{\rm ds}}\frac{(1+z_s)}{c}\left( \frac{1}{2}(\vec{\theta}-\vec{\beta})^2-\psi(\theta)\right)\quad,
\end{equation}
where the first and second terms are known as the geometric and Shapiro contributions to the time delay.  
For the popular singular isothermal sphere (SIS) model, the time delay is simply $\Delta T \propto (\theta_1^2- \theta_2^2)$, where $\theta_1$ and $\theta_2$ are the observed positions of the two counterimages. A simple rotation to make the deflection field parallel to the line connecting the two counterimages gives $\theta_1 = C + d_{\rm cc}$ and $\theta_2 = C - d_{\rm cc}$, where  $C$ is the position of the CC. That makes  $\Delta T \propto d_{\rm cc}$. Because near a CC $\mu \propto d_{\rm cc}^{-1}$,  the SIS model gives $T_0 \propto \mu_0^{-1}$.

The observed relation between $T_0$ and $\mu_0$, shown in Fig.~\ref{Fig_Mu_vs_dT}, follows the simple expectation but with some dispersion. The simple scaling $T_o=10/<\mu_0>$ reproduces to first order the relation between the amplitudes $T_0$ and $\mu_0$. This comes from the irregular and dynamically perturbed nature of MACS0416. 
There is an interesting trend with redshift. Arcs with lower redshift tend to have larger time delays and smaller magnification factors. This is a consequence of low-$z$ images appearing closer to the inner regions of the cluster. In these regions, the lensing potential changes more rapidly with distance. Because the magnification near CCs is inversely proportional to the gradient of the lensing potential, caustic-crossing arcs near the center of the cluster are generally less magnified. On the other hand, the larger potential near the center increases the Shapiro term in the time delay, resulting in counterimages from these galaxies having relatively longer time delays. The largest normalization factor $T_0$ is observed for System~1 at $z=0.9397$. This system is very close to the main BCG of the cluster, where the Shapiro delay is largest.

\section{The relation between mass and globular clusters}\label{s:GC}
%%%%%%%%%%%%%%%%%%%%%%%%%%%%%%%%%%%%%%%%%%%%%%%%%%%%%%%%%

%The lens model can be used also to study the correlation between the distribution of mass in the lens model and the spatial distribution of GCs. 
A scaling between the number of GCs and the mass of the halo hosting them has been established both from observations at lower redshifts and from $N$-body simulations \citep{Blakeslee1997,Harris2017,Burkert2020,Valenzuela2021,Dornan2023}. This relation is a powerful proxy for halo masses when lensing constraints or spectroscopic measurements (dynamical mass) are not available. Another use is to set prior masses for cluster galaxies and DM halos before starting to optimize a lens model. The relation can also be used as a sanity check of lens models because the derived lensing mass for each halo is expected to correlate with the observed number of GCs within the halo.  

\begin{figure} 
   \includegraphics[width=9cm]{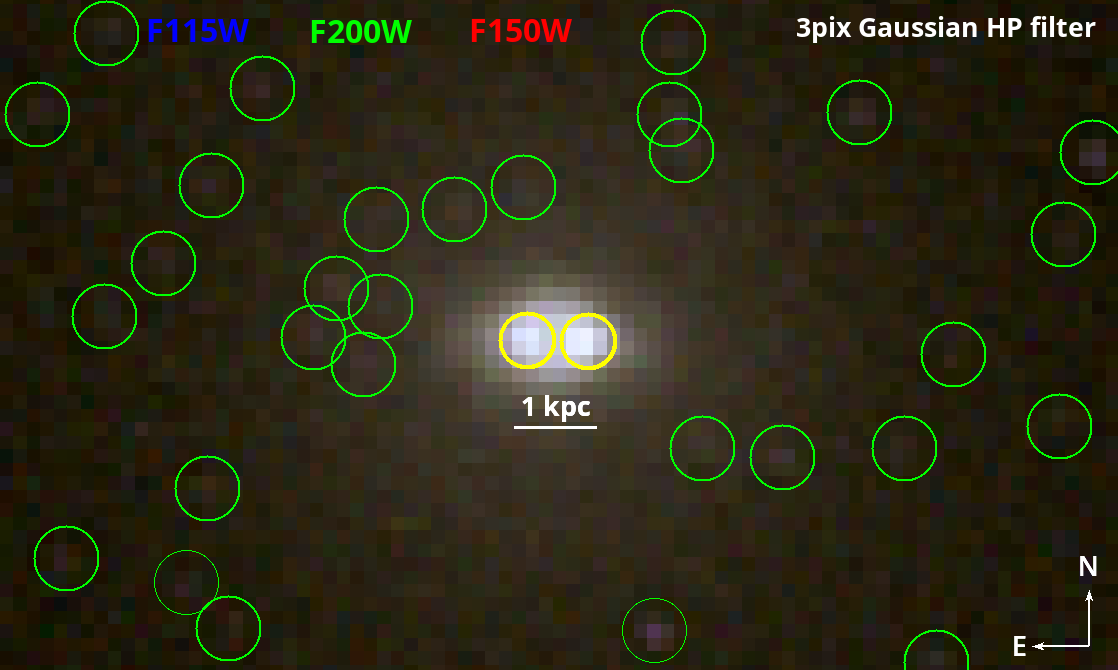}
      \caption{Central region of Halo~1 after high-pass spatial filtering. Scale and orientation are indicated. Green circles mark sources identified as GCs, and yellow circles mark the components of an apparent binary AGN. The two candidate SMBHs are separated by less than 1\,kpc in projection.
      The image was created by applying a 3-pixel Gaussian high-pass filter to three JWST images as marked at the top of the figure before combining them into the color composite.  
         }
         \label{Fig_DoubleBH}
\end{figure}

%Our lens model provides masses in the region constrained by multiple lensed galaxies, and 
The JWST data have the resolution and sensitivity needed to see many GCs in MACS0416. The smallest ones (in terms of mass and hence luminosity) will remain undetected, but there should still be a scaling between the observed number of GCs and the halo masses given by our lens model.
Appendix~\ref{Ap_B} gives details about the definition of the GC sample, and
Fig.~\ref{Fig_GC_All} shows the GCs detected in the region constrained by multiple lensed galaxies. This region contains  $N_{\rm GC}\approx 7000$ GCs. As expected, the GCs concentrate around the massive halos. Fig.~\ref{Fig_GC_All} also shows a good correspondence between the distribution of GCs, the distribution of projected mass (contours), and the ICL. 
Fig.~\ref{Fig_DoubleBH} shows, as one example,  GCs detected around Halo~1, which is particularly challenging for GC identification because it corresponds to a massive and luminous galaxy (with increased photon noise at its center). The image also shows what appears to be a binary supermassive black hole (SMBH) in the center of Halo 1. The binary becomes evident only after high-pass filtering. 
The compact core surrounding the binary nucleus is intriguing. Free-floating stars in the central region are expected to be scattered away by the scouring action of the putative binary SMBH \cite{Postman2012}.  Additional examples of GCs detected in JWST images around other massive halos are shown in Figs.~\ref{Fig_GC2a} and ~\ref{Fig_GC2b}. \\

%%%Figure made by "screenshot". DS9 files in Dell /JWST/MACS0416 
%\begin{figure} 
%   \includegraphics[width=9cm]{Figs/Gal_ClusterSystem.png}
%      \caption{A halo with a rich system of satellite unresolved clumps. The nature of these clumps remains unclear but they could be globular cluster or remnant galactic cores. The light profile is steep with a projected surface brightness scaling as $r^{-2}$ from the central point to the outskirts, and corresponding to a large Sersic index and small very effective radius. The center of this galaxy is likely hosting a SMBH.  
%         }
%         \label{Fig_Halo1}
%\end{figure}

%%Figure made by "screenshot". DS9 files in Dell /JWST/MACS0416 

\begin{figure} 
   \includegraphics[width=9cm]{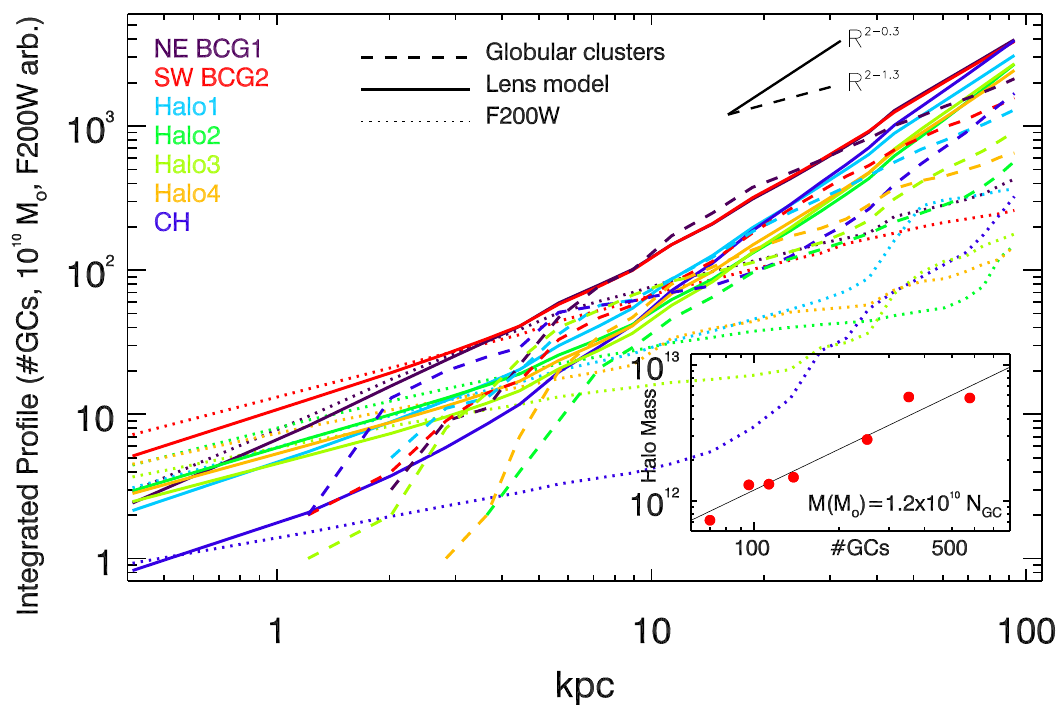}
      \caption{Profiles of the lensing mass, number of GCs, and light around seven halos. The solid lines show the projected mass profiles from the lens model as a function of radius, the dashed lines show the corresponding number-density profile of GCs, and the dotted lines show the F200W surface brightness. The profiles are centered on each of the seven halos identified in Fig.~\ref{Fig_GC_All} and plotted in different colors as shown in the legend.
      The inset shows the total mass versus number of GCs for each halo integrated up to the radius of the halo's yellow circle in Fig.~\ref{Fig_GC_All}. The solid line in the inset shows the relation $M=1.2\times10^{10}N_{\rm GC}$.
         }
         \label{Fig_Profiles_GC}
\end{figure}

At a more quantitative level, Fig.~\ref{Fig_Profiles_GC} shows the integrated profiles for seven well-constrained halos (marked by big yellow circles in Fig.~\ref{Fig_GC_All}). 
At distances $\la$30\,kpc, the number density of GCs scales as the projected mass with a slope in the number (or mass) surface density $\alpha \approx -0.3$. At larger radii, the surface number density of GCs falls faster, with  $\alpha \approx -1.3$, that is $\approx $1\,dex faster than the mass surface density,  in good agreement with results from simulations \citep{Schaller2015,Pillepich2018,Alonso2020} as well as earlier observational results from JWST in galaxy-cluster lenses \citep{Diego2023d}. 
The number of GCs found within each of the seven correlates with the halo projected mass within the same radius. The halo mass is reasonably well approximated by the law $M /\Msun \approx 1.2 \times10^{10} N_{\rm GC}$. 
(The circular regions where we computed the halo mass and $N_{\rm GC}$ are shown in Fig.~\ref{Fig_GC_All}.) We took smaller circles for smaller halos, but the circle size makes little difference provided the radius is in the range where the slopes of the solid-line and dashed-line profiles in Fig.~\ref{Fig_Profiles_GC} are similar.

%%Figure made by Dell /JWST/MACS0416/GlobularClusters/Profile_GC.pro

\section{Discussion}\label{sect_discussion}
\begin{figure} 
      \includegraphics[width=9cm]{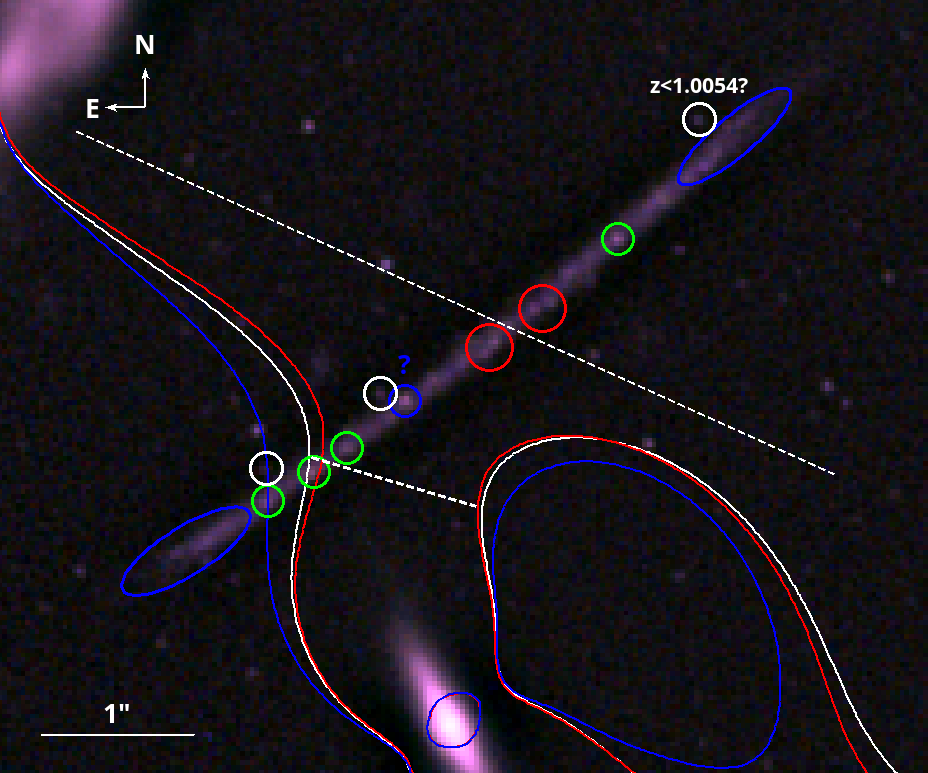} 
      \caption{High-pass filtered image of the Spock arc with CCs and possible interpretation of multiply imaged sources.  The white curve is the CC predicted by the full model (systems with rank A+B and adaptive grid with 495 grid points) at $z=1.0054$. The image shows multiple unresolved features along the Spock arc. Sources marked with the same color circle are assumed to be the same source that is multiply imaged, and the dashed lines mark the positions of the CC based on these sources.  The white circles mark a lensed galaxy 
 candidate at a slightly different redshift.
 If the mass of the smooth component is reduced slightly in this portion of the lens plane, the main CC would move towards the NW up to the position of the long dashed line ($\approx$0\farcs5 away).  The critical line around the member galaxy in the south of the image would then close (smaller dashed line), naturally producing the triplet marked with three green circles. The red and blue CCs are from two alternative models where the grid configuration is changed allowing a slightly different distribution for the smooth mass component. Because we did not use multiple knots in this arc as constraints, the lensing constraints are insufficient to properly reproduce the multiple crossing positions of the CC. 
         }
         \label{Fig_Spock}
\end{figure}

%%%%%%%%%%%%%%%%%%%%%%%%%%%%%%%%%%%%%%%%%%%%%
Despite the large number of available lensing constraints, the lens model still struggles to offer a satisfactory answer for some lensed galaxies. A remarkable example is the Spock arc (System~2), where the lens model is unable to place the CC at a viable symmetry point. In part this is because only two positions (marked by two blue ellipses in Fig.~\ref{Fig_Spock}) are available as constraints. These ellipses mark the only easily recognizable pair of counterimages in this arc. JWST images show several unresolved features along the arc, but the correspondence (in terms of pairs of counterimages) between them is unclear. Applying a high-pass filter to the data reveals a possible symmetry plane close to the midpoint of the arc (long dashed line in Fig.~\ref{Fig_Spock}). 
However, our lens model (solid white curve) does not place the CC close enough to that point. 
The same plot shows two alternative CCs derived by changing the grid configuration (one of the free variables in our reconstruction algorithm). 
The CC (white, red, and blue curves in Fig.~\ref{Fig_Spock}) intersects the arc just once, but a loop-like structure forms at $\approx$0\farcs5 from the arc and south of the possible symmetry point. 

A small modification in the lens model near the Spock arc could bring its loop-like CC to an intersection point with the arc at the presumed symmetry point. 
This would require a slight reduction in
%The lens model could place the CC through that point if 
the mass in the diffuse component (parameterized by the grid of Gaussian points). This may be possible if the number of constraints nearby, or in the arc itself, is increased to force the CC to pass through the intersection between the larger dashed line and the arc. However, these constraints can not be confirmed valid at the moment, so they are not used. Reducing the mass of the smooth component would also close the loop-like CC around the cluster galaxy at the bottom of Fig.~\ref{Fig_Spock}. This naturally produces a CC that grazes the SE portion of the arc and runs between the triplet marked by three green circles. This triplet's image system would then include the counterimage in the fourth green circle in the NW portion of the arc.  What would have been a single counterimage in the SE gets split into three counterimages because of the cluster galaxy to the south. In this interpretation, where the true CCs pass through the dashed lines, only two of the three white circles could be counterimages of a single source, and the third would be a different source seen in projection. The object marked with a blue circle and a question mark has no obvious counterpart. It could also be a projection effect (one of the many GCs) or a small structure (possibly an EMO-star) being magnified locally by a microlens or millilens.  
Additional lens models based on the available and future spectroscopic confirmation of some of the lensed galaxy candidates will provide further evidence in support of the single versus multiple caustic crossings of this arc. 

In conjunction with the lens model, the number of transient events reported by \cite{Yan2023} constrains the massive-star population in the Spock galaxy.  The observed number agrees with the low-end expectation of \cite{Diego2023b}, corresponding to a situation where SG stars above the Humphreys--Davidson \citep[HD,][]{HD1979} limit do not exist. The HD limit is an observational upper limit for the luminosity of red SG stars derived in the local Universe. In other words, the results from \cite{Yan2023} are consistent with a scenario where the HD limit is also valid at $z=1$. This limit may be a consequence of wind-driven mass loss in red SG stars \citep{Vink2023}, a possibility that can be tested with observations of low-metallicity EMO-stars at high-$z$. \\

%------------------------------------------------------------------------
%%Figure made by "screenshot". DS9 files in Dell /JWST/MACS0416 and JWST/MACS0416/CritCurves_JWST.odp

%In addition to the Spock arc, we have discussed the presence of other additional caustic crossing arcs. 
The amplitudes listed in Table~\ref{Tab_1} provide context to past transient events observed in arcs such as Spock, Warhol, and Mothra as well as for future transient events in these arcs.   JWST observes an average of $\approx$3 events per epoch at a depth of AB 29 magnitude in F200W (or $\approx$1 hour of integration time, \citealt{Yan2023}). 
One of the most intriguing events found by \cite{Yan2023} was D21\_W1 in the Warhol galaxy (panel C in Fig.~\ref{Fig_CritCurves}). 
Both D21\_W1 and its counterimage,  D21\_W1$'$, were detected in previous HST images from the HFF program as well as in the new JWST images.  Given the moderate macromodel magnification factor predicted from our lens model, $\mu^{-}=28$ and $\mu^{+}=46$ for D21\_W1 and  D21\_W1$'$, respectively, the source's intrinsic luminosity must be significantly larger than the luminosity of SG stars and more consistent with a GC or small star-forming region. 
If this source is indeed a GC at $z\approx 1$ (or similar object containing multiple stars), that would explain its brightness in previous HST images and during the JWST observations (in epoch~1 and before the transient was observed). However, this would make it difficult to explain the flux increase of $\approx 0.5$ magnitude in F200W between epoch~1 and epoch~2 of PEARLS, separated by 80~days).

A single star in a GC being microlensed results in a relatively small change in the observed flux because the increase in flux from the microlensed star still represents a relatively small fraction of the total flux from the entire GC\citep{Dai2021}. This is a similar argument to the one  that applies to larger sources, with radii much larger than the Einstein radius of a microlens (typically much smaller than 1 parsec) and for which microlensing effects are very small. 
Thus any source that shows significant change due to microlensing must be much smaller than 1 parsec. If the source is a single star, it must have an extraordinary brightness. At the redshift of Warhol, the distance modulus is 43.96. At macromodel magnification $\mu=30$, the source must have absolute magnitude ${\approx} {-12}$ in order to appear as $\rm AB \approx 28$ magnitude in JWST F200W. This would make D21\_W1 comparable to the brightest stars known in our local Universe, such as V4650 Sagittarii, a luminous blue variable (LBV). However, D21\_W1 exhibits a SED more compatible with a cooler star (typically fainter), making it difficult to explain its luminosity. A long-lasting (years) outburst episode in a cooler SG star could explain the high luminosity. This would be similar to the case of Godzilla, another EMO-star at $z=2.38$, that is suspected to be undergoing a major outburst lasting several years, similar to the Great Eruption of Eta Carinae in the 19$^{\rm th}$ century \citep{Diego2022_Godzilla}. 

Considering both luminosity and light curve, the transient event D21\_W1 \citep{Yan2023} can be explained in two alternative ways: 1) a secondary minor outburst within a major stellar-outburst event, or 2) a microlensing event of a single luminous star within a group of stars. 
In the first scenario, the outburst should be observed in both counterimages because it is intrinsic to the source. The counterimage, D21\_W1$'$, showed no variability in JWST data. However, this is consistent with our model prediction of  $\Delta T= 1.35$~years ($d_{\rm cc}=0\farcs43$) while the JWST observations span approximately 5~months. According to our model, if an outburst took place D21\_W1 during the second epoch of JWST observations, it would have appeared in D21\_W1$'$ $\approx$1.15 years before JWST first pointed to this cluster. In this case, the fact that no variability was observed in the counterimage during the JWST observations  implies that the minor outburst lasted less than $\approx$0.5 years in the rest frame of the source. 
The explanation of the second scenario above is much simpler. If a microlens temporarily boosted the magnification of  D21\_W1, we expect no correlation between  flux changes in the two images, and D21\_W1$'$ may maintain constant flux for long periods of time. Future observations spanning a period longer than $\Delta T$ are  needed to establish correlations between the two counterimages' variability.  Also, as discussed earlier, if the microlensing interpretation is correct, the source must be a small group of stars with the star undergoing microlensing being one of the dominant stars in that group, so its relative change in flux due to microlensing is significant when compared to the total flux of the group.
 The asymmetry in the light curves \citep{Yan2023} favors the outburst scenario because microlensing events tend to produce  symmetric light curves. However, the evidence in favor of the outburst scenario is not conclusive, and more data in future campaigns are needed.\\

%-----------------------------------------------------------------

Another interesting issue is the validity of the predicted correlation between the number of GCs and the lensing mass. Before JWST, GCs in galaxy clusters could be observed only in clusters at relatively low redshifts ($z\la0.2$). At these redshifts, the optical depth for lensing is also very small, so reliable lensing masses are not available for such low-$z$ halos. The notable exception is Abell 1689, for which HST was able to  observe both GCs in the cluster and a relatively large number (30) of lensed galaxies \cite{Broadhurst2005,AlamoMartinez2013}.  JWST can extend this type of study to larger redshifts and simultaneously detect hundreds of lensed galaxies behind massive clusters and thousands of GCs in the lensing cluster. 
As described in Section~\ref{s:GC}, there is a linear correlation between halo mass and number of GCs. 
This relation is similar to the one obtained from observations, $M_{\rm vir} = 5\times10^9\times N_{\rm GC}$\,\Msun\ \citep{Burkert2020} but with an amplitude a factor 2.4 times higher. The larger amplitude can be interpreted as a result of the fraction of fainter GCs that are missed in our sample due to the relatively high redshift of the cluster and the limited depth of the observations. This would imply that the total number of GCs in this cluster is $\approx$2.4 times higher than the observed number. This estimate is relevant for discussions of the lensing effect of GCs near the CC region because one can estimate the true number density of GCs near a certain region by simply counting the number of GCs in that region and correcting for a factor 2.4. The actual correction factor depends on the redshift of the lens and depth of the observations. Thus one can assess the probability that a GC is aligned with a particular feature in a lensed arc. For instance,  Mothra  was interpreted  \citep{Diego2023c} as a star at $z=2.091$ being magnified by a millilens in the cluster. The most likely candidate for the millilens is one of the multiple GCs around the NE BCG. The probability of a chance alignment of a GC with the background star was estimated as  $8\%$ based on the observed local number density of GCs near the Mothra position ($\approx$1.6 GCs per arcsecond$^2$). Correcting for the factor of 2.4, this probability increases to $\approx$20\%, making a GC an even more likely interpretation for the source millilensing Mothra. 

The logic for the abundance of GCs can be applied to other arcs. For instance, the Spock arc has a local number density of GCs similar to that of Mothra, so the probability of finding a GC millilens in a given position in that arc should also be $\approx$20\%. For Warhol, the probability should be a bit higher because the local number density of observed GCs is also higher because Warhol is closer to the BCG.  

The linear scaling between the number of GCs and halo mass is not maintained at larger radii. This is because the number density of GCs falls off more sharply beyond $\approx$30 kpc, and their density cannot keep up with the DM density, which falls more slowly with distance. A similar trend was observed in Abell 1689 \cite{AlamoMartinez2013}. Inside the regions shown in  Fig.~\ref{Fig_GC_All},  $M = 2.7 \times10^{14}$\,\Msun, and $N_{\rm GC}=6500$, while the $M$--$N_{\rm GC}$ law above would predict  $M = 7.8 \times10^{13}$ \Msun, that is $\approx$3.4 times less mass.  
The scaling between $M$ and $N_{\rm GC}$ is thus a good proxy for halo mass only up to a few tens of kpc from the center of the halo. As found previously, the DM halo for MACS0416 is shallower than the profile of the distribution of GCs. This is fully consistent  with the CDM model, where DM is composed of much lighter particles than the baryonic structures they interact with (gravitationally) such as stars, GCs, or even member galaxies, hence acquiring larger velocities in the constant 3-body interactions. Another way to look at the same process is that dynamical friction causes the most massive objects, such as GCs, to migrate towards the center of the halo. 

\section{Conclusions}\label{s:concl}
%%%%%%%%%%%%%%%%%%%%%%%%%%%%%%%%%%%%%%%%%
The new data from JWST reveal a wealth of new lensed galaxy candidates behind MACS0416.  The data bring the total number of lensed background galaxies to 119 galaxies (including candidates, a few of which may turn out to be false positives) producing a record number of 343 multiple images. With such a large number of constraints, free-form algorithms can produce reliable lens models while making a minimum number of assumptions regarding the distribution of the invisible dark matter. The free-form WSLAP+ model presented here was based on all the spectrocopic systems and the full set of constraints that includes also system candidates without spectroscopic confirmation. The redshifts for the systems without spectroscopic redshift were derived from the lens model (geo-$z$), which shows a good correlation with independent photo-$z$ estimates. The model also provides magnification and time-delay estimates for all lensed images. The derived lens model consists of two large halos with nearly identical mass profiles, in agreement with earlier results derived using parametric models and a similar set of spectroscopic systems. 

MACS0416 is exceptional in its number of caustic-crossing galaxies. We modelled the magnification and time delay for 12 of them, paying special attention to possible EMO-stars in these arcs. The amplitudes of the time-delay and magnification scaling relations are anticorrelated, as expected for smooth analytical models, but there are departures from perfect anti-correlation. These can be attributed to galactic-scale contributions to the deflection angle and the non-relaxed nature of the cluster. \\ 

For the Spock arc, the large-scale lens model fails to produce a CC at a viable symmetry point. Small changes to the lens model can make the CC intersect the arc at three points, thus explaining unresolved features seen in the new JWST data around these points. 
The candidate EMO-star, D21\_W1, is an intriguing and relatively bright transient found by \cite{Yan2023} in the Warhol arc at $z=0.9397$. 
Based on the model's time-delay prediction between this source and its counterimage  D21\_W1$'$, the source may be better described by an intrinsically variable  source than by a microlensing event. Spectral information rules out a hot LBV.  A cooler variable star that temporarily increased in brightness may offer a better explanation, but further monitoring is needed to rule out micro-lensing.

For the most massive halos in MACS0416,
there is a simple linear correlation between the lensing mass and number of GCs. This is the same as found at lower redshifts although the  normalization is higher (i.e., more mass for a given number of GCs). This is likely related to the incompleteness of our sample of GCs. The correlation is valid up to a radius of a few tens of kpc but not beyond.

\begin{acknowledgements}
%%%%%%%%%%%%%%%%%%%%%%%%%
We dedicate this paper to the memory of our dear PEARLS colleague Mario Nonino, who was a gifted 
and dedicated scientist, and a generous person.
This work is based on observations made with the NASA/ESA/CSA \textit{James Webb Space
Telescope}. The data were obtained from the Mikulski Archive for Space
Telescopes at the Space Telescope Science Institute, which is operated by the
Association of Universities for Research in Astronomy, Inc., under NASA
contract NAS 5-03127 for JWST. These observations are associated with JWST
programs 1176 and 2738.
 We thank the anonymous referee for useful comments and suggestions.
J.M.D. acknowledges the support of project PID2022-138896NB-C51 (MCIU/AEI/MINECO/FEDER, UE) Ministerio de Ciencia, Investigaci\'on y Universidades. 
RAW, SHC, and RAJ acknowledge support from NASA JWST Interdisciplinary
Scientist grants NAG5-12460, NNX14AN10G and 80NSSC18K0200 from GSFC. Work by
CJC acknowledges support from the European Research Council (ERC) Advanced
Investigator Grant EPOCHS (788113). BLF thanks the Berkeley Center for
Theoretical Physics for their hospitality during the writing of this paper.
MAM acknowledges the support of a National Research Council of Canada Plaskett
Fellowship, and the Australian Research Council Centre of Excellence for All
Sky Astrophysics in 3 Dimensions (ASTRO 3D), through project number CE17010001.
CNAW acknowledges funding from the JWST/NIRCam contract NASS-0215 to the
University of Arizona.
We thank the CANUCS team for generously providing early access to their proprietary data of MACS0416. 
AZ acknowledges support by Grant No. 2020750 from the United States-Israel Binational Science Foundation (BSF) and Grant No. 2109066 from the United States National Science Foundation (NSF); by the Ministry of Science \& Technology, Israel; and by the Israel Science Foundation Grant No. 864/23.
We also acknowledge the indigenous peoples of Arizona, including the Akimel
O'odham (Pima) and Pee Posh (Maricopa) Indian Communities, whose care and
keeping of the land has enabled us to be at ASU's Tempe campus in the Salt
River Valley, where much of our work was conducted.
%%% \end{acknowledgements}

%\sn 
%\software{ 
{\small SOFTWARE: }
Astropy: \url{http://www.astropy.org} \citep{Robitaille2013, Astropy2018};
IDL Astronomy Library: \url{https://idlastro.gsfc.nasa.gov} \citep{Landsman1993};
Photutils:
\url{https://photutils.readthedocs.io/en/stable/} \citep{Bradley20}; 
ProFound: \url{https://github.com/asgr/ProFound} \citep{Robotham2018}; 
ProFit: \url{https://github.com/ICRAR/ProFit} \citep{Robotham2017}; 
SExtractor: SourceExtractor:
\url{https://www.astromatic.net/software/sextractor/} or
\url{https://sextractor.readthedocs.io/en/latest/} \citep{Bertin1996}. 
%}

%\facilities{ 
{\small FACILITIES:} Hubble and James Webb Space Telescope Mikulski Archive
\url{https://archive.stsci.edu}\ 
%}

\end{acknowledgements}

\bibliographystyle{aa} %use aa.bst
\bibliography{MyBiblio} % References in MyBiblio.bib run bibtex after latex/pdflatex 

\begin{thebibliography}{85}
\expandafter\ifx\csname natexlab\endcsname\relax\def\natexlab#1{#1}\fi

\bibitem[{{Adams} {et~al.}(2024){Adams}, {Conselice}, {Austin}, {Harvey},
  {Ferreira}, {Trussler}, {Juod{\v{z}}balis}, {Li}, {Windhorst}, {Cohen},
  {Jansen}, {Summers}, {Tompkins}, {Driver}, {Robotham}, {D'Silva}, {Yan},
  {Coe}, {Frye}, {Grogin}, {Koekemoer}, {Marshall}, {Pirzkal}, {Ryan},
  {Maksym}, {Rutkowski}, {Willmer}, {Hammel}, {Nonino}, {Bhatawdekar},
  {Wilkins}, {Bradley}, {Broadhurst}, {Cheng}, {Dole}, {Hathi}, \&
  {Zitrin}}]{Adams2023}
{Adams}, N.~J., {Conselice}, C.~J., {Austin}, D., {et~al.} 2024, \apj, 965, 169

\bibitem[{{Alamo-Mart{\'\i}nez} {et~al.}(2013){Alamo-Mart{\'\i}nez},
  {Blakeslee}, {Jee}, {C{\^o}t{\'e}}, {Ferrarese},
  {Gonz{\'a}lez-L{\'o}pezlira}, {Jord{\'a}n}, {Meurer}, {Peng}, \&
  {West}}]{AlamoMartinez2013}
{Alamo-Mart{\'\i}nez}, K.~A., {Blakeslee}, J.~P., {Jee}, M.~J., {et~al.} 2013,
  \apj, 775, 20

\bibitem[{{Alonso Asensio} {et~al.}(2020){Alonso Asensio}, {Dalla Vecchia},
  {Bah{\'e}}, {Barnes}, \& {Kay}}]{Alonso2020}
{Alonso Asensio}, I., {Dalla Vecchia}, C., {Bah{\'e}}, Y.~M., {Barnes}, D.~J.,
  \& {Kay}, S.~T. 2020, \mnras, 494, 1859

\bibitem[{{Astropy Collaboration} {et~al.}(2018){Astropy Collaboration},
  {Price-Whelan}, {Sip{\H{o}}cz}, {G{\"u}nther}, {Lim}, {Crawford}, {Conseil},
  {Shupe}, {Craig}, {Dencheva}, {Ginsburg}, {VanderPlas}, {Bradley},
  {P{\'e}rez-Su{\'a}rez}, {de Val-Borro}, {Aldcroft}, {Cruz}, {Robitaille},
  {Tollerud}, {Ardelean}, {Babej}, {Bach}, {Bachetti}, {Bakanov}, {Bamford},
  {Barentsen}, {Barmby}, {Baumbach}, {Berry}, {Biscani}, {Boquien}, {Bostroem},
  {Bouma}, {Brammer}, {Bray}, {Breytenbach}, {Buddelmeijer}, {Burke},
  {Calderone}, {Cano Rodr{\'\i}guez}, {Cara}, {Cardoso}, {Cheedella}, {Copin},
  {Corrales}, {Crichton}, {D'Avella}, {Deil}, {Depagne}, {Dietrich}, {Donath},
  {Droettboom}, {Earl}, {Erben}, {Fabbro}, {Ferreira}, {Finethy}, {Fox},
  {Garrison}, {Gibbons}, {Goldstein}, {Gommers}, {Greco}, {Greenfield},
  {Groener}, {Grollier}, {Hagen}, {Hirst}, {Homeier}, {Horton}, {Hosseinzadeh},
  {Hu}, {Hunkeler}, {Ivezi{\'c}}, {Jain}, {Jenness}, {Kanarek}, {Kendrew},
  {Kern}, {Kerzendorf}, {Khvalko}, {King}, {Kirkby}, {Kulkarni}, {Kumar},
  {Lee}, {Lenz}, {Littlefair}, {Ma}, {Macleod}, {Mastropietro}, {McCully},
  {Montagnac}, {Morris}, {Mueller}, {Mumford}, {Muna}, {Murphy}, {Nelson},
  {Nguyen}, {Ninan}, {N{\"o}the}, {Ogaz}, {Oh}, {Parejko}, {Parley}, {Pascual},
  {Patil}, {Patil}, {Plunkett}, {Prochaska}, {Rastogi}, {Reddy Janga},
  {Sabater}, {Sakurikar}, {Seifert}, {Sherbert}, {Sherwood-Taylor}, {Shih},
  {Sick}, {Silbiger}, {Singanamalla}, {Singer}, {Sladen}, {Sooley},
  {Sornarajah}, {Streicher}, {Teuben}, {Thomas}, {Tremblay}, {Turner},
  {Terr{\'o}n}, {van Kerkwijk}, {de la Vega}, {Watkins}, {Weaver}, {Whitmore},
  {Woillez}, {Zabalza}, \& {Astropy Contributors}}]{Astropy2018}
{Astropy Collaboration}, {Price-Whelan}, A.~M., {Sip{\H{o}}cz}, B.~M., {et~al.}
  2018, \aj, 156, 123

\bibitem[{{Astropy Collaboration} {et~al.}(2013){Astropy Collaboration},
  {Robitaille}, {Tollerud}, {Greenfield}, {Droettboom}, {Bray}, {Aldcroft},
  {Davis}, {Ginsburg}, {Price-Whelan}, {Kerzendorf}, {Conley}, {Crighton},
  {Barbary}, {Muna}, {Ferguson}, {Grollier}, {Parikh}, {Nair}, {Unther},
  {Deil}, {Woillez}, {Conseil}, {Kramer}, {Turner}, {Singer}, {Fox}, {Weaver},
  {Zabalza}, {Edwards}, {Azalee Bostroem}, {Burke}, {Casey}, {Crawford},
  {Dencheva}, {Ely}, {Jenness}, {Labrie}, {Lim}, {Pierfederici}, {Pontzen},
  {Ptak}, {Refsdal}, {Servillat}, \& {Streicher}}]{Robitaille2013}
{Astropy Collaboration}, {Robitaille}, T.~P., {Tollerud}, E.~J., {et~al.} 2013,
  \aap, 558, A33

\bibitem[{{Bergamini} {et~al.}(2023){Bergamini}, {Grillo}, {Rosati},
  {Vanzella}, {Me{\v{s}}tri{\'c}}, {Mercurio}, {Acebron}, {Caminha}, {Granata},
  {Meneghetti}, {Angora}, \& {Nonino}}]{Bergamini2022}
{Bergamini}, P., {Grillo}, C., {Rosati}, P., {et~al.} 2023, \aap, 674, A79

\bibitem[{{Berkheimer} {et~al.}(2024){Berkheimer}, {Carleton}, {Windhorst},
  {Keel}, {Holwerda}, {Nonino}, {Cohen}, {Jansen}, {Coe}, {Conselice},
  {Driver}, {Frye}, {Grogin}, {Koekemoer}, {Lucas}, {Marshall}, {Pirzkal},
  {Robertson}, {Robotham}, {Ryan}, {Smith}, {Summers}, {Tompkins}, {Willmer},
  \& {Yan}}]{Berkheimer2024}
{Berkheimer}, J.~M., {Carleton}, T., {Windhorst}, R.~A., {et~al.} 2024, \apjl,
  964, L29

\bibitem[{{Bertin} \& {Arnouts}(1996)}]{Bertin1996}
{Bertin}, E. \& {Arnouts}, S. 1996, \aaps, 117, 393

\bibitem[{{Blakeslee} {et~al.}(1997){Blakeslee}, {Tonry}, \&
  {Metzger}}]{Blakeslee1997}
{Blakeslee}, J.~P., {Tonry}, J.~L., \& {Metzger}, M.~R. 1997, \aj, 114, 482

\bibitem[{{Bradley} {et~al.}(2020){Bradley}, {Sip{\H{o}}cz}, {Robitaille},
  {Tollerud}, {Vin{\'\i}cius}, {Deil}, {Barbary}, {Wilson}, {Busko},
  {G{\"u}nther}, {Cara}, {Conseil}, {Bostroem}, {Droettboom}, {Bray}, {Andersen
  Bratholm}, {Lim}, {Barentsen}, {Craig}, {Pascual}, {Perren}, {Greco},
  {Donath}, {De Val-Borro}, {Kerzendorf}, {Bach}, {Weaver}, {D'Eugenio},
  {Souchereau}, \& {Ferreira}}]{Bradley20}
{Bradley}, L., {Sip{\H{o}}cz}, B., {Robitaille}, T., {et~al.} 2020,
  {astropy/photutils: 1.0.0}, Zenodo

\bibitem[{{Brammer} {et~al.}(2008){Brammer}, {van Dokkum}, \&
  {Coppi}}]{Brammer2008}
{Brammer}, G.~B., {van Dokkum}, P.~G., \& {Coppi}, P. 2008, \apj, 686, 1503

\bibitem[{{Broadhurst} {et~al.}(2005){Broadhurst}, {Ben{\'\i}tez}, {Coe},
  {Sharon}, {Zekser}, {White}, {Ford}, {Bouwens}, {Blakeslee}, {Clampin},
  {Cross}, {Franx}, {Frye}, {Hartig}, {Illingworth}, {Infante}, {Menanteau},
  {Meurer}, {Postman}, {Ardila}, {Bartko}, {Brown}, {Burrows}, {Cheng},
  {Feldman}, {Golimowski}, {Goto}, {Gronwall}, {Herranz}, {Holden}, {Homeier},
  {Krist}, {Lesser}, {Martel}, {Miley}, {Rosati}, {Sirianni}, {Sparks},
  {Steindling}, {Tran}, {Tsvetanov}, \& {Zheng}}]{Broadhurst2005}
{Broadhurst}, T., {Ben{\'\i}tez}, N., {Coe}, D., {et~al.} 2005, \apj, 621, 53

\bibitem[{{Burkert} \& {Forbes}(2020)}]{Burkert2020}
{Burkert}, A. \& {Forbes}, D.~A. 2020, \aj, 159, 56

\bibitem[{{Caminha} {et~al.}(2016){Caminha}, {Grillo}, {Rosati}, {Balestra},
  {Karman}, {Lombardi}, {Mercurio}, {Nonino}, {Tozzi}, {Zitrin}, {Biviano},
  {Girardi}, {Koekemoer}, {Melchior}, {Meneghetti}, {Munari}, {Suyu}, {Umetsu},
  {Annunziatella}, {Borgani}, {Broadhurst}, {Caputi}, {Coe}, {Delgado-Correal},
  {Ettori}, {Fritz}, {Frye}, {Gobat}, {Maier}, {Monna}, {Postman}, {Sartoris},
  {Seitz}, {Vanzella}, \& {Ziegler}}]{Caminha2016}
{Caminha}, G.~B., {Grillo}, C., {Rosati}, P., {et~al.} 2016, \aap, 587, A80

\bibitem[{{Caminha} {et~al.}(2017){Caminha}, {Grillo}, {Rosati}, {Balestra},
  {Mercurio}, {Vanzella}, {Biviano}, {Caputi}, {Delgado-Correal}, {Karman},
  {Lombardi}, {Meneghetti}, {Sartoris}, \& {Tozzi}}]{Caminha2017}
{Caminha}, G.~B., {Grillo}, C., {Rosati}, P., {et~al.} 2017, \aap, 600, A90

\bibitem[{{Caminha} {et~al.}(2023){Caminha}, {Grillo}, {Rosati}, {Liu},
  {Acebron}, {Bergamini}, {Caputi}, {Mercurio}, {Tozzi}, {Vanzella}, {Demarco},
  {Frye}, {Rosani}, \& {Sharon}}]{Caminha2023}
{Caminha}, G.~B., {Grillo}, C., {Rosati}, P., {et~al.} 2023, \aap, 678, A3

\bibitem[{{Chen} {et~al.}(2019){Chen}, {Kelly}, {Diego}, {Oguri}, {Williams},
  {Zitrin}, {Treu}, {Smith}, {Broadhurst}, {Kaiser}, {Foley}, {Filippenko},
  {Salo}, {Hjorth}, \& {Selsing}}]{Chen2019}
{Chen}, W., {Kelly}, P.~L., {Diego}, J.~M., {et~al.} 2019, \apj, 881, 8

\bibitem[{{Dai}(2021)}]{Dai2021}
{Dai}, L. 2021, \mnras, 501, 5538

\bibitem[{{Diego} {et~al.}(2016{\natexlab{a}}){Diego}, {Broadhurst}, {Chen},
  {Lim}, {Zitrin}, {Chan}, {Coe}, {Ford}, {Lam}, \& {Zheng}}]{Diego2016}
{Diego}, J.~M., {Broadhurst}, T., {Chen}, C., {et~al.} 2016{\natexlab{a}},
  \mnras, 456, 356

\bibitem[{{Diego} {et~al.}(2015){Diego}, {Broadhurst}, {Molnar}, {Lam}, \&
  {Lim}}]{Diego2015b}
{Diego}, J.~M., {Broadhurst}, T., {Molnar}, S.~M., {Lam}, D., \& {Lim}, J.
  2015, \mnras, 447, 3130

\bibitem[{{Diego} {et~al.}(2016{\natexlab{b}}){Diego}, {Broadhurst}, {Wong},
  {Silk}, {Lim}, {Zheng}, {Lam}, \& {Ford}}]{Diego2016b}
{Diego}, J.~M., {Broadhurst}, T., {Wong}, J., {et~al.} 2016{\natexlab{b}},
  \mnras, 459, 3447

\bibitem[{{Diego} {et~al.}(2018{\natexlab{a}}){Diego}, {Kaiser}, {Broadhurst},
  {Kelly}, {Rodney}, {Morishita}, {Oguri}, {Ross}, {Zitrin}, {Jauzac},
  {Richard}, {Williams}, {Vega-Ferrero}, {Frye}, \& {Filippenko}}]{Diego2018}
{Diego}, J.~M., {Kaiser}, N., {Broadhurst}, T., {et~al.} 2018{\natexlab{a}},
  \apj, 857, 25

\bibitem[{{Diego} {et~al.}(2024{\natexlab{a}}){Diego}, {Li}, {Amruth}, {Meena},
  {Broadhurst}, {Kelly}, {Filippenko}, {Williams}, {Zitrin}, {Harris},
  {Reina-Campos}, {Giocoli}, {Dai}, {Struble}, {Treu}, {Fudamoto}, {Gilman},
  {Koekemoer}, {Lim}, {Palencia}, {Sun}, \& {Windhorst}}]{Diego2024b}
{Diego}, J.~M., {Li}, S.~K., {Amruth}, A., {et~al.} 2024{\natexlab{a}},
  accepted by \aap, arXiv:2404.08033

\bibitem[{{Diego} {et~al.}(2024{\natexlab{b}}){Diego}, {Li}, {Meena},
  {Niemiec}, {Acebron}, {Jauzac}, {Struble}, {Amruth}, {Broadhurst}, {Cerny},
  {Ebeling}, {Filippenko}, {Jullo}, {Kelly}, {Koekemoer}, {Lagattuta}, {Lim},
  {Limousin}, {Mahler}, {Patel}, {Remolina}, {Richard}, {Sharon}, {Steinhardt},
  {Umetsu}, {Williams}, {Zitrin}, {Palencia}, {Dai}, {Ji}, \&
  {Pascale}}]{Diego2023b}
{Diego}, J.~M., {Li}, S.~K., {Meena}, A.~K., {et~al.} 2024{\natexlab{b}}, \aap,
  681, A124

\bibitem[{{Diego} {et~al.}(2023{\natexlab{a}}){Diego}, {Meena}, {Adams},
  {Broadhurst}, {Dai}, {Coe}, {Frye}, {Kelly}, {Koekemoer}, {Pascale},
  {Willner}, {Zackrisson}, {Zitrin}, {Windhorst}, {Cohen}, {Jansen}, {Summers},
  {Tompkins}, {Conselice}, {Driver}, {Yan}, {Grogin}, {Marshall}, {Pirzkal},
  {Robotham}, {Ryan}, {Willmer}, {Bradley}, {Caminha}, {Caputi}, {Carleton}, \&
  {Kamieneski}}]{Diego2023a}
{Diego}, J.~M., {Meena}, A.~K., {Adams}, N.~J., {et~al.} 2023{\natexlab{a}},
  \aap, 672, A3

\bibitem[{{Diego} {et~al.}(2020){Diego}, {Molnar}, {Cerny}, {Broadhurst},
  {Windhorst}, {Zitrin}, {Bouwens}, {Coe}, {Conselice}, \&
  {Sharon}}]{Diego2020}
{Diego}, J.~M., {Molnar}, S.~M., {Cerny}, C., {et~al.} 2020, \apj, 904, 106

\bibitem[{{Diego} {et~al.}(2023{\natexlab{b}}){Diego}, {Pascale}, {Frye},
  {Zitrin}, {Broadhurst}, {Mahler}, {Caminha}, {Jauzac}, {Lee}, {Bae}, {Jang},
  \& {Montes}}]{Diego2023d}
{Diego}, J.~M., {Pascale}, M., {Frye}, B., {et~al.} 2023{\natexlab{b}}, \aap,
  679, A159

\bibitem[{{Diego} {et~al.}(2022){Diego}, {Pascale}, {Kavanagh}, {Kelly}, {Dai},
  {Frye}, \& {Broadhurst}}]{Diego2022_Godzilla}
{Diego}, J.~M., {Pascale}, M., {Kavanagh}, B.~J., {et~al.} 2022, \aap, 665,
  A134

\bibitem[{{Diego} {et~al.}(2005a){Diego}, {Protopapas}, {Sandvik}, \&
  {Tegmark}}]{Diego2005}
{Diego}, J.~M., {Protopapas}, P., {Sandvik}, H.~B., \& {Tegmark}, M. 2005a,
  \mnras, 360, 477

\bibitem[{{Diego} {et~al.}(2018{\natexlab{b}}){Diego}, {Schmidt}, {Broadhurst},
  {Lam}, {Vega-Ferrero}, {Zheng}, {Lee}, {Morishita}, {Bernstein}, {Lim},
  {Silk}, \& {Ford}}]{Diego2018b}
{Diego}, J.~M., {Schmidt}, K.~B., {Broadhurst}, T., {et~al.}
  2018{\natexlab{b}}, \mnras, 473, 4279

\bibitem[{{Diego} {et~al.}(2023{\natexlab{c}}){Diego}, {Sun}, {Yan}, {Furtak},
  {Zackrisson}, {Dai}, {Kelly}, {Nonino}, {Adams}, {Meena}, {Willner},
  {Zitrin}, {Cohen}, {D'Silva}, {Jansen}, {Summers}, {Windhorst}, {Coe},
  {Conselice}, {Driver}, {Frye}, {Grogin}, {Koekemoer}, {Marshall}, {Pirzkal},
  {Robotham}, {Rutkowski}, {Ryan}, {Tompkins}, {Willmer}, \&
  {Bhatawdekar}}]{Diego2023c}
{Diego}, J.~M., {Sun}, B., {Yan}, H., {et~al.} 2023{\natexlab{c}}, \aap, 679,
  A31

\bibitem[{{Diego} {et~al.}(2007){Diego}, {Tegmark}, {Protopapas}, \&
  {Sandvik}}]{Diego2007}
{Diego}, J.~M., {Tegmark}, M., {Protopapas}, P., \& {Sandvik}, H.~B. 2007,
  \mnras, 375, 958

\bibitem[{{Dornan} \& {Harris}(2023)}]{Dornan2023}
{Dornan}, V. \& {Harris}, W.~E. 2023, \apj, 950, 179

\bibitem[{{Ebeling} {et~al.}(2001){Ebeling}, {Edge}, \& {Henry}}]{Ebeling2001}
{Ebeling}, H., {Edge}, A.~C., \& {Henry}, J.~P. 2001, ApJ, 553, 668

\bibitem[{{Faisst} {et~al.}(2022){Faisst}, {Chary}, {Brammer}, \&
  {Toft}}]{Faisst2022}
{Faisst}, A.~L., {Chary}, R.~R., {Brammer}, G., \& {Toft}, S. 2022, \apjl, 941,
  L11

\bibitem[{{Furtak} {et~al.}(2024){Furtak}, {Meena}, {Zackrisson}, {Zitrin},
  {Brammer}, {Coe}, {Diego}, {Eldridge}, {Jim{\'e}nez-Teja}, {Kokorev},
  {Ricotti}, {Welch}, {Windhorst}, {Abdurro'uf}, {Andrade-Santos},
  {Bhatawdekar}, {Bradley}, {Broadhurst}, {Chen}, {Conselice}, {Dayal}, {Frye},
  {Fujimoto}, {Hsiao}, {Kelly}, {Mahler}, {Mandelker}, {Norman}, {Oguri},
  {Pirzkal}, {Postman}, {Ravindranath}, {Vanzella}, \& {Wilkins}}]{Furtak2024}
{Furtak}, L.~J., {Meena}, A.~K., {Zackrisson}, E., {et~al.} 2024, \mnras, 527,
  L7

\bibitem[{{Golubchik} {et~al.}(2023){Golubchik}, {Zitrin}, {Pierel}, {Furtak},
  {Meena}, {Graur}, {Kelly}, {Coe}, {Andrade-Santos}, {Asif}, {Bradley},
  {Chen}, {Frye}, {Gomez}, {Jha}, {Mahler}, {Nonino}, {Strolger}, \&
  {Su}}]{Golubchik2023}
{Golubchik}, M., {Zitrin}, A., {Pierel}, J., {et~al.} 2023, \mnras, 522, 4718

\bibitem[{{Harris} {et~al.}(2017){Harris}, {Blakeslee}, \&
  {Harris}}]{Harris2017}
{Harris}, W.~E., {Blakeslee}, J.~P., \& {Harris}, G. L.~H. 2017, \apj, 836, 67

\bibitem[{{Harris} \& {Reina-Campos}(2023)}]{Harris2023}
{Harris}, W.~E. \& {Reina-Campos}, M. 2023, \mnras, 526, 2696

\bibitem[{{Hernquist}(1990)}]{Hernquist1990}
{Hernquist}, L. 1990, \apj, 356, 359

\bibitem[{{Hoag} {et~al.}(2016){Hoag}, {Huang}, {Treu}, {Brada{\v{c}}},
  {Schmidt}, {Wang}, {Brammer}, {Broussard}, {Amorin}, {Castellano}, {Fontana},
  {Merlin}, {Schrabback}, {Trenti}, \& {Vulcani}}]{Hoag2016}
{Hoag}, A., {Huang}, K.~H., {Treu}, T., {et~al.} 2016, \apj, 831, 182

\bibitem[{{Humphreys} \& {Davidson}(1979)}]{HD1979}
{Humphreys}, R.~M. \& {Davidson}, K. 1979, \apj, 232, 409

\bibitem[{{Ishigaki} {et~al.}(2018){Ishigaki}, {Kawamata}, {Ouchi}, {Oguri},
  {Shimasaku}, \& {Ono}}]{Ishigaki2018}
{Ishigaki}, M., {Kawamata}, R., {Ouchi}, M., {et~al.} 2018, \apj, 854, 73

\bibitem[{{Jauzac} {et~al.}(2014){Jauzac}, {Cl{\'e}ment}, {Limousin},
  {Richard}, {Jullo}, {Ebeling}, {Atek}, {Kneib}, {Knowles}, {Natarajan},
  {Eckert}, {Egami}, {Massey}, \& {Rexroth}}]{Jauzac2014}
{Jauzac}, M., {Cl{\'e}ment}, B., {Limousin}, M., {et~al.} 2014, \mnras, 443,
  1549

\bibitem[{{Jauzac} {et~al.}(2015){Jauzac}, {Jullo}, {Eckert}, {Ebeling},
  {Richard}, {Limousin}, {Atek}, {Kneib}, {Cl{\'e}ment}, {Egami}, {Harvey},
  {Knowles}, {Massey}, {Natarajan}, {Neichel}, \& {Rexroth}}]{Jauzac2015}
{Jauzac}, M., {Jullo}, E., {Eckert}, D., {et~al.} 2015, \mnras, 446, 4132

\bibitem[{{Johnson} {et~al.}(2014){Johnson}, {Sharon}, {Bayliss}, {Gladders},
  {Coe}, \& {Ebeling}}]{Johnson2014}
{Johnson}, T.~L., {Sharon}, K., {Bayliss}, M.~B., {et~al.} 2014, \apj, 797, 48

\bibitem[{{Kaurov} {et~al.}(2019){Kaurov}, {Dai}, {Venumadhav},
  {Miralda-Escud{\'e}}, \& {Frye}}]{Kaurov2019}
{Kaurov}, A.~A., {Dai}, L., {Venumadhav}, T., {Miralda-Escud{\'e}}, J., \&
  {Frye}, B. 2019, \apj, 880, 58

\bibitem[{{Kawamata} {et~al.}(2016){Kawamata}, {Oguri}, {Ishigaki},
  {Shimasaku}, \& {Ouchi}}]{Kawamata2016}
{Kawamata}, R., {Oguri}, M., {Ishigaki}, M., {Shimasaku}, K., \& {Ouchi}, M.
  2016, \apj, 819, 114

\bibitem[{{Kelly} {et~al.}(2022){Kelly}, {Chen}, {Alfred}, {Broadhurst},
  {Diego}, {Emami}, {Filippenko}, {Keen}, {Kei Li}, {Lim}, {Meena}, {Oguri},
  {Scarlata}, {Treu}, {Williams}, {Williams}, {Zhou}, {Zitrin}, {Foley}, {Jha},
  {Kaiser}, {Mehta}, {Rieck}, {Salo}, {Smith}, \& {Weisz}}]{Kelly2023}
{Kelly}, P.~L., {Chen}, W., {Alfred}, A., {et~al.} 2022, arXiv e-prints,
  arXiv:2211.02670

\bibitem[{{Kelly} {et~al.}(2018){Kelly}, {Diego}, {Rodney}, {Kaiser},
  {Broadhurst}, {Zitrin}, {Treu}, {P{\'e}rez-Gonz{\'a}lez}, {Morishita},
  {Jauzac}, {Selsing}, {Oguri}, {Pueyo}, {Ross}, {Filippenko}, {Smith},
  {Hjorth}, {Cenko}, {Wang}, {Howell}, {Richard}, {Frye}, {Jha}, {Foley},
  {Norman}, {Bradac}, {Zheng}, {Brammer}, {Benito}, {Cava}, {Christensen}, {de
  Mink}, {Graur}, {Grillo}, {Kawamata}, {Kneib}, {Matheson}, {McCully},
  {Nonino}, {P{\'e}rez-Fournon}, {Riess}, {Rosati}, {Schmidt}, {Sharon}, \&
  {Weiner}}]{Kelly2018}
{Kelly}, P.~L., {Diego}, J.~M., {Rodney}, S., {et~al.} 2018, Nature Astronomy,
  2, 334

\bibitem[{{Lagattuta} {et~al.}(2017){Lagattuta}, {Richard}, {Cl{\'e}ment},
  {Mahler}, {Patr{\'\i}cio}, {Pell{\'o}}, {Soucail}, {Schmidt}, {Wisotzki},
  {Martinez}, \& {Bina}}]{Lagattuta2017}
{Lagattuta}, D.~J., {Richard}, J., {Cl{\'e}ment}, B., {et~al.} 2017, \mnras,
  469, 3946

\bibitem[{{Landsman}(1993)}]{Landsman1993}
{Landsman}, W.~B. 1993, in Astronomical Society of the Pacific Conference
  Series, Vol.~52, Astronomical Data Analysis Software and Systems II, ed.
  R.~J. {Hanisch}, R.~J.~V. {Brissenden}, \& J.~{Barnes}, 246

\bibitem[{{Larson} {et~al.}(2023){Larson}, {Hutchison}, {Bagley},
  {Finkelstein}, {Yung}, {Somerville}, {Hirschmann}, {Brammer}, {Holwerda},
  {Papovich}, {Morales}, \& {Wilkins}}]{Larson2022}
{Larson}, R.~L., {Hutchison}, T.~A., {Bagley}, M., {et~al.} 2023, \apj, 958,
  141

\bibitem[{{Lee} {et~al.}(2022){Lee}, {Bae}, \& {Jang}}]{Lee2022}
{Lee}, M.~G., {Bae}, J.~H., \& {Jang}, I.~S. 2022, \apjl, 940, L19

\bibitem[{{Lotz} {et~al.}(2017){Lotz}, {Koekemoer}, {Coe}, {Grogin}, {Capak},
  {Mack}, {Anderson}, {Avila}, {Barker}, {Borncamp}, {Brammer}, {Durbin},
  {Gunning}, {Hilbert}, {Jenkner}, {Khandrika}, {Levay}, {Lucas}, {MacKenty},
  {Ogaz}, {Porterfield}, {Reid}, {Robberto}, {Royle}, {Smith},
  {Storrie-Lombardi}, {Sunnquist}, {Surace}, {Taylor}, {Williams}, {Bullock},
  {Dickinson}, {Finkelstein}, {Natarajan}, {Richard}, {Robertson}, {Tumlinson},
  {Zitrin}, {Flanagan}, {Sembach}, {Soifer}, \& {Mountain}}]{Lotz2017}
{Lotz}, J.~M., {Koekemoer}, A., {Coe}, D., {et~al.} 2017, \apj, 837, 97

\bibitem[{{Mann} \& {Ebeling}(2012)}]{Mann2012}
{Mann}, A.~W. \& {Ebeling}, H. 2012, \mnras, 420, 2120

\bibitem[{{McLeod} {et~al.}(2015){McLeod}, {McLure}, {Dunlop}, {Robertson},
  {Ellis}, \& {Targett}}]{McLeod2015}
{McLeod}, D.~J., {McLure}, R.~J., {Dunlop}, J.~S., {et~al.} 2015, \mnras, 450,
  3032

\bibitem[{{Meena} {et~al.}(2023){Meena}, {Zitrin}, {Jim{\'e}nez-Teja},
  {Zackrisson}, {Chen}, {Coe}, {Diego}, {Dimauro}, {Furtak}, {Kelly}, {Oguri},
  {Welch}, {Abdurro'uf}, {Andrade-Santos}, {Adamo}, {Bhatawdekar},
  {Brada{\v{c}}}, {Bradley}, {Broadhurst}, {Conselice}, {Dayal}, {Donahue},
  {Frye}, {Fujimoto}, {Hsiao}, {Kokorev}, {Mahler}, {Vanzella}, \&
  {Windhorst}}]{Meena2023}
{Meena}, A.~K., {Zitrin}, A., {Jim{\'e}nez-Teja}, Y., {et~al.} 2023, \apjl,
  944, L6

\bibitem[{{Meneghetti} {et~al.}(2017){Meneghetti}, {Natarajan}, {Coe},
  {Contini}, {De Lucia}, {Giocoli}, {Acebron}, {Borgani}, {Bradac}, {Diego},
  {Hoag}, {Ishigaki}, {Johnson}, {Jullo}, {Kawamata}, {Lam}, {Limousin},
  {Liesenborgs}, {Oguri}, {Sebesta}, {Sharon}, {Williams}, \&
  {Zitrin}}]{Meneghetti2017}
{Meneghetti}, M., {Natarajan}, P., {Coe}, D., {et~al.} 2017, \mnras, 472, 3177

\bibitem[{{Miralda-Escude}(1991)}]{MiraldaEscude1991}
{Miralda-Escude}, J. 1991, \apj, 379, 94

\bibitem[{{Montes} \& {Trujillo}(2022)}]{Montes2022}
{Montes}, M. \& {Trujillo}, I. 2022, \apjl, 940, L51

\bibitem[{{Nagai} {et~al.}(2007){Nagai}, {Kravtsov}, \&
  {Vikhlinin}}]{Nagai2007}
{Nagai}, D., {Kravtsov}, A.~V., \& {Vikhlinin}, A. 2007, \apj, 668, 1

\bibitem[{{Navarro} {et~al.}(1997){Navarro}, {Frenk}, \& {White}}]{NFW}
{Navarro}, J.~F., {Frenk}, C.~S., \& {White}, S. D.~M. 1997, \apj, 490, 493

\bibitem[{{Oesch} {et~al.}(2015){Oesch}, {Bouwens}, {Illingworth}, {Franx},
  {Ammons}, {van Dokkum}, {Trenti}, \& {Labb{\'e}}}]{Oesch2015}
{Oesch}, P.~A., {Bouwens}, R.~J., {Illingworth}, G.~D., {et~al.} 2015, \apj,
  808, 104

\bibitem[{{Oke} \& {Gunn}(1983)}]{Oke1983}
{Oke}, J.~B. \& {Gunn}, J.~E. 1983, \apj, 266, 713

\bibitem[{{Pillepich} {et~al.}(2018){Pillepich}, {Nelson}, {Hernquist},
  {Springel}, {Pakmor}, {Torrey}, {Weinberger}, {Genel}, {Naiman}, {Marinacci},
  \& {Vogelsberger}}]{Pillepich2018}
{Pillepich}, A., {Nelson}, D., {Hernquist}, L., {et~al.} 2018, \mnras, 475, 648

\bibitem[{{Postman} {et~al.}(2012){Postman}, {Lauer}, {Donahue}, {Graves},
  {Coe}, {Moustakas}, {Koekemoer}, {Bradley}, {Ford}, {Grillo}, {Zitrin},
  {Lemze}, {Broadhurst}, {Moustakas}, {Ascaso}, {Medezinski}, \&
  {Kelson}}]{Postman2012}
{Postman}, M., {Lauer}, T.~R., {Donahue}, M., {et~al.} 2012, \apj, 756, 159

\bibitem[{{Richard} {et~al.}(2021){Richard}, {Claeyssens}, {Lagattuta},
  {Guaita}, {Bauer}, {Pello}, {Carton}, {Bacon}, {Soucail}, {Lyon}, {Kneib},
  {Mahler}, {Cl{\'e}ment}, {Mercier}, {Variu}, {Tamone}, {Ebeling}, {Schmidt},
  {Nanayakkara}, {Maseda}, {Weilbacher}, {Bouch{\'e}}, {Bouwens}, {Wisotzki},
  {de la Vieuville}, {Martinez}, \& {Patr{\'\i}cio}}]{Richard2021}
{Richard}, J., {Claeyssens}, A., {Lagattuta}, D., {et~al.} 2021, \aap, 646, A83

\bibitem[{{Rihtar{\v{s}}i{\v{c}}} {et~al.}(2024){Rihtar{\v{s}}i{\v{c}}},
  {Brada{\v{c}}}, {Desprez}, {Harshan}, {Noirot}, {Estrada-Carpenter},
  {Martis}, {Abraham}, {Asada}, {Brammer}, {Iyer}, {Matharu}, {Mowla},
  {Muzzin}, {Sarrouh}, {Sawicki}, {Strait}, {Willott}, {Gledhill}, {Markov}, \&
  {Tripodi}}]{Grego2024}
{Rihtar{\v{s}}i{\v{c}}}, G., {Brada{\v{c}}}, M., {Desprez}, G., {et~al.} 2024,
  arXiv e-prints, arXiv:2406.10332

\bibitem[{{Robotham} {et~al.}(2018){Robotham}, {Davies}, {Driver}, {Koushan},
  {Taranu}, {Casura}, \& {Liske}}]{Robotham2018}
{Robotham}, A.~S.~G., {Davies}, L.~J.~M., {Driver}, S.~P., {et~al.} 2018,
  \mnras, 476, 3137

\bibitem[{{Robotham} {et~al.}(2017){Robotham}, {Taranu}, {Tobar}, {Moffett}, \&
  {Driver}}]{Robotham2017}
{Robotham}, A.~S.~G., {Taranu}, D.~S., {Tobar}, R., {Moffett}, A., \& {Driver},
  S.~P. 2017, \mnras, 466, 1513

\bibitem[{{Rodney} {et~al.}(2018){Rodney}, {Balestra}, {Bradac}, {Brammer},
  {Broadhurst}, {Caminha}, {Chiriv{\i}}, {Diego}, {Filippenko}, {Foley},
  {Graur}, {Grillo}, {Hemmati}, {Hjorth}, {Hoag}, {Jauzac}, {Jha}, {Kawamata},
  {Kelly}, {McCully}, {Mobasher}, {Molino}, {Oguri}, {Richard}, {Riess},
  {Rosati}, {Schmidt}, {Selsing}, {Sharon}, {Strolger}, {Suyu}, {Treu},
  {Weiner}, {Williams}, \& {Zitrin}}]{Rodney2018}
{Rodney}, S.~A., {Balestra}, I., {Bradac}, M., {et~al.} 2018, Nature Astronomy,
  2, 324

\bibitem[{{Schaller} {et~al.}(2015){Schaller}, {Frenk}, {Bower}, {Theuns},
  {Trayford}, {Crain}, {Furlong}, {Schaye}, {Dalla Vecchia}, \&
  {McCarthy}}]{Schaller2015}
{Schaller}, M., {Frenk}, C.~S., {Bower}, R.~G., {et~al.} 2015, \mnras, 452, 343

\bibitem[{{Sendra} {et~al.}(2014){Sendra}, {Diego}, {Broadhurst}, \&
  {Lazkoz}}]{Sendra2014}
{Sendra}, I., {Diego}, J.~M., {Broadhurst}, T., \& {Lazkoz}, R. 2014, \mnras,
  437, 2642

\bibitem[{{Steinhardt} {et~al.}(2020){Steinhardt}, {Jauzac}, {Acebron}, {Atek},
  {Capak}, {Davidzon}, {Eckert}, {Harvey}, {Koekemoer}, {Lagos}, {Mahler},
  {Montes}, {Niemiec}, {Nonino}, {Oesch}, {Richard}, {Rodney}, {Schaller},
  {Sharon}, {Strolger}, {Allingham}, {Amara}, {Bah{\'e}}, {B{\oe}hm}, {Bose},
  {Bouwens}, {Bradley}, {Brammer}, {Broadhurst}, {Ca{\~n}as}, {Cen},
  {Cl{\'e}ment}, {Clowe}, {Coe}, {Connor}, {Darvish}, {Diego}, {Ebeling},
  {Edge}, {Egami}, {Ettori}, {Faisst}, {Frye}, {Furtak}, {G{\'o}mez-Guijarro},
  {Remolina Gonz{\'a}lez}, {Gonzalez}, {Graur}, {Gruen}, {Harvey}, {Hensley},
  {Hovis-Afflerbach}, {Jablonka}, {Jha}, {Jullo}, {Kneib}, {Kokorev},
  {Lagattuta}, {Limousin}, {von der Linden}, {Linzer}, {Lopez}, {Magdis},
  {Massey}, {Masters}, {Maturi}, {McCully}, {McGee}, {Meneghetti}, {Mobasher},
  {Moustakas}, {Murphy}, {Natarajan}, {Neyrinck}, {O'Connor}, {Oguri}, {Pagul},
  {Rhodes}, {Rich}, {Robertson}, {Sereno}, {Shan}, {Smith}, {Sneppen},
  {Squires}, {Tam}, {Tchernin}, {Toft}, {Umetsu}, {Weaver}, {van Weeren},
  {Williams}, {Wilson}, {Yan}, \& {Zitrin}}]{Steinhardt2020}
{Steinhardt}, C.~L., {Jauzac}, M., {Acebron}, A., {et~al.} 2020, \apjs, 247, 64

\bibitem[{{Valenzuela} {et~al.}(2021){Valenzuela}, {Moster}, {Remus},
  {O'Leary}, \& {Burkert}}]{Valenzuela2021}
{Valenzuela}, L.~M., {Moster}, B.~P., {Remus}, R.-S., {O'Leary}, J.~A., \&
  {Burkert}, A. 2021, \mnras, 505, 5815

\bibitem[{{Vanzella} {et~al.}(2021){Vanzella}, {Caminha}, {Rosati}, {Mercurio},
  {Castellano}, {Meneghetti}, {Grillo}, {Sani}, {Bergamini}, {Calura},
  {Caputi}, {Cristiani}, {Cupani}, {Fontana}, {Gilli}, {Grazian}, {Gronke},
  {Mignoli}, {Nonino}, {Pentericci}, {Tozzi}, {Treu}, {Balestra}, \&
  {Dijkstra}}]{Vanzella2021}
{Vanzella}, E., {Caminha}, G.~B., {Rosati}, P., {et~al.} 2021, \aap, 646, A57

\bibitem[{{Vink} \& {Sabhahit}(2023)}]{Vink2023}
{Vink}, J.~S. \& {Sabhahit}, G.~N. 2023, \aap, 678, L3

\bibitem[{{Welch} {et~al.}(2022){Welch}, {Coe}, {Diego}, {Zitrin},
  {Zackrisson}, {Dimauro}, {Jim{\'e}nez-Teja}, {Kelly}, {Mahler}, {Oguri},
  {Timmes}, {Windhorst}, {Florian}, {de Mink}, {Avila}, {Anderson}, {Bradley},
  {Sharon}, {Vikaeus}, {McCandliss}, {Brada{\v{c}}}, {Rigby}, {Frye}, {Toft},
  {Strait}, {Trenti}, {Sharma}, {Andrade-Santos}, \& {Broadhurst}}]{Welch2022}
{Welch}, B., {Coe}, D., {Diego}, J.~M., {et~al.} 2022, \nat, 603, 815

\bibitem[{{Willott} {et~al.}(2017){Willott}, {Abraham}, {Albert}, {Bradac},
  {Brammer}, {Chayer}, {Dixon}, {Doyon}, {Dupuis}, {Ferrarese}, {Goudfrooij},
  {Hutchings}, {Martel}, {Pacifici}, {Ravindranath}, \& {Sawicki}}]{Willot2017}
{Willott}, C.~J., {Abraham}, R.~G., {Albert}, L., {et~al.} 2017, {CANUCS: The
  CAnadian NIRISS Unbiased Cluster Survey}, JWST Proposal. Cycle 1, ID. \#1208

\bibitem[{{Windhorst} {et~al.}(2023){Windhorst}, {Cohen}, {Jansen}, {Summers},
  {Tompkins}, {Conselice}, {Driver}, {Yan}, {Coe}, {Frye}, {Grogin},
  {Koekemoer}, {Marshall}, {O'Brien}, {Pirzkal}, {Robotham}, {Ryan}, {Willmer},
  {Carleton}, {Diego}, {Keel}, {Porto}, {Redshaw}, {Scheller}, {Wilkins},
  {Willner}, {Zitrin}, {Adams}, {Austin}, {Arendt}, {Beacom}, {Bhatawdekar},
  {Bradley}, {Broadhurst}, {Cheng}, {Civano}, {Dai}, {Dole}, {D'Silva},
  {Duncan}, {Fazio}, {Ferrami}, {Ferreira}, {Finkelstein}, {Furtak}, {Gim},
  {Griffiths}, {Hammel}, {Harrington}, {Hathi}, {Holwerda}, {Honor}, {Huang},
  {Hyun}, {Im}, {Joshi}, {Kamieneski}, {Kelly}, {Larson}, {Li}, {Lim}, {Ma},
  {Maksym}, {Manzoni}, {Meena}, {Milam}, {Nonino}, {Pascale}, {Petric},
  {Pierel}, {del Carmen Polletta}, {R{\"o}ttgering}, {Rutkowski}, {Smail},
  {Straughn}, {Strolger}, {Swirbul}, {Trussler}, {Wang}, {Welch}, {B. Wyithe},
  {Yun}, {Zackrisson}, {Zhang}, \& {Zhao}}]{Windhorst2023}
{Windhorst}, R.~A., {Cohen}, S.~H., {Jansen}, R.~A., {et~al.} 2023, \aj, 165,
  13

\bibitem[{{Wyithe} {et~al.}(2001){Wyithe}, {Turner}, \& {Spergel}}]{Wyithe2001}
{Wyithe}, J.~S.~B., {Turner}, E.~L., \& {Spergel}, D.~N. 2001, \apj, 555, 504

\bibitem[{{Yan} {et~al.}(2023){Yan}, {Ma}, {Sun}, {Wang}, {Kelly}, {Diego},
  {Cohen}, {Windhorst}, {Jansen}, {Grogin}, {Beacom}, {Conselice}, {Driver},
  {Frye}, {Coe}, {Marshall}, {Koekemoer}, {Willmer}, {Robotham}, {D'Silva},
  {Summers}, {Nonino}, {Pirzkal}, {Ryan}, {Ortiz}, {Tompkins}, {Bhatawdekar},
  {Cheng}, {Zitrin}, \& {Willner}}]{Yan2023}
{Yan}, H., {Ma}, Z., {Sun}, B., {et~al.} 2023, \apjs, 269, 43

\bibitem[{{Zhao}(1996)}]{Zhao1996}
{Zhao}, H. 1996, \mnras, 278, 488

\bibitem[{{Zitrin} {et~al.}(2013){Zitrin}, {Meneghetti}, {Umetsu},
  {Broadhurst}, {Bartelmann}, {Bouwens}, {Bradley}, {Carrasco}, {Coe}, {Ford},
  {Kelson}, {Koekemoer}, {Medezinski}, {Moustakas}, {Moustakas}, {Nonino},
  {Postman}, {Rosati}, {Seidel}, {Seitz}, {Sendra}, {Shu}, {Vega}, \&
  {Zheng}}]{Zitrin2013a}
{Zitrin}, A., {Meneghetti}, M., {Umetsu}, K., {et~al.} 2013, \apjl, 762, L30

\end{thebibliography}

\begin{appendix}
%%%%%%%%%%%%%%%%%%%%%%%%%%%%%%%%%

%%%%%%%%%%%%%%%%%%%%%%%%%%%%%%%%%%%%%%%%%%%%%%%%%%%%%%%%%%%%%
\section{Globular clusters}\label{Ap_B}
%%%%%%%%%%%%%%%%%%%%%%%%%%%%%%%%%%%%%%%%%%%%%%%%%%%%%%%%%%%%%
To search for GCs, we first high-pass filter the three bands F115W, F150W, and F200W. Bands in the LW channel lack the resolution for GCs while the F090W is less sensitive than F115W. The high-pass filter is defined as the difference between the direct image (PSF convolved) and a filtered image with a Gaussian kernel with 3 pixels FWHM. This results in a filter very similar to an A Trous wavelet. After filtering the three bands, we visually search for unresolved sources in the combination color image. Since visual inspection is most sensitive to green-yellow colors, we set the filtered F200W as the green channel, since it offers the best combination between resolution and  sensitivity to detected GCs. Blue and red channels are set to F115W and F150W, respectively. Most GCs in our sample are seen in all three bands, but some are seen in just two. We reject candidates that are only seen in one band. Some of these single-band candidates can be noisy fluctuations but also faint GCs or very small high-redshift dropout galaxies \citep[see][for a discussion of contaminants in JWST images]{Berkheimer2024}. 

\begin{figure} 
   \includegraphics[width=9.1cm]{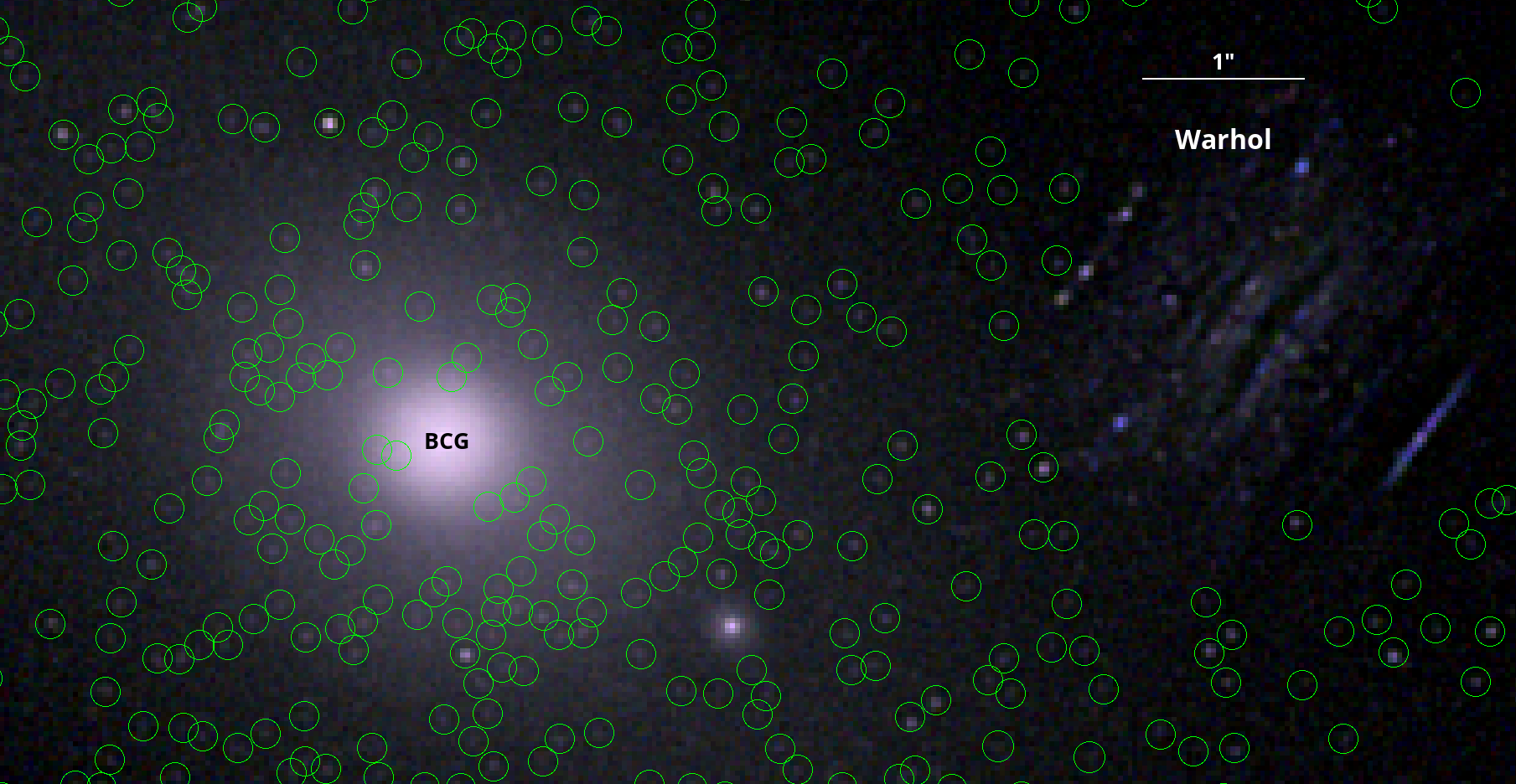}
      \caption{Detail of globular clusters found in MACS0416 (marked with green circles) detected around the NE BCG. The galaxy Warhol is shown in the top right. Unresolved features in Warhol and other resolved galaxies are not included in the sample of GCs. The image is a composition made from two filtered images (using a Gaussian Kernel with 3 and 15 pixels) and the raw data. The red channel is the F150W filter, the blue channel is the F115W. The most sensitive F200W is used for the green channel in order to maximize the effectiveness of the visual inspection.  
         }
         \label{Fig_GC2a}
\end{figure}

Detection of GCs is more challenging near crowded areas. For instance, around the BCGs, the larger fluctuations from photon noise make it harder to detect point sources. Detection of the same source in at least two bands is used to reduce false positives. The presence of background lensed galaxies adds another complication since these galaxies may contain unresolved features on their own that can be confused with GCs. This is shown in Fig.~\ref{Fig_GC2a}, where the lensed galaxy Warhol overlaps with a region containing a relatively high number density of GCs near the NE BCG. We avoid these regions and as a consequence we detect fewer GCs near background galaxies.  

\begin{figure} 
\begin{center}
   \includegraphics[width=8.2cm]{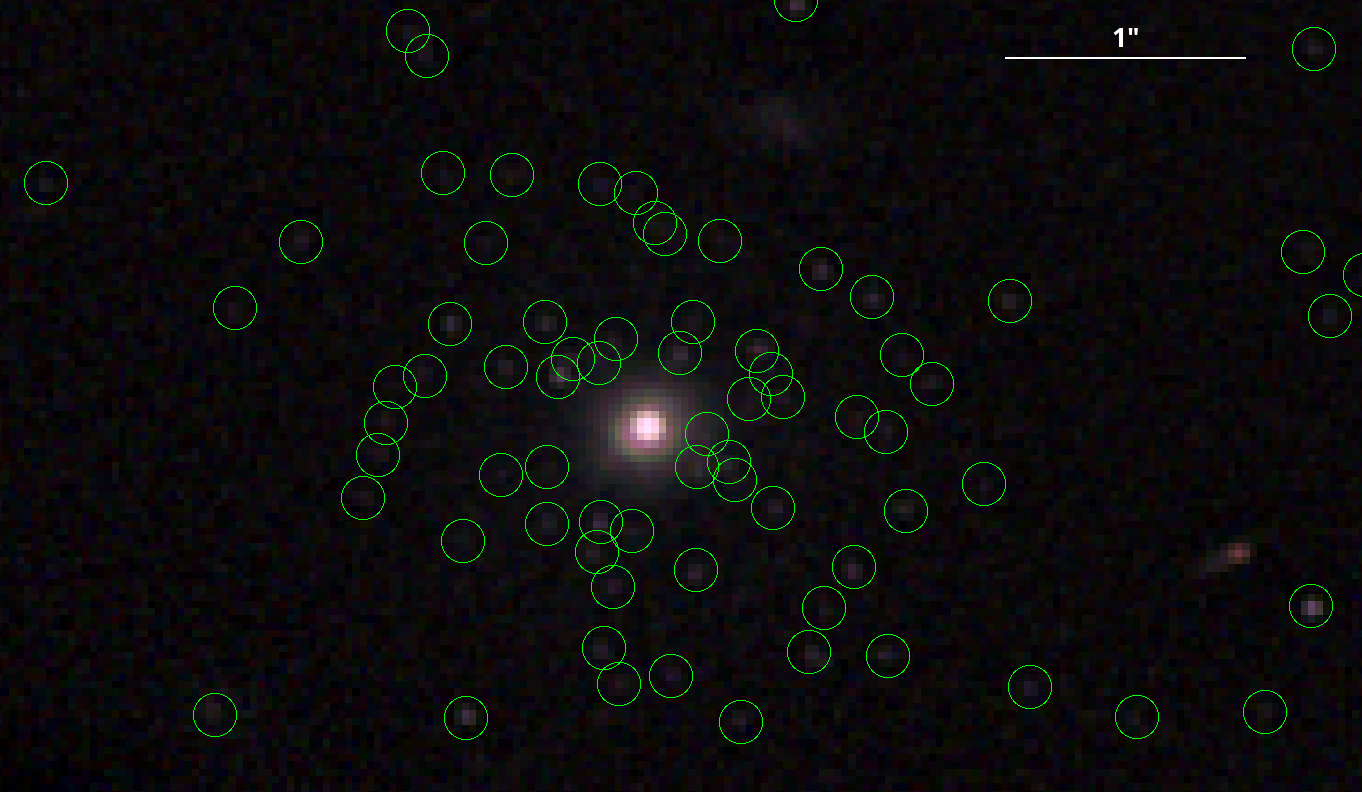}
      \caption{Detail of the rich GC structure found around the compact halo at RA=4:16:08.84, DEC=-24:04:20.12  (CG in Fig. 6). The image has been processed in a similar way to Fig.~\ref{Fig_GC2a}.    
         }
         \label{Fig_GC2b}
    
\end{center}         
\end{figure}

A different example is shown in Fig.~\ref{Fig_GC2b}, this time around the compact halo (labelled CH in Fig.~\ref{Fig_GC_All} and with steep profiles for the ICL and GCs, as shown in Fig.~\ref{Fig_Profiles_GC}). This halo is not affected by foreground lensed galaxies and due to its compact nature we can detect GCs much closer to the center of the halo. 

%%Figure made by "screenshot". DS9 files in Dell /JWST/MACS0416/Data/CropImages/ 
%% Use Combo_*_V2.fits files for the RGB image.

%%Figure made by "screenshot". DS9 files in Dell /JWST/MACS0416/Data/CropImages/ 
%% Use Combo_*_V2.fits files for the RGB image.

%\newpage

%%%%%%%%%%%%%%%%%%%%%%%%%%%%%%%%%
\section{Lensing constraints}\label{Ap_A}
%%%%%%%%%%%%%%%%%%%%%%%%%%%%%%%%%
%This appendix presents the strong lensing constraints used to derive the lens model. 
The lensing constraints used in this work are summarized in Table~\ref{tab_arcs}.\footnote{A region file with arc positions can be downloaded from https://zenodo.org/records/13304629}
The first 77 systems in our sample are the spectroscopic galaxies and are the same as the systems used in \cite{Bergamini2022,Diego2023d}. System 77 has the same redshift as System 33 forming a pair of lensed galaxies.  
Systems with ID 78 and above are candidate systems. Most of them are found in the new JWST data but some where also proposed as system candidates from previous HST data \citep[][]{Jauzac2014,Jauzac2015,Johnson2014, Diego2015b, Kawamata2016}. Arcs in the spectroscopic sample that are marked with \dag \, are new counterimages found in JWST data. Columns four, five, and six show the spectroscopic redshift, geometric redshift, and photometric redshift,  respectively for each system or arc. Images with negative photometric redshift are too faint to produce meaningful redshift estimates.
The geometric redshift for systems 113, 114, 116, and 117 are all above 10 and unreliable. These systems are either poorly fitted by the spectroscopic lens model (systems 113 and 114) or the geometric redshift prediction is also poor because the system has only two counterimages, with a relatively small angular separation between counterimages (systems 116 and 117). However, for System 117, both the photo-z and geo-z predict this galaxy to be at high redshift. Some of the photo-z estimates may be affected by contamination from the member galaxies ($z\approx 0.4$) resulting in photo-z estimates close to the cluster redshift. 
Column seven shows the magnification predicted by the full model and for each arc at the RA and DEC positions listed in columns 2 and 3, while column eight shows the time delay (at the same RA and DEC positions) relative to the first image to arrive from the same model. Finally, the last column indicates the rank of the images. Rank A is for images that are spectroscopically confirmed. Rank B are for images which are good candidates but lack spectroscopic confirmation (color match, morphology and parity, and good consistency with the spectroscopic lens model). Rank C are similar to Rank B images but usually lack morphological confirmation because they are unresolved. Finally images with rank D (two systems) are the least reliable because they are in tension with the spectroscopic lens model. 
System 113 (Rank D) is an interesting candidate for follow up since it could consist of a  background galaxy being lensed by an AGN or QSO at RA=4:16:08.5462, DEC=-24:05:22.004. It is not used as a lensing constraint but is included here as a possible interesting candidate. System 114 (D) appears lensed by a member galaxy but in a region of the cluster that is poorly constrained. The small separation between the counterimages of System 114 make it a relatively poor lensing constraint. 
Other interesting candidates, not included in the list are a small Einstein ring of a potentially high-redshift background galaxy lensed by a galaxy at RA=4:16:11.1176, DEC=-24:05:24.916. Several galaxy-galaxy lenses are also easily identifiable in the new JWST data and beyond the region constrained by the multiply lensed galaxies.  
Errors for geo-z are derived from the lens model that relies in rank A systems only. 
%Errors on photo-z are omitted because they are derived from a preliminary analysis. 
Errors in the time delay and magnification are not included because they are derived from the lens model including rank B systems, which lack spectroscopic confirmation. Hence, these magnification and time delays should be taken only as indicative values.

\begin{figure} 
   \includegraphics[width=9.1cm]{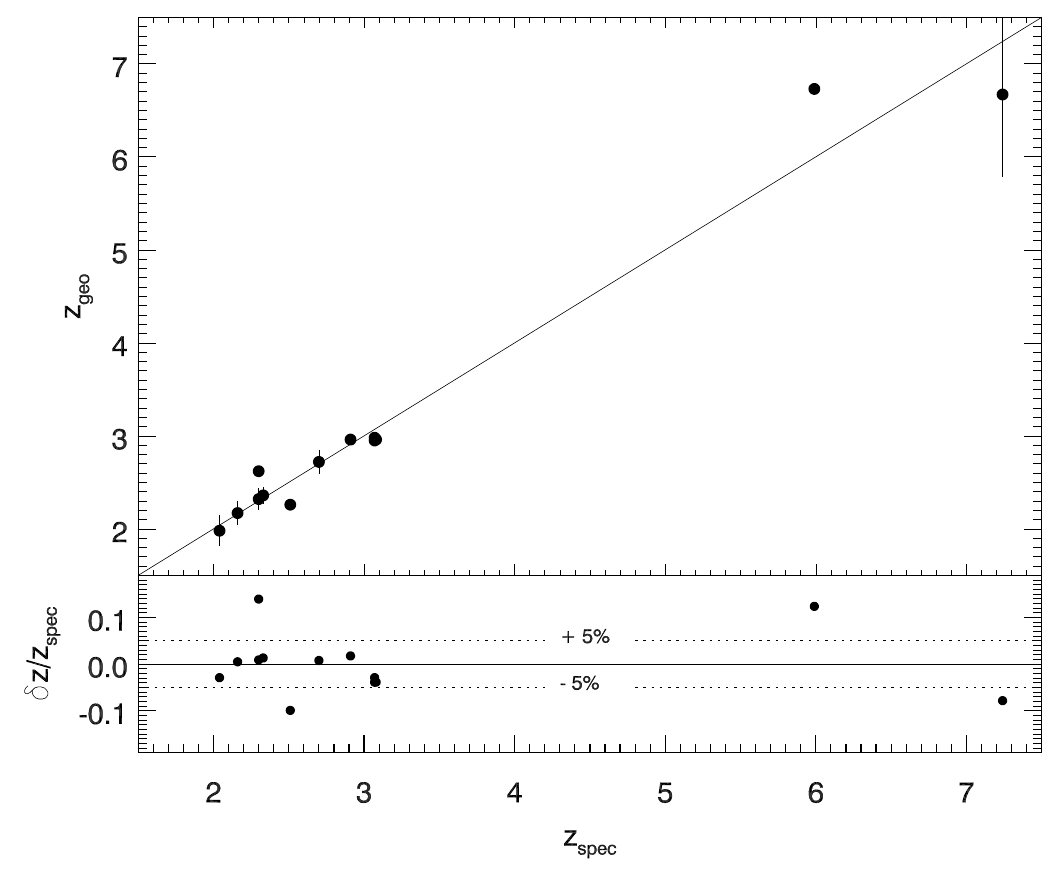}
      \caption{A posteriori confirmation of $z_{geo}$ with the new MUSE redshifts in \cite{Grego2024}. Uncertainties have been multiplied by a factor 4 to better show some of the error bars. The bottom panel shows $(z_{geo}-z_{spec})/z_{spec}$. 
      The two dotted lines in the bottom panel show the 5\% relative change. 
         }
         \label{Fig_Zgeo_vs_MUSE}
\end{figure}

During the last phases of the refereeing process, new spectroscopic redshifts for 13 of our system candidates were obtained by MUSE in \cite{Grego2024}. The new redshifts confirm these systems but more importantly demonstrates the accuracy of the geometric redshifts obtained from the well calibrated lens model. The comparison between our $z_{geo}$ and the new MUSE $z_{spec}$ is shown in Figure~\ref{Fig_Zgeo_vs_MUSE}. The 13 systems are 82, 87, 89, 90, 91, 95, 97, 98, 99, 100, 106, 109, and 115 in Table~\ref{tab_arcs} below. The bottom panel of the figure shows the relative difference between $z_{geo}$ and $z_{spec}$. Most systems fall within 5\% and five of them are within 2\% of the correct value. The worst outlier is 15\% above the correct value. The two high-z systems (97 and 100) have $z_{geo}$ within 12\% and 8\% of the correct value, respectively.  
%%Figure made by Toshiba /WSLAPplus/scripts/Plots_Zgeo_vs_Zspect_MACS0416_NewMUSE.pro

\clearpage

% Table B1 below removed following Editorial instructions. Available at CDS

%=====================================
% Files are made by Dell /JWST/MACS0416/PhotoZ/Match_NathanCatalog.pro & PlotZ_V4.pro
% Also see Toshiba /WSLAPplus/scripts/MakeTable4Appendix_MACS0416Paper_JWST.pro
%=====================================
\onecolumn
\begin{longtable}{rccccccrc}
\caption{Lensed families and multiple images \label{tab_arcs}}\\
\hline\hline
  \textbf{ID}
& \textbf{RA}
& \textbf{DEC}
& \textbf{\zs}
& \textbf{\zg}
& \textbf{\zp}
& \textbf{$\mu$}
& \textbf{$\delta T$ (yr)}
& \textbf{Rank} \\
\hline
\endfirsthead
\caption{Continued on next page.}\\
\hline\hline
  \textbf{ID}
& \textbf{RA}
& \textbf{DEC}
& \textbf{\zs}
& \textbf{\zg}
& \textbf{\zp}
& \textbf{$\mu$}
& \textbf{$\delta T$ (yr)}
& \textbf{Rank} \\
\hline
\endhead
\hline
\endfoot
\hline
\endlastfoot
%INSERT THE DATA HERE DIRECTLY, DO NOT USE INPUT{TABLE}
%\input{MyTable_MACS0416_JWST}
%     ID  &     RA    &    DEC     &    Zs   &   Zg   & Zphot &   Mu1 &    dT1 & Rank  \\
      1.a & 64.036575 & -24.067059 &  0.9397 &  0.923$\pm 0.007$ &  0.78 &   9.2 &     0  & A  \\
      1.b & 64.036919 & -24.067490 &    -    &    -              &  0.56 &  31.7 &   2.99 & A  \\
      1.c & 64.040962 & -24.071217 &    -    &    -              &  0.78 &   5.8 &   1.64 & A  \\
\hline
      2.a & 64.039337 & -24.070425 &  1.0054 &  1.005$\pm 0.010$ &  0.99 &  46.1 &  23.15 & A  \\
      2.b & 64.038368 & -24.069754 &    -    &    -              &  0.72 &  99.9 &  23.25 & A  \\
      2.c & 64.034309 & -24.066050 &    -    &    -              &  0.77 &   4.4 &     0  & A  \\
\hline
      3.a & 64.033348 & -24.067499 &  1.1472 &  1.146$\pm 0.001$ &  0.73 &   7.2 &     0  & A  \\
      3.b & 64.040443 & -24.073088 &    -    &    -              &  1.05 &   5.8 &   6.91 & A  \\
  3.c\dag & 64.035667 & -24.069759 &    -    &    -              &  0.59 &   3.3 &  11.67 & C  \\
\hline
      4.a & 64.026321 & -24.074366 &  1.6333 &  1.663$\pm 0.016$ &  1.63 &   4.2 &     0  & A  \\
      4.b & 64.031113 & -24.078981 &    -    &    -              &  1.70 &   3.5 &  12.87 & A  \\
      4.c & 64.035912 & -24.081350 &    -    &    -              &  1.62 &   5.7 &   6.39 & A  \\
\hline
      5.a & 64.023895 & -24.077621 &  1.8178 &  1.910$\pm 0.021$ &  1.70 &   3.4 &     0  & A  \\
      5.b & 64.030571 & -24.082706 &    -    &    -              &  0.56 &   6.1 &  11.35 & A  \\
      5.c & 64.032509 & -24.083782 &    -    &    -              &  1.80 &  10.8 &   9.86 & A  \\
\hline
      6.a & 64.041161 & -24.061836 &  1.8950 &  1.915$\pm 0.017$ &  1.93 &   9.1 &   0.89 & A  \\
      6.b & 64.043152 & -24.063120 &    -    &    -              &  1.89 &   8.5 &   2.25 & A  \\
      6.c & 64.047531 & -24.068851 &    -    &    -              &  1.90 &   4.0 &     0  & A  \\
\hline
      7.a & 64.040871 & -24.061646 &  1.8960 &  1.911$\pm 0.024$ &  1.86 &   7.5 &     0  & A  \\
      7.b & 64.043541 & -24.063534 &    -    &    -              &  1.78 &  20.7 &   2.39 & A  \\
      7.c & 64.047440 & -24.068699 &    -    &    -              &  1.71 &   4.2 &   0.69 & A  \\
\hline
      8.a & 64.029121 & -24.066780 &  1.9530 &  1.979$\pm 0.036$ &  2.36 &   3.6 &     0  & A  \\
      8.b & 64.037506 & -24.073175 &    -    &    -              &  1.18 &  12.3 &  29.54 & A  \\
      8.c & 64.038620 & -24.073906 &    -    &    -              &  1.27 &  68.9 &  29.20 & A  \\
\hline
      9.a & 64.024155 & -24.080921 &  1.9656 &  1.970$\pm 0.023$ &  1.78 &   7.5 &     0  & A  \\
      9.b & 64.028412 & -24.084572 &    -    &    -              &  1.84 &   9.9 &   1.82 & A  \\
      9.c & 64.031693 & -24.085789 &    -    &    -              &  1.98 &   6.4 &   0.72 & A  \\
\hline
     10.a & 64.030846 & -24.067158 &  1.9894 &  2.031$\pm 0.022$ &  2.29 &   5.4 &   1.25 & A  \\
     10.b & 64.035332 & -24.071020 &    -    &    -              &  2.28 &   4.7 &  12.13 & A  \\
     10.c & 64.041878 & -24.075760 &    -    &    -              &  2.29 &   4.8 &     0  & A  \\
\hline
     11.a & 64.039871 & -24.063120 &  2.0881 &  2.045$\pm 0.032$ &  2.29 &  31.9 &  11.77 & A  \\
     11.b & 64.040741 & -24.063623 &    -    &    -              &  1.98 &  14.3 &  12.07 & A  \\
     11.c & 64.047173 & -24.071136 &    -    &    -              &  2.09 &   3.1 &     0  & A  \\
\hline
     12.a & 64.036674 & -24.066154 &  2.0910 &  2.281$\pm 0.044$ &  2.00 &  33.7 &  37.35 & A  \\
     12.b & 64.036919 & -24.066378 &    -    &    -              &  1.97 &  41.3 &  37.50 & A  \\
 12.c\dag & 64.046150 & -24.075211 &    -    &    -              &  2.07 &   3.4 &     0  & B  \\
\hline
     13.a & 64.034409 & -24.063763 &  2.0943 &  2.061$\pm 0.002$ &  1.93 &   4.0 &     0  & A  \\
     13.b & 64.039650 & -24.066660 &    -    &    -              &  1.90 &   1.1 &  11.03 & A  \\
     13.c & 64.044632 & -24.072124 &    -    &    -              &  2.00 &   4.0 &   2.43 & A  \\
\hline
     14.a & 64.032478 & -24.068443 &  2.0948 &  2.147$\pm 0.019$ & -1.00 &  99.9 &  29.13 & A  \\
     14.b & 64.032730 & -24.068697 &    -    &    -              &  2.07 &  44.9 &  29.40 & A  \\
     14.c & 64.033592 & -24.069477 &    -    &    -              &  2.07 &  28.1 &  30.53 & A  \\
     14.d & 64.043633 & -24.076994 &    -    &    -              &  2.04 &   4.3 &     0  & A  \\
\hline
     15.a & 64.042274 & -24.060577 &  2.1067 &  2.054$\pm 0.002$ &  1.85 &   6.0 &     0  & A  \\
     15.b & 64.047539 & -24.066078 &    -    &    -              &  2.24 &  49.7 &   3.48 & A  \\
     15.c & 64.048187 & -24.066992 &    -    &    -              &  1.99 &   7.2 &   3.36 & A  \\
\hline
     16.a & 64.023552 & -24.081749 &  2.2182 &  2.203$\pm 0.039$ &  2.28 &   7.6 &     0  & A  \\
     16.b & 64.028671 & -24.086002 &    -    &    -              &  0.04 &  99.9 &   1.18 & A  \\
     16.c & 64.029900 & -24.086395 &    -    &    -              &  2.27 &  17.3 &   1.24 & A  \\
\hline
     17.a & 64.025124 & -24.083340 &  2.2210 &  2.009$\pm 0.038$ &  2.28 &  99.9 &   1.39 & A  \\
     17.b & 64.026176 & -24.084280 &    -    &    -              &  2.28 &  18.7 &   1.51 & A  \\
     17.c & 64.030945 & -24.086756 &    -    &    -              &  2.31 &   6.3 &     0  & A  \\
\hline
     18.a & 64.026718 & -24.074705 &  2.2300 &  2.286$\pm 0.036$ &  2.28 &   6.0 &  12.50 & A  \\
     18.b & 64.030228 & -24.078606 &    -    &    -              &  0.55 &   1.6 &  16.98 & A  \\
     18.c & 64.038177 & -24.082342 &    -    &    -              &  2.27 &   3.6 &     0  & A  \\
\hline
     19.a & 64.040993 & -24.062986 &  2.2430 &  2.168$\pm 0.038$ &  2.30 &  64.7 &  11.12 & A  \\
     19.b & 64.041161 & -24.063095 &    -    &    -              &  2.32 &  28.1 &  11.17 & A  \\
 19.c\dag & 64.047653 & -24.070810 &    -    &    -              &  1.76 &   3.1 &     0  & B  \\
\hline
     20.a & 64.025841 & -24.072264 &  2.2800 &  2.409$\pm 0.002$ &  1.89 &   6.3 &     0  & A  \\
     20.b & 64.032974 & -24.077028 &    -    &    -              &  0.53 &   2.2 &  20.27 & A  \\
     20.c & 64.037712 & -24.080503 &    -    &    -              &  2.29 &   6.4 &  15.50 & A  \\
\hline
     21.a & 64.034332 & -24.063032 &  2.2810 &  2.199$\pm 0.002$ &  2.17 &   3.1 &     0  & A  \\
     21.b & 64.040550 & -24.066360 &    -    &    -              &  2.01 &   6.9 &  14.06 & A  \\
     21.c & 64.044701 & -24.071520 &    -    &    -              &  2.16 &   4.6 &   9.90 & A  \\
\hline
     22.a & 64.026131 & -24.077278 &  2.2982 &  2.262$\pm 0.027$ &  1.94 &   8.9 &  13.33 & A  \\
     22.b & 64.028481 & -24.079771 &    -    &    -              &  2.24 &   6.9 &  14.64 & A  \\
     22.c & 64.036781 & -24.083941 &    -    &    -              & -1.00 &   3.1 &     0  & A  \\
\hline
     23.a & 64.026939 & -24.075771 &  2.3340 &  2.396$\pm 0.028$ &  2.31 &   8.8 &  18.28 & A  \\
     23.b & 64.029526 & -24.078621 &    -    &    -              &  0.52 &   4.6 &  19.99 & A  \\
     23.c & 64.038307 & -24.083006 &    -    &    -              &  2.31 &   3.2 &     0  & A  \\
\hline
     24.a & 64.024704 & -24.071487 &  2.5425 &  2.604$\pm 0.002$ &  2.70 &   3.6 &     0  & A  \\
     24.b & 64.032715 & -24.078537 &    -    &    -              &  2.68 &  12.0 &  29.50 & A  \\
     24.c & 64.035736 & -24.079952 &    -    &    -              &  2.62 &  13.4 &  29.42 & A  \\
\hline
     25.a & 64.036278 & -24.060682 &  2.9220 &  2.978$\pm 0.002$ &  2.68 &   2.5 &     0  & A  \\
     25.b & 64.044838 & -24.066723 &    -    &    -              &  0.29 &  99.9 &  14.83 & A  \\
     25.c & 64.046104 & -24.068829 &    -    &    -              &  2.72 &   8.5 &  14.91 & A  \\
\hline
     26.a & 64.026299 & -24.076046 &  2.9259 &  3.029$\pm 0.002$ &  3.19 &   7.7 &  20.77 & A  \\
     26.b & 64.028923 & -24.079155 &    -    &    -              &  2.79 &   4.9 &  21.32 & A  \\
 26.c\dag & 64.038338 & -24.083937 &    -    &    -              &  2.29 &   2.9 &     0  & B  \\
\hline
     27.a & 64.037872 & -24.068842 &  2.9911 &  2.836$\pm 0.002$ &  0.55 &   2.8 &  15.52 & A  \\
     27.b & 64.043831 & -24.073641 &    -    &    -              &  0.38 &   5.9 &   6.86 & A  \\
 27.c\dag & 64.032242 & -24.064041 &    -    &    -              &  3.07 &   3.4 &     0  & C  \\
\hline
     28.a & 64.028976 & -24.075569 &  3.0750 &  3.098$\pm 0.944$ &  4.08 &  99.9 &   0.13 & A  \\
     28.b & 64.029701 & -24.076210 &    -    &    -              &  4.67 &  43.0 &     0  & A  \\
\hline
     29.a & 64.025269 & -24.073608 &  3.0773 &  3.086$\pm 0.087$ &  2.76 &   4.0 &     0  & A  \\
     29.b & 64.030563 & -24.079248 &    -    &    -              &  3.31 &   3.1 &   9.78 & A  \\
     29.c & 64.037796 & -24.082417 &    -    &    -              &  2.76 &   4.1 &   0.41 & A  \\
\hline
     30.a & 64.025520 & -24.073679 &  3.1103 &  3.130$\pm 0.079$ &  2.70 &   4.3 &   3.36 & A  \\
     30.b & 64.030449 & -24.079046 &    -    &    -              &  0.57 &   2.7 &  11.77 & A  \\
     30.c & 64.038147 & -24.082426 &    -    &    -              &  2.72 &   3.9 &     0  & A  \\
\hline
     31.a & 64.027649 & -24.072704 &  3.2175 &  3.142$\pm 0.050$ &  2.71 &   8.6 &  20.35 & A  \\
     31.b & 64.032234 & -24.075134 &    -    &    -              &  3.00 &   4.7 &  22.84 & A  \\
     31.c & 64.040428 & -24.081505 &    -    &    -              & -1.00 &   4.3 &     0  & A  \\
\hline
     32.a & 64.026863 & -24.069242 &  3.2195 &  3.225$\pm 0.002$ &  2.73 &   4.6 &     0  & A  \\
     32.b & 64.034866 & -24.074615 &    -    &    -              &  3.01 &   5.7 &  17.74 & A  \\
     32.c & 64.039764 & -24.078896 &    -    &    -              &  2.77 &   6.1 &  13.39 & A  \\
\hline
     33.a & 64.034569 & -24.066988 &  3.2215 &  3.213$\pm 0.002$ &  2.78 &  99.9 &  43.08 & A  \\
     33.b & 64.034264 & -24.066517 &    -    &    -              &  0.33 &   7.0 &  43.19 & A  \\
     33.c & 64.034042 & -24.066940 &    -    &    -              &  0.45 &   2.6 &  43.21 & A  \\
     33.d & 64.046127 & -24.076797 &    -    &    -              &  2.71 &   4.3 &     0  & A  \\
     33.e & 64.034081 & -24.066479 &    -    &    -              &  2.79 &  19.2 &  43.18 & A  \\
\hline
     34.a & 64.046532 & -24.060431 &  3.2355 &  3.331$\pm 0.345$ &  3.17 &  34.6 &   0.08 & A  \\
     34.b & 64.047035 & -24.060827 &    -    &    -              &  2.74 &  99.9 &   0.09 & A  \\
     34.c & 64.049156 & -24.062895 &    -    &    -              &  3.19 &  99.9 &     0  & A  \\
\hline
     35.a & 64.038429 & -24.084162 &  3.2526 &  3.206$\pm 0.078$ &  2.71 &   2.9 &     0  & A  \\
     35.b & 64.026405 & -24.076729 &    -    &    -              &  3.06 &  13.5 &  23.47 & A  \\
     35.c & 64.028404 & -24.079033 &    -    &    -              &  2.94 &   7.3 &  23.30 & A  \\
\hline
     36.a & 64.040154 & -24.066771 &  3.2882 &  3.161$\pm 0.125$ &  2.69 &   2.4 &   5.71 & A  \\
     36.b & 64.045181 & -24.072372 &    -    &    -              &  3.06 &   4.4 &     0  & A  \\
\hline
     37.a & 64.040070 & -24.066681 &  3.2883 &  3.098$\pm 0.098$ & -1.00 &   1.7 &   7.19 & A  \\
     37.b & 64.045425 & -24.072557 &    -    &    -              &  2.79 &   4.3 &     0  & A  \\
\hline
     38.a & 64.041634 & -24.060032 &  3.2885 &  3.319$\pm 0.105$ &  3.04 &   5.7 &   1.04 & A  \\
     38.b & 64.045334 & -24.062792 &    -    &    -              &  3.02 &  15.9 &   2.09 & A  \\
     38.c & 64.049309 & -24.068213 &    -    &    -              &  2.71 &   4.2 &     0  & A  \\
\hline
     39.a & 64.023926 & -24.075027 &  3.2910 &  3.206$\pm 0.042$ &  3.13 &   3.4 &     0  & A  \\
     39.b & 64.030174 & -24.080952 &    -    &    -              &  4.19 &   4.4 &  11.39 & A  \\
     39.c & 64.035744 & -24.083611 &    -    &    -              &  0.28 &   4.5 &   7.29 & A  \\
\hline
     40.a & 64.045586 & -24.072701 &  3.2920 &  3.079$\pm 0.082$ &  3.00 &   4.3 &     0  & A  \\
     40.b & 64.040001 & -24.066645 &    -    &    -              &  2.70 &   1.4 &   8.29 & A  \\
\hline
     41.a & 64.046913 & -24.075411 &  3.2922 &  3.316$\pm 0.080$ &  2.98 &   3.7 &     0  & A  \\
     41.b & 64.035301 & -24.064764 &    -    &    -              &  2.72 &  10.3 &  35.99 & A  \\
     41.c & 64.038589 & -24.065998 &    -    &    -              &  0.55 &  10.2 &  36.25 & A  \\
\hline
     42.a & 64.022774 & -24.074625 &  3.4406 &  3.493$\pm 0.002$ &  2.72 &   2.8 &     0  & A  \\
     42.b & 64.031319 & -24.081928 &    -    &    -              &  2.72 &   6.2 &  21.65 & A  \\
     42.c & 64.033714 & -24.083220 &    -    &    -              &  2.72 &  14.4 &  21.88 & A  \\
\hline
     43.a & 64.025024 & -24.075050 &  3.4909 &  3.417$\pm 0.056$ &  3.55 &   4.4 &   6.99 & A  \\
     43.b & 64.029488 & -24.079918 &    -    &    -              &  3.56 &   4.3 &  12.01 & A  \\
     43.c & 64.037552 & -24.083683 &    -    &    -              &  3.65 &   3.3 &     0  & A  \\
\hline
     44.a & 64.037750 & -24.060776 &  3.6065 &  3.574$\pm 0.065$ &  0.37 &   3.4 &   0.51 & A  \\
     44.b & 64.043709 & -24.064419 &    -    &    -              &  2.98 &   6.0 &   4.67 & A  \\
     44.c & 64.047882 & -24.070196 &    -    &    -              & -1.00 &   3.7 &     0  & A  \\
\hline
     45.a & 64.023621 & -24.073181 &  3.7153 &  4.005$\pm 0.002$ &  0.15 &   3.2 &     0  & A  \\
     45.b & 64.031158 & -24.080338 &    -    &    -              &  3.49 &   4.5 &  19.40 & A  \\
     45.c & 64.035950 & -24.082600 &    -    &    -              &  2.67 &   6.0 &  18.18 & A  \\
\hline
     46.a & 64.026932 & -24.069977 &  3.8710 &  3.651$\pm 0.086$ &  0.20 &   5.3 &     0  & A  \\
     46.b & 64.034012 & -24.074596 &    -    &    -              &  0.58 &   4.6 &  11.16 & A  \\
     46.c & 64.040260 & -24.079895 &    -    &    -              &  0.87 &   5.6 &   1.19 & A  \\
\hline
     47.a & 64.023819 & -24.078503 &  3.9228 &  3.598$\pm 0.002$ &  0.13 &   5.3 &   5.27 & A  \\
     47.b & 64.027702 & -24.082638 &    -    &    -              &  4.63 &   7.2 &   7.00 & A  \\
 47.c\dag & 64.035065 & -24.085680 &    -    &    -              &  0.91 &   3.4 &     0  & B  \\
\hline
     48.a & 64.027199 & -24.073603 &  3.9230 &  3.393$\pm 0.254$ &  3.83 &   8.2 &     0  & A  \\
     48.b & 64.031067 & -24.077208 &    -    &    -              &  0.57 &  85.8 &   0.36 & A  \\
\hline
     49.a & 64.027245 & -24.068253 &  3.9680 &  3.997$\pm 0.035$ &  4.09 &   4.7 &     0  & A  \\
     49.b & 64.035187 & -24.073887 &    -    &    -              &  2.98 &   6.0 &  16.59 & A  \\
     49.c & 64.040565 & -24.078417 &    -    &    -              &  0.63 &   6.8 &  12.22 & A  \\
\hline
     50.a & 64.031136 & -24.064331 &  4.0690 &  4.039$\pm 0.033$ &  0.67 &   3.4 &     0  & A  \\
     50.b & 64.037308 & -24.069706 &    -    &    -              &  0.63 &   4.1 &  12.43 & A  \\
     50.c & 64.043991 & -24.075085 &    -    &    -              &  0.65 &   4.9 &   2.88 & A  \\
\hline
     51.a & 64.033760 & -24.065819 &  4.0707 &  4.068$\pm 0.084$ &  4.89 &  12.0 &  41.64 & A  \\
     51.b & 64.035835 & -24.067743 &    -    &    -              &  0.62 &   4.0 &  41.17 & A  \\
     51.c & 64.046516 & -24.076773 &    -    &    -              &  2.97 &   4.4 &     0  & A  \\
\hline
     52.a & 64.026611 & -24.070532 &  4.1032 &  3.947$\pm 0.002$ &  0.65 &   5.2 &   0.59 & A  \\
     52.b & 64.033760 & -24.074852 &    -    &    -              & -1.00 &   4.4 &  10.79 & A  \\
     52.c & 64.040154 & -24.080339 &    -    &    -              &  4.60 &   5.4 &     0  & A  \\
\hline
     53.a & 64.037743 & -24.061060 &  4.1138 &  4.116$\pm 0.113$ &  0.92 &   3.8 &   6.72 & A  \\
     53.b & 64.042961 & -24.063931 &    -    &    -              &  4.12 &   7.8 &   9.11 & A  \\
     53.c & 64.048248 & -24.070921 &    -    &    -              &  0.38 &   3.3 &     0  & A  \\
\hline
     54.a & 64.037834 & -24.059858 &  4.1150 &  4.105$\pm 0.091$ & -1.00 &   2.7 &     0  & A  \\
     54.b & 64.045944 & -24.065847 &    -    &    -              & -1.00 &  10.2 &   8.95 & A  \\
     54.c & 64.047768 & -24.068672 &    -    &    -              &  0.63 &   5.7 &   9.23 & A  \\
\hline
     55.a & 64.023491 & -24.076160 &  4.1218 &  3.952$\pm 0.086$ & -1.00 &   3.5 &     0  & A  \\
     55.b & 64.029312 & -24.081839 &    -    &    -              &  4.35 &   5.0 &   7.55 & A  \\
     55.c & 64.035568 & -24.084707 &    -    &    -              &  4.34 &   3.8 &   2.50 & A  \\
\hline
     56.a & 64.039207 & -24.069830 &  4.2980 &  4.207$\pm 0.002$ & 17.10 &   4.1 &  27.04 & A  \\
     56.b & 64.042809 & -24.072178 &    -    &    -              &  3.53 &   8.6 &  27.33 & A  \\
 56.c\dag & 64.030678 & -24.063269 &    -    &    -              &  0.55 &   2.6 &     0  & C  \\
\hline
     57.a & 64.036659 & -24.063314 &  4.2990 &  4.385$\pm 0.002$ &  6.36 &   8.3 &  31.55 & A  \\
     57.b & 64.039703 & -24.064278 &    -    &    -              &  4.79 &   7.5 &  30.76 & A  \\
     57.c & 64.048080 & -24.074223 &    -    &    -              &  0.41 &   3.2 &     0  & A  \\
\hline
     58.a & 64.025612 & -24.071329 &  4.5018 &  4.511$\pm 0.002$ &  4.48 &   4.2 &     0  & A  \\
     58.b & 64.033104 & -24.076052 &    -    &    -              &  0.57 &   2.1 &  11.43 & A  \\
     58.c & 64.039330 & -24.081331 &    -    &    -              & -1.00 &   4.9 &   2.02 & A  \\
\hline
     59.a & 64.024651 & -24.077213 &  4.5300 &  6.107$\pm 1.112$ &  0.23 &   6.0 &  15.25 & B  \\
     59.b & 64.028671 & -24.081221 &    -    &    -              & -1.00 &   5.4 &  14.95 & B  \\
 59.c\dag & 64.037582 & -24.085730 &    -    &    -              &  0.61 &   2.5 &     0  & C  \\
\hline
     60.a & 64.036697 & -24.068043 &  4.6090 &  4.453$\pm 0.122$ &  0.45 &   2.2 &  35.24 & A  \\
     60.b & 64.046585 & -24.076204 &    -    &    -              &  0.56 &   4.3 &     0  & A  \\
 60.c\dag & 64.033569 & -24.065058 &    -    &    -              &  4.85 &   7.5 &  35.31 & C  \\
\hline
     61.a & 64.040672 & -24.059166 &  5.0996 &  5.195$\pm 0.003$ &  0.23 &   3.5 &     0  & A  \\
     61.b & 64.046875 & -24.063866 &    -    &    -              &  4.64 &  11.2 &   2.38 & A  \\
     61.c & 64.049652 & -24.068056 &    -    &    -              &  0.30 &   4.5 &   1.63 & A  \\
\hline
     62.a & 64.029366 & -24.073355 &  5.1060 &  5.001$\pm 0.131$ &  5.15 &  68.2 &  38.54 & A  \\
     62.b & 64.030869 & -24.074207 &    -    &    -              &  5.15 &  12.3 &  37.72 & A  \\
 62.c\dag & 64.042152 & -24.082333 &    -    &    -              &  5.14 &   4.5 &     0  & B  \\
\hline
     63.a & 64.026344 & -24.068735 &  5.1060 &  5.016$\pm 0.054$ &  5.13 &   4.6 &     0  & A  \\
     63.b & 64.034988 & -24.074608 &    -    &    -              &  0.28 &   5.5 &  17.73 & A  \\
     63.c & 64.040131 & -24.079134 &    -    &    -              &  5.12 &   6.8 &  14.83 & A  \\
\hline
     64.a & 64.023064 & -24.077299 &  5.3650 &  4.874$\pm 0.147$ &  5.48 &   3.8 &     0  & A  \\
     64.b & 64.028481 & -24.083029 &    -    &    -              &  4.86 &   7.9 &   4.90 & A  \\
     64.c & 64.035126 & -24.085533 &    -    &    -              &  5.48 &   3.6 &   0.20 & A  \\
\hline
     65.a & 64.022209 & -24.077309 &  5.6380 &  5.246$\pm 0.002$ &  5.72 &   3.0 &     0  & A  \\
     65.b & 64.028748 & -24.084246 &    -    &    -              &  0.60 &   4.4 &   9.43 & A  \\
     65.c & 64.033806 & -24.085730 &    -    &    -              &  3.89 &   5.3 &   7.95 & A  \\
\hline
     66.a & 64.030907 & -24.083735 &  5.9729 &  4.348$\pm 1.517$ &  6.01 &  24.7 &     0  & A  \\
     66.b & 64.032112 & -24.084261 &    -    &    -              &  6.05 &  99.9 &   0.70 & A  \\
\hline
     67.a & 64.039940 & -24.066923 &  5.9980 &  6.033$\pm 0.390$ &  0.35 &   1.7 &   9.10 & A  \\
     67.b & 64.046043 & -24.074070 &    -    &    -              &  4.67 &   4.7 &     0  & A  \\
\hline
     68.a & 64.036499 & -24.062275 &  6.0665 &  6.107$\pm 0.208$ & -1.00 &   5.0 &  24.54 & A  \\
     68.b & 64.041115 & -24.064133 &    -    &    -              & -1.00 &   3.7 &  23.71 & A  \\
     68.c & 64.048477 & -24.073648 &    -    &    -              &  5.71 &   3.1 &     0  & A  \\
\hline
     69.a & 64.043648 & -24.059034 &  6.1452 &  6.277$\pm 0.409$ &  6.19 &   6.2 &   1.39 & A  \\
     69.b & 64.047913 & -24.062090 &    -    &    -              &  6.07 &  14.8 &   0.88 & A  \\
     69.c & 64.050941 & -24.066565 &    -    &    -              &  5.97 &   5.7 &     0  & A  \\
\hline
     70.a & 64.050858 & -24.066435 &  6.1450 &  6.060$\pm 0.422$ &  5.65 &   6.1 &     0  & A  \\
     70.b & 64.043488 & -24.058945 &    -    &    -              &  6.38 &   5.6 &   0.64 & A  \\
     70.c & 64.048248 & -24.062433 &    -    &    -              &  5.92 &  16.2 &   0.47 & A  \\
\hline
     71.a & 64.038963 & -24.060680 &  6.1470 &  6.491$\pm 0.281$ &  1.05 &   4.8 &  12.44 & A  \\
     71.b & 64.043365 & -24.062984 &    -    &    -              &  0.58 &   4.9 &  11.70 & A  \\
 71.c\dag & 64.049370 & -24.071024 &    -    &    -              & -1.00 &   3.0 &     0  & C  \\
\hline
     72.a & 64.051147 & -24.066538 &  6.1490 &  5.783$\pm 0.438$ &  1.09 &   5.4 &     0  & A  \\
     72.b & 64.047218 & -24.061197 &    -    &    -              &  1.75 &  16.7 &   1.65 & A  \\
     72.c & 64.044563 & -24.059338 &    -    &    -              & -1.00 &   9.8 &   2.27 & A  \\
\hline
     73.a & 64.041496 & -24.062630 &  6.1490 &  6.037$\pm 0.167$ &  0.58 &  20.8 &  22.24 & A  \\
     73.b & 64.040092 & -24.061878 &    -    &    -              &  6.57 &  24.4 &  22.60 & A  \\
 73.c\dag & 64.049477 & -24.072273 &    -    &    -              &  0.12 &   2.8 &     0  & C  \\
\hline
     74.a & 64.050758 & -24.065859 &  6.1490 &  6.762$\pm 2.167$ &  3.82 &   8.4 &     0  & A  \\
     74.b & 64.049301 & -24.063375 &    -    &    -              &  6.23 &  24.2 &   0.17 & A  \\
\hline
     75.a & 64.045898 & -24.060200 &  6.6290 & 11.108$\pm 7.257$ &  4.67 &  99.9 &     0  & A  \\
     75.b & 64.045601 & -24.060043 &    -    &    -              & -1.00 &  99.9 &   0.02 & A  \\
\hline
     76.a & 64.030846 & -24.067251 &  1.9900 &  2.031$\pm 0.022$ &  1.92 &   5.5 &   2.64 & A  \\
     76.b & 64.035240 & -24.071032 &    -    &    -              &  1.88 &   4.4 &  12.84 & A  \\
     76.c & 64.041916 & -24.075859 &    -    &    -              &  1.94 &   4.8 &     0  & A  \\
\hline
     77.a & 64.034386 & -24.066576 &  3.2215 &  3.246$\pm 0.031$ &  2.78 &   7.2 &  42.82 & A  \\
     77.b & 64.034546 & -24.066895 &    -    &    -              &  2.78 &  69.8 &  42.83 & A  \\
     77.c & 64.034004 & -24.066929 &    -    &    -              &  0.44 &   2.9 &  43.00 & A  \\
     77.d & 64.046150 & -24.076826 &    -    &    -              & -1.00 &   4.3 &     0  & A  \\
\hline
     78.a & 64.037956 & -24.074450 &   --    &  2.458$\pm 0.053$ &  2.75 &  56.4 &  31.96 & B  \\
     78.b & 64.038078 & -24.074560 &    -    &    -              &  1.79 &  99.9 &  31.99 & B  \\
     78.c & 64.027893 & -24.067070 &    -    &    -              &  0.97 &   3.6 &     0  & B  \\
\hline
     79.a & 64.036247 & -24.074194 &   --    &  1.524$\pm 0.002$ & -1.00 &  13.5 &  23.94 & B  \\
     79.b & 64.037827 & -24.075367 &    -    &    -              &  1.59 &  15.3 &  23.11 & B  \\
     79.c & 64.028877 & -24.069141 &    -    &    -              &  1.64 &   4.3 &     0  & C  \\
\hline
     80.a & 64.043884 & -24.072388 &   --    &  3.269$\pm 0.051$ &  2.69 &   5.4 &  17.83 & B  \\
     80.b & 64.039352 & -24.067961 &    -    &    -              &  0.56 &  17.1 &  20.66 & B  \\
     80.c & 64.032013 & -24.063265 &    -    &    -              &  2.72 &   2.8 &     0  & B  \\
\hline
     81.a & 64.032738 & -24.081612 &   --    & 11.299$\pm 0.544$ &  2.56 &   9.2 &  33.32 & B  \\
     81.b & 64.033119 & -24.081804 &    -    &    -              &  2.35 &  13.0 &  33.70 & B  \\
     81.c & 64.022079 & -24.071712 &    -    &    -              &  0.39 &   3.0 &     0  & C  \\
\hline
     82.a & 64.027115 & -24.078629 &   --    &  1.980$\pm 0.041$ &  1.91 &  99.9 &  15.26 & B  \\
     82.b & 64.027580 & -24.079121 &    -    &    -              &  1.96 &  23.4 &  15.60 & B  \\
     82.c & 64.036530 & -24.084009 &    -    &    -              &  2.12 &   3.0 &     0  & C  \\
\hline
     83.a & 64.026840 & -24.075186 &   --    & 11.669$\pm 0.004$ & -1.00 &  14.9 &  36.46 & B  \\
     83.b & 64.029076 & -24.077995 &    -    &    -              &  1.06 &   7.0 &  34.10 & B  \\
     83.c & 64.041054 & -24.084436 &    -    &    -              &  7.48 &   3.3 &     0  & C  \\
\hline
     84.a & 64.036873 & -24.071026 &   --    &  1.114$\pm 0.052$ & -1.00 &  25.6 &   0.06 & B  \\
     84.b & 64.037666 & -24.071342 &    -    &    -              &  0.97 &  99.9 &     0  & B  \\
\hline
     85.a & 64.024178 & -24.081656 &   --    &  1.870$\pm 0.030$ & -1.00 &   9.0 &     0  & B  \\
     85.b & 64.030502 & -24.085911 &    -    &    -              &  0.60 &   8.4 &   0.85 & B  \\
     85.c & 64.029388 & -24.085510 &    -    &    -              &  0.17 &  20.5 &   1.12 & C  \\
\hline
     86.a & 64.025436 & -24.080927 &   --    &  1.248$\pm 0.002$ &  0.70 &   7.1 &     0  & B  \\
     86.b & 64.026184 & -24.081728 &    -    &    -              &  1.53 &  11.7 &   0.55 & B  \\
     86.c & 64.030518 & -24.084610 &    -    &    -              &  0.64 &   8.5 &   1.68 & C  \\
\hline
     87.a & 64.028076 & -24.067270 &   --    &  2.961$\pm 0.002$ &  2.71 &   4.1 &     0  & B  \\
     87.b & 64.036201 & -24.073383 &    -    &    -              & -1.00 &   7.1 &  22.48 & B  \\
     87.c & 64.040428 & -24.076496 &    -    &    -              &  0.37 &   8.4 &  20.90 & B  \\
\hline
     88.a & 64.025635 & -24.076721 &   --    &  3.644$\pm 0.002$ & -1.00 &   7.8 &  20.22 & B  \\
     88.b & 64.028183 & -24.079634 &    -    &    -              &  2.71 &   6.2 &  20.26 & B  \\
     88.c & 64.038002 & -24.084518 &    -    &    -              &  3.39 &   2.8 &     0  & C  \\
\hline
     89.a & 64.028015 & -24.070890 &   --    &  2.964$\pm 0.002$ & -1.00 &   6.8 &  13.59 & B  \\
     89.b & 64.033173 & -24.074184 &    -    &    -              &  2.72 &   5.4 &  18.48 & B  \\
     89.c & 64.040741 & -24.080198 &    -    &    -              & -1.00 &   4.9 &     0  & B  \\
\hline
     90.a & 64.028214 & -24.070808 &   --    &  2.952$\pm 0.002$ & -1.00 &   7.2 &  15.32 & B  \\
     90.b & 64.033134 & -24.073980 &    -    &    -              &  3.06 &   5.7 &  19.51 & B  \\
     90.c & 64.040901 & -24.080122 &    -    &    -              & 13.80 &   5.0 &     0  & B  \\
\hline
     91.a & 64.027756 & -24.070908 &   --    &  2.976$\pm 0.002$ &  2.69 &   6.4 &  10.85 & B  \\
     91.b & 64.033272 & -24.074343 &    -    &    -              &  0.52 &   5.1 &  16.94 & B  \\
     91.c & 64.040520 & -24.080225 &    -    &    -              &  0.23 &   4.9 &     0  & B  \\
\hline
     92.a & 64.027428 & -24.070860 &   --    &  3.071$\pm 0.053$ &  1.91 &   5.9 &   8.14 & B  \\
     92.b & 64.040314 & -24.080204 &    -    &    -              &  0.15 &   5.0 &     0  & B  \\
\hline
     93.a & 64.045799 & -24.064848 &   --    &  1.708$\pm 0.032$ &  1.53 &  79.0 &   3.64 & B  \\
     93.b & 64.047104 & -24.066744 &    -    &    -              &  0.61 &  96.7 &   4.19 & B  \\
     93.c & 64.041931 & -24.061298 &    -    &    -              &  6.76 &   6.6 &     0  & C  \\
\hline
     94.a & 64.038376 & -24.072420 &   --    &  1.265$\pm 0.013$ & -1.00 &  99.9 &  25.70 & B  \\
     94.b & 64.038704 & -24.072590 &    -    &    -              &  1.72 &  22.0 &  25.65 & B  \\
     94.c & 64.031525 & -24.066914 &    -    &    -              &  0.18 &   4.0 &     0  & C  \\
\hline
     95.a & 64.037231 & -24.062462 &   --    &  2.359$\pm 0.022$ &  1.97 &   4.7 &   7.53 & B  \\
     95.b & 64.041443 & -24.064547 &    -    &    -              &  1.98 &   4.3 &  10.71 & B  \\
     95.c & 64.046837 & -24.071337 &    -    &    -              &  2.34 &   3.3 &     0  & B  \\
\hline
     96.a & 64.043442 & -24.067177 &   --    &  2.059$\pm 0.149$ &  1.73 &  11.8 &     0  & B  \\
     96.b & 64.044884 & -24.068945 &    -    &    -              & -1.00 &  16.3 &   0.69 & B  \\
\hline
     97.a & 64.037560 & -24.061295 &   --    &  6.731$\pm 0.002$ &  6.00 &   4.4 &  15.15 & B  \\
     97.b & 64.042343 & -24.063738 &    -    &    -              &  0.61 &   2.6 &  15.23 & B  \\
     97.c & 64.048927 & -24.071964 &    -    &    -              &  2.42 &   3.0 &     0  & B  \\
\hline
     98.a & 64.033585 & -24.071421 &   --    &  2.318$\pm 0.029$ &  2.34 &  11.5 &  24.27 & B  \\
     98.b & 64.030800 & -24.068945 &    -    &    -              & -1.00 &  10.5 &  21.63 & B  \\
     98.c & 64.042534 & -24.077715 &    -    &    -              & -1.00 &   5.0 &     0  & B  \\
\hline
     99.a & 64.040581 & -24.075516 &   --    &  2.255$\pm 0.002$ &  1.99 &   6.9 &  16.18 & B  \\
     99.b & 64.036140 & -24.072449 &    -    &    -              &  1.91 &   8.1 &  20.09 & B  \\
     99.c & 64.029343 & -24.067066 &    -    &    -              &  2.32 &   4.2 &     0  & C  \\
\hline
    100.a & 64.041115 & -24.078957 &   --    &  6.674$\pm 0.222$ &  7.40 &   7.3 &   7.39 & B  \\
    100.b & 64.034920 & -24.073774 &    -    &    -              &  7.39 &   5.4 &  13.17 & C  \\
    100.c & 64.026932 & -24.068048 &    -    &    -              &  7.39 &   5.0 &     0  & B  \\
\hline
    101.a & 64.037834 & -24.073202 &   --    &  1.262$\pm 0.016$ &  2.05 &  77.4 &  24.11 & B  \\
    101.b & 64.037445 & -24.072962 &    -    &    -              &  1.86 &  32.3 &  24.33 & B  \\
    101.c & 64.030815 & -24.067913 &    -    &    -              &  0.29 &   4.4 &     0  & C  \\
\hline
    102.a & 64.033752 & -24.066456 &   --    &  2.623$\pm 0.002$ &  0.45 &  15.8 &  34.50 & B  \\
    102.b & 64.045410 & -24.076340 &    -    &    -              &  0.04 &   4.2 &     0  & B  \\
    102.c & 64.035576 & -24.068159 &    -    &    -              &  0.63 &  17.3 &  35.38 & C  \\
\hline
    103.a & 64.029175 & -24.067690 &   --    &  1.851$\pm 0.003$ &  6.27 &   4.1 &     0  & B  \\
    103.b & 64.035934 & -24.072861 &    -    &    -              &  1.79 &   8.5 &  21.15 & B  \\
    103.c & 64.040009 & -24.075634 &    -    &    -              &  0.56 &   7.0 &  17.98 & B  \\
\hline
    104.a & 64.043694 & -24.067207 &   --    &  2.003$\pm 0.002$ & -1.00 &  15.2 &  16.98 & B  \\
    104.b & 64.044983 & -24.068687 &    -    &    -              &  0.31 &  99.9 &  17.81 & B  \\
    104.c & 64.036156 & -24.061491 &    -    &    -              &  0.88 &   2.7 &     0  & C  \\
\hline
    105.a & 64.033630 & -24.066181 &   --    &  3.207$\pm 0.052$ &  2.71 &  12.7 &  38.98 & B  \\
    105.b & 64.045914 & -24.076571 &    -    &    -              & -1.00 &   4.3 &     0  & B  \\
\hline
    106.a & 64.033600 & -24.066488 &   --    &  2.716$\pm 0.033$ &  2.68 &  14.0 &  35.31 & B  \\
    106.b & 64.035408 & -24.068256 &    -    &    -              &  0.43 &   5.2 &  35.82 & B  \\
    106.c & 64.045441 & -24.076439 &    -    &    -              &  2.66 &   4.2 &     0  & B  \\
\hline
    107.a & 64.035553 & -24.066431 &   --    &  0.907$\pm 0.002$ & -1.00 &   5.1 &     0  & B  \\
    107.b & 64.039993 & -24.070284 &    -    &    -              &  0.75 &   7.2 &  17.48 & B  \\
    107.c & 64.037453 & -24.068464 &    -    &    -              &  0.56 &   3.9 &  17.00 & C  \\
\hline
    108.a & 64.027473 & -24.077440 &   --    &  4.137$\pm 2.423$ &  0.31 &  48.6 &   0.40 & C  \\
    108.b & 64.027969 & -24.077969 &    -    &    -              &  1.74 &  13.7 &     0  & C  \\
\hline
    109.a & 64.031990 & -24.085476 &   --    &  2.173$\pm 0.033$ &  1.90 &   7.1 &   3.17 & C  \\
    109.b & 64.028961 & -24.084475 &    -    &    -              &  2.30 &   8.4 &   4.14 & C  \\
    109.c & 64.023598 & -24.079748 &    -    &    -              &  2.21 &   4.7 &     0  & C  \\
\hline
    110.a & 64.035530 & -24.067764 &   --    &  1.414$\pm 0.372$ &  0.38 &   2.9 &     0  & C  \\
    110.b & 64.035255 & -24.067951 &    -    &    -              &  0.37 &   4.9 &   0.17 & C  \\
\hline
    111.a & 64.035690 & -24.067989 &   --    &  1.460$\pm 1.112$ &  0.56 &   8.4 &   0.07 & C  \\
    111.b & 64.035553 & -24.068079 &    -    &    -              &  0.60 &  18.9 &     0  & C  \\
\hline
    112.a & 64.033035 & -24.064110 &   --    &  1.541$\pm 0.002$ &  1.74 &   3.0 &     0  & C  \\
    112.b & 64.041008 & -24.070484 &    -    &    -              &  2.36 &  17.8 &  26.88 & C  \\
    112.c & 64.041237 & -24.070604 &    -    &    -              &  1.60 &  15.4 &  26.78 & C  \\
\hline
    113.a & 64.035179 & -24.089832 &   --    & 10.697$\pm 6.500$ &  2.02 &   1.3 &   2.50 & D  \\
    113.b & 64.035690 & -24.089853 &    -    &    -              &  2.04 &   1.4 &     0  & D  \\
    113.c & 64.036652 & -24.089678 &    -    &    -              &  2.35 &   1.3 &   0.54 & D  \\
\hline
    114.a & 64.022339 & -24.082636 &   --    & 16.671$\pm 1.419$ &  0.40 &  99.9 &   0.04 & D  \\
    114.b & 64.022301 & -24.082575 &    -    &    -              &  0.45 &  99.9 &   0.10 & D  \\
    114.c & 64.022270 & -24.082546 &    -    &    -              &  0.51 &  99.9 &     0  & D  \\
\hline
    115.a & 64.044930 & -24.061071 &   --    &  2.622$\pm 0.002$ &  2.43 &  31.1 &   1.16 & B  \\
    115.b & 64.045525 & -24.061422 &    -    &    -              &  2.27 &  15.1 &   1.20 & B  \\
    115.c & 64.049713 & -24.066822 &    -    &    -              &  1.04 &   4.8 &     0  & C  \\
    115.d & 64.048317 & -24.064520 &    -    &    -              &  0.60 &  16.7 &   0.77 & C  \\
\hline
    116.a & 64.036842 & -24.077229 &   --    & 16.694$\pm 1.858$ &  3.11 &  38.2 &     0  & C  \\
    116.b & 64.037270 & -24.077747 &    -    &    -              &  0.56 &  99.9 &   0.96 & C  \\
\hline
    117.a & 64.041512 & -24.058384 &   --    & 10.599$\pm 0.002$ &  0.34 &   3.4 &     0  & C  \\
    117.b & 64.049667 & -24.064619 &    -    &    -              &  8.26 &   8.6 &   3.31 & C  \\
    117.c & 64.050392 & -24.066420 &    -    &    -              & -1.00 &   8.1 &   2.99 & C  \\
\hline
    118.a & 64.029228 & -24.075890 &   --    &  2.742$\pm 2.668$ &  4.06 &  84.5 &   0.04 & B  \\
    118.b & 64.029411 & -24.076067 &    -    &    -              &  4.06 &  54.0 &     0  & B  \\
\hline
    119.a & 64.025459 & -24.081047 &   --    &  2.678$\pm 2.489$ &  3.00 &  99.9 &   0.09 & B  \\
    119.b & 64.025673 & -24.081308 &    -    &    -              &  6.60 &  49.2 &     0  & B  \\

\end{longtable}
\twocolumn

\end{appendix}

\end{document}